\documentclass[leqno,draft,11pt,a4paper]{article}
\usepackage{amssymb,amsmath,latexsym,theorem,multirow}
\usepackage[includeheadfoot,margin=1in]{geometry}

\allowdisplaybreaks[2]

\newcommand\M{\mathit}

\newcommand\et{\mathrel\&}
\newcommand\eq{\leftrightarrow}
\newcommand\EQ{\Leftrightarrow}
\renewcommand\iff{\quad\text{iff}\quad}
\newcommand\LOR{\bigvee}
\newcommand\ET{\bigwedge}
\newcommand\TO{\Rightarrow}
\newcommand\model{\vDash}
\newcommand\nmodel{\nvDash}
\newcommand\Par{\mathrm{Par}}
\newcommand\Var{\mathrm{Var}}

\newcommand\fii{\varphi}
\newcommand\roo{\varrho}

\newcommand\p[1]{\langle#1\rangle}
\newcommand\lh[1]{\lvert#1\rvert}
\newcommand\bez{\smallsetminus}
\newcommand\sset{\subseteq}
\newcommand\nsset{\nsubseteq}
\newcommand\ssset{\subsetneq}
\newcommand\Sset{\supseteq}
\newcommand\sSset{\supsetneq}
\newcommand\onto{\twoheadrightarrow}
\newcommand\bigcupd{\mathop{\dot\bigcup}}
\newcommand\pw[1]{\mathcal P(#1)}
\newcommand\nul{\varnothing}
\newcommand\two{\mathbf2}
\newcommand\id{\mathrm{id}}

\newcommand\dia{\Diamond}
\newcommand\diadot{{\centdot\dia}}
\newcommand\centdot{\mathpalette\docentdot}
\ifx\symlasy\undefined
  
  \renewcommand\boxdot{{\origboxdot}}
  \newcommand\docentdot[2]{%
    \setbox0\hbox{$#1\mathop{#2}$}\dimen0 \ht0
    \setbox0\hbox{$#1#2$}\advance\dimen0 -\ht0
    \setbox2\hbox to\wd0{\hss$#1\mathop{\cdot}$\hss}\wd2=0pt
    \lower\dimen0\box2\box0 }
\else
  \renewcommand\boxdot{{\centdot\Box}}
  \newcommand\docentdot[2]{%
     \setbox0\hbox{$#1#2$}%
     \raise0.206\ht0\hbox to\wd0{\hss$#1\cdot$\hss}%
     \kern-\wd0 \box0 }
\fi
\newcommand\T{\mathsf T}
\newcommand\bdtr{\mathsf B}
\newcommand\reltr{\mathsf R}
\newcommand\ru{\mathrel/}
\newcommand\Ru{\Bigm/}
\newcommand\up{\mathord\uparrow}
\newcommand\down{\mathord\downarrow}
\newcommand\Up{{\setbox0\hbox{$\uparrow$}%
         \lower\dp0\hbox to\wd0{\hss\vrule width4pt height.4pt\hss}%
         \kern-\wd0\box0}}
\newcommand\Down{{\setbox0\hbox{$\downarrow$}%
         \raise\ht0\hbox to\wd0{\hss\vrule width4pt depth.4pt\hss}%
         \kern-\wd0\box0}}
\newcommand\sgen{\subseteq\mathrel\cdot\!}
\newcommand\frx{\mathrm{Fr}}
\newcommand\ECI{\EC^\infty}
\newcommand\EC{\M{EC}}
\newcommand\unifr[3][]{U_{#2}\mathopen#1(#3\mathclose#1)}

\newif\ifnadm
\DeclareRobustCommand*\adm{\nadmfalse\doadm}
\DeclareRobustCommand*\nadm{\nadmtrue\doadm}
\newcommand\doadm{\mathrel{%
   \setbox0 \hbox{$\mathop\vdash$}\dimen0 \ht0
   \setbox0 \hbox{$\vdash$}\advance\dimen0 -\ht0
   \vrule width.8\fontdimen8 \textfont3 height\ht0 depth\dp0
   \mkern-1mu
   \lower\dimen0 \hbox{$\vcenter{%
      \ifnadm
        \setbox0 \hbox{$\scriptstyle\sim\mathstrut$}%
        \hbox{\hbox to\wd0{\hss$\scriptstyle/$\hss}\kern-\wd0 \box0 }%
      \else
        \hbox{$\scriptstyle\sim\mathstrut$}%
      \fi}$}}}

\newcommand\extmathpalette[1]{{\mathchoice
     {#1\displaystyle\scriptstyle}%
     {#1\textstyle\scriptstyle}%
     {#1\scriptstyle\scriptscriptstyle}%
     {#1\scriptscriptstyle\scriptscriptstyle}}}
\newcommand\nr[1]{\extmathpalette{\nrstyle{#1}}}
\newcommand\nrstyle[3]{%
  \setbox0\hbox{$#2\bigcirc$}%
  \vcenter{\hbox to\wd0{\hss$#3#1$\hss}}%
  \kern-\wd0\box0 }

\DeclareMathOperator\dom{dom}
\DeclareMathOperator\cls{cl}
\DeclareMathOperator\Mod{Mod}
\DeclareMathOperator\Form{Form}
\DeclareMathOperator\Ext{Ext}
\DeclareMathOperator\NExt{NExt}
\DeclareMathOperator\rcl{rcl}
\DeclareMathOperator\tp{tp}
\DeclareMathOperator\base{bas}
\DeclareMathOperator\Sub{Sub}
\DeclareMathOperator\sat{Sat}
\DeclareMathOperator\refl{refl}

\newcommand\cxt[1]{\mathrm{#1}}
\newcommand\np{\cxt{NP}}
\newcommand\conp{\cxt{coNP}}
\newcommand\ptime{\cxt P}
\newcommand\EXP{\cxt{EXP}}
\newcommand\NEXP{\cxt{NEXP}}
\newcommand\coNEXP{\cxt{coNEXP}}
\DeclareMathOperator\poly{poly}
\newcommand\psp{\cxt{PSPACE}}
\newcommand\Ac{\cxt{AC}^0}
\newcommand\task[1]{{\normalfont\textsc{#1}}}

\newcommand\lgc[1]{\mathbf{#1}}
\newcommand\CPC{\lgc{CPC}}
\newcommand\IPC{\lgc{IPC}}
\newcommand\kiv{\lgc{K4}}
\newcommand\I{{\bullet}}
\newcommand\R{{\circ}}
\newcommand\RI{{*}}
\newcommand\DP{\mathrm{DP}}

\newcommand\ob[1]{\overline{#1}}
\newcommand\txto{${}\to{}$}
\makeatletter
\newcommand\iddots{\mathinner{\mkern1mu\raise\p@\hbox{.}\mkern2mu
        \raise4\p@\hbox{.}\mkern2mu\raise7\p@\vbox{\kern7\p@\hbox{.}}\mkern1mu}}
\makeatother

\def\bme{\hskip.75em\relax}
\def\noproof{\leavevmode\unskip\bme\vadjust{}\nobreak\hfill$\qed$\par}
\newcommand\qed{\Box}
\newenvironment{Pf}
  {\par\noindent\textit{Proof:}\bme\ignorespaces}
  {\noproof\pagebreak[2]\vskip\medskipamount\ignorespacesafterend}
\def\qedhere{\relax\ifmmode\eqno\qed\expandafter\aftergroup
                   \else\noproof\fi\noqed}
\def\noqed{\let\noproof\relax}

\theoremstyle{plain}
\newtheorem{Thm}{Theorem}[section]
\newtheorem{Cor}[Thm]{Corollary}
\newtheorem{Lem}[Thm]{Lemma}
\newtheorem{Obs}[Thm]{Observation}
\newtheorem{Cl}{Claim}[Thm]

\theorembodyfont\upshape
\newtheorem{Def}[Thm]{Definition}
\newtheorem{Rem}[Thm]{Remark}
\newtheorem{Exm}[Thm]{Example}
\newenvironment{Pf*}{\let\qed\qedCl\Pf}\endPf

\def\allowhyphens{\nobreak\hskip0pt\relax}
\DeclareRobustCommand*\magiclparen{\ifmmode(\else\textup(\allowhyphens\fi}
\DeclareRobustCommand*\magicrparen{\ifmmode)\else\textup)\fi}
\let\lparen=(  \let\rparen=)
\def\magicparon{\catcode`\(\active\catcode`\)\active}
\def\magicparoff{\catcode`\(12 \catcode`\)12 }
\magicparon
\let (=\magiclparen  \let )=\magicrparen

\newcommand\citi[1]{\ifx\relax#1\relax\dociti{}\else\dociti{[#1]}\fi}
\newcommand\dociti[1]{\cite#1{ej:modparami}}

\author{Emil Je\v r\'abek\\[\medskipamount]
The Czech Academy of Sciences, Institute of Mathematics\\
\small \v Zitn\'a 25,
115\:67 Praha 1,
Czech Republic,
email: \texttt{jerabek@math.cas.cz}
}

\title{Rules with parameters in modal logic~II}

\begin{document}
\maketitle

\begin{abstract}
We analyze the computational complexity of admissibility and unifiability with parameters in
transitive modal logics. The class of cluster-extensible (clx) logics was introduced in the first part of this
series of papers~\citi{}. We completely classify the complexity of unifiability or inadmissibility in any clx logic as
being complete for one of $\Sigma^{\exp}_2$, $\NEXP$, $\coNEXP$, $\psp$, or $\Pi^p_2$. In addition to the main case where
arbitrary parameters are allowed, we consider restricted problems with the number of parameters bounded by a
constant, and the parameter-free case.

Our upper bounds are specific to clx logics, but we also include similar results for logics of bounded depth and width.
In contrast, our lower bounds are very general: they apply each to a class of all transitive logics whose frames allow
occurrence of certain finite subframes.

We also discuss the baseline problem of complexity of derivability: it is $\conp$-complete or $\psp$-complete for each
clx logic. In particular, we prove $\psp$-hardness of derivability for a broad class of transitive logics that includes
all logics with the disjunction property.


\end{abstract}

\section{Introduction}\label{sec:introduction}

In this paper, we continue the analysis of admissible rules with parameters (constants) in transitive modal logics
satisfying certain extension properties started in Je\v r\'abek~\citi{}. We recall that the first part was devoted to
structural results. We introduced the class of \emph{cluster-extensible (clx)} logics, encompassing the most common
transitive modal logics such as $\kiv$, $\lgc{S4}$, $\lgc{GL}$, $\lgc{S4Grz}$, $\lgc{S4.3}$, and many others. We proved
that in the setting of rules with parameters, all formulas have projective approximations in any clx logic~$L$, whence
$L$-admissibility is decidable, and we can compute finite complete sets of $L$-unifiers to any given formula. We
provided semantic characterization of $L$-admissibility in terms of certain classes of frames (called $L$-extensible),
and axiomatizations of $L$-admissible rules by explicit bases.

The topic of this second part is the \emph{computational complexity} of admissibility with parameters, and of the closely
related problem of unifiability, in clx and other transitive logics. We mention that the complexity of admissibility in
transitive modal logics was previously studied in~\cite{ej:admcomp}: the main results were that admissibility in
certain logics (called extensible) is either $\conp$-complete or $\coNEXP$-complete depending on if the logic is linear
(i.e., of width~$1$) or not; the lower bound, stating that admissibility is $\coNEXP$-hard, was proved under a weak
hypothesis applicable to a larger class of logics. The class of extensible logics of~\cite{ej:admcomp} is
incomparable with the clx logics of~\citi{}: on the one hand, the condition of extensibility only constraints frames
with a one-element root cluster, hence in this sense it is less restrictive; on the other hand, the definition of
extensibility in~\cite{ej:admcomp} did not accommodate nonlinear logics of bounded branching. However, the principal
difference is that \cite{ej:admcomp} only considered admissibility \emph{without parameters}.

As we will see, the introduction of parameters leads to a richer and more complicated landscape: we will encounter
several more complexity classes than just $\conp$ and $\coNEXP$, and while linearity of the logic will remain an
important dividing line, the complexity of the problem will also be influenced by other factors, the most important
being if the logic allows clusters of unbounded size.

On the other hand, the usage of parameters makes our results on complexity more robust, and simpler to formulate. This
is most clearly seen for lower bounds: first, they apply already to the special case of unifiability rather than to the
full (in)admissibility problem, and second, they only require weak and easily checkable assumptions on the logics, such
as being of width $\ge2$, or having unbounded cluster size. (In contrast, the parameter-free $\coNEXP$ lower bound
from~\cite{ej:admcomp} only applies to admissibility, and relies on a peculiar extensibility condition on the logic.)
In general, our lower bounds will have the form that $L$-unifiability is $\mathcal C$-hard (for a particular complexity
class $\mathcal C$) whenever certain finite frames may be embedded into $L$-frames, or more precisely, when there exist
$L$-frames that subreduce or weakly subreduce to said finite frames; see below for definitions.

Wherever possible, we also include results on the complexity of restricted unifiability or admissibility
problems in which only a constant number of parameters are allowed, though in this case the conditions on logics get
more complicated, and do not seem optimal.

In contrast to lower bounds, decent upper bounds can be proved only for well-behaved logics, as random transitive
modal logics may be quite wild, with already the set of tautologies being of high complexity or even undecidable.
We are primarily interested in the case of clx logics, and one of the goals of this paper is to find a complete
classification of the complexity of unifiability and admissibility in clx logics, both in the regime with arbitrary
parameters, and with only constantly many parameters.

Additionally, we present upper bounds on the complexity of admissibility and unifiability in \emph{logics of bounded
depth and width}. To this end, we need first to prove a
few structural results on admissibility in logics of bounded depth---in particular, semantical characterizations---since
such logics are not covered by the framework from~\citi{}.

Similar to clx logics, logics of bounded depth and width are tame and well behaved, which allows us to prove interesting
general results about their complexity. However, the class of these logics is structurally quite different from the
class of clx logics. Clx logics are few and far between, they are particularly nice logics cherry-picked from the
lattice $\NExt\kiv$ of all transitive logics, both weak and strong. In contrast, logics of bounded depth and width form
an \emph{ideal} in $\NExt\kiv$: in particular, all extensions of any logic from this class are also in the class.
Thus, logics of bounded depth and width may be construed as a toy model that is representative of all
well-behaved logics, whereas it may easily happen that there are classes of nice logics whose properties are quite
different from clx logics. In fact, as we will see in the sequel, there \emph{are} reasonable classes of nice non-clx
logics, including logics with a single top cluster such as $\lgc{S4.2}$, and some of these have different complexity of
unifiability and admissibility from what we will encounter in this paper. These cases duly manifest already for logics
of bounded depth and width, and for this reason we cannot give a complete classification of complexity of
unifiability in these logics in the present paper.

We acknowledge that this guiding principle should not be taken too seriously: for example, we
will also see that there is an interesting class of logics for which unifiability and admissibility are $\psp$-complete,
which cannot happen for logics of bounded depth and width, and likewise for all nontrivial results concerning the
setting with constantly many parameters.

The main focus of the paper is on the complexity of admissibility and unifiability, but as a starting point, we also
settle the complexity of \emph{derivability} in clx logics. Generalizing results of Ladner~\cite{ladner}, we show that
nonlinear clx logics are $\psp$-complete. For the lower bound, we prove $\psp$-hardness for a broad class of transitive
logics that includes all logics with the disjunction property, and many other logics such as $\lgc{S4.2}$: similar to
our lower bounds on unifiability, the result applies to all transitive logics $L$ such that, roughly speaking, all
finite trees may be embedded into $L$-frames.

In the planned third part of this series of papers, we will adapt the set-up of clx logics to logics with a single top
cluster such as $\lgc{K4.2}$ and~$\lgc{S4.2}$, including both structural results as in~\citi{}, and computation
complexity results as here. Among other things, this will allow us to complete the classification of the complexity of
unifiability and admissibility for logics of bounded depth and width, and to obtain a complete classification of
\emph{hereditary} properties of transitive logics that guarantee hardness of unifiability for some complexity
class. Both of these suggest that our results on complexity are in a certain sense optimal. We also plan to adapt our
results to superintuitionistic logics.

\subsection{Overview of results}\label{sec:results}

\begin{table}
\centering
\renewcommand\arraystretch{1.2}
\begin{tabular}{|c|c||c|c|c|c||c|}
\hline
\multicolumn{2}{|c||}{logic}&\multirow{3}{*}{$\nvdash_L$}&\multicolumn{3}{c||}{unifiability, $\nadm_L$}&\multirow{3}{*}{examples}\\
\cline{1-2}\cline{4-6}
\multirow{2}{*}{branching}&cluster&&\multicolumn{3}{c||}{parameters:}&\\ 
&size&&no&$O(1)$&any&\\
\hline\hline
\multirow{2}{*}{$0$}&$<\infty$&\multicolumn{3}{c|}{}&$\Pi_2^p$&$\lgc{S5}\oplus\lgc{Alt}_k$, $\lgc{Verum}$\\
\cline{2-2}\cline{6-7}
&$\infty$&\multicolumn{2}{c}{\multirow{2}{*}{$\np$}}&&$\coNEXP$&$\lgc{S5}$, $\lgc{K4B}$\\
\cline{1-2}\cline{5-7}
\multirow{2}{*}{$1$}&$<\infty$&\multicolumn{2}{c|}{}&\multicolumn{2}{l||}{$\psp$}&$\lgc{GL.3}$, $\lgc{S4Grz.3}$\\
\cline{2-2}\cline{6-7}
&$\infty$&\multicolumn{2}{c|}{}&&$\coNEXP$&$\lgc{S4.3}$, $\lgc{K4.3}$\\
\hline
\multirow{2}{*}{$\ge2$}&$<\infty$&\multirow{2}{*}{$\psp$}&\multicolumn{2}{c}{\multirow{2}{*}{$\NEXP$}}&&$\lgc{GL}$,
$\lgc{K4Grz}$, $\lgc{S4Grz}$\\
\cline{2-2}\cline{6-7}
&$\infty$&&\multicolumn{2}{c|}{}&$\Sigma_2^{\exp}$&$\kiv$, $\lgc{S4}$, $\lgc{K4BB}_k$\\
\hline
\end{tabular}
\caption{Complexity of nonderivability, unifiability, and inadmissibility problems for clx logics. Note: results in
the parameter-free column only apply to inadmissibility, not unifiability.}
\label{tab:clx}
\end{table}

The linear organization of the paper is as follows. Section~\ref{sec:preliminaries} contains a few preliminary
definitions and facts; its Subsection~\ref{sec:complexity-classes} reviews the needed background in complexity theory,
including a proof of completeness of certain problems for levels of the exponential hierarchy.

Section~\ref{sec:derivability} deals with the complexity of derivability.

Section~\ref{sec:upper-bounds} is devoted to upper bounds on the complexity of admissibility and unifiability in
certain logics: nonlinear and linear clx logics in Subsections \ref{sec:nonl-clust-extens}
and~\ref{sec:line-clust-extens} (respectively), and logics of bounded depth (and, apart from structural results,
bounded width) in Subsection~\ref{sec:logics-bounded-depth}.

Section~\ref{sec:lower-bounds} presents lower bounds on the complexity of unifiability and admissibility. Hardness
results for levels of the exponential hierarchy ($\NEXP$, $\coNEXP$, $\Sigma^{\exp}_2$) appear in
Subsection~\ref{sec:expon-hier}, except for results in the setting with a constant number of parameters, which are more
complicated and are relegated to Subsection~\ref{sec:nexp-const-param}. Hardness results for $\psp$ and levels of
the polynomial hierarchy are in Subsection~\ref{sec:psp-below}.

Section~\ref{sec:conclusion} concludes the paper.

A summary of results on the complexity of derivability, unifiability, and admissibility in consistent clx
logics is given in Table~\ref{tab:clx}. Each entry in the table should be understood so that the stated problem is
\emph{complete} for the complexity class indicated, for every clx logic that meets the description.
Notice that for a given logic, unifiability and admissibility usually have dual
complexity, because unifiability is a special case of inadmissibility; in order to avoid confusing switching between
dual complexity classes, we adopt in this overview the convention to indicate the complexity of \emph{unifiability} and
\emph{inadmissibility} rather than admissibility. In accordance with this, we also indicate the complexity of
\emph{nonderivability} (or equivalently: local satisfiability) rather than derivability. However, detailed statements
of theorems later in the paper will often mention both.

The referee expressed surprise that some of these classes are existential and some universal. On the most basic level,
unifiability and inadmissibility are existential properties as they amount to the existence of certain substitutions,
but this is an unbounded existential quantifier, and even though our results also imply computable bounds, this usually
makes a quite inefficient test. We rely instead on semantic criteria characterizing inadmissibility by the existence of
certain finite models, and depending on properties of the logic, different quantifiers in this characterization become
dominant.

Roughly speaking, in the $\Sigma^{\exp}_2$~case, the outer~$\exists$ quantifies over a model, and the inner~$\forall$
comes from verification of pseudoextensibility of the model, quantifying over valuations of parameters in a cluster. If
cluster size or the number of parameters is bounded, the $\forall$ becomes polynomial-size and disappears,
leading to~$\NEXP$. If the logic is linear, the model is an upside-down tree of clusters, and rather than fixing it
beforehand, we can explore it one branch at a time; this leads to~$\coNEXP$ as the outer~$\exists$ disappears. If both
restrictions are met, we are left with only polynomial-size quantifiers: basically, alternating $\forall$~over
parameter valuations with $\exists$~over variable valuations. This gives $\psp$; furthermore, if the logic has
bounded depth, the number of alternations is likewise bounded, leading to~$\Pi^p_{2d}$.

Here is a cross-reference of our complexity results sorted according to complexity classes:
\begin{itemize}
\item $\Sigma^{\exp}_2$
\begin{itemize}
\item Upper bounds: inadmissibility in clx logics (Thm.~\ref{thm:ub-clx})
and in logics of bounded depth and width (Thm.~\ref{thm:ub-bddp-bdwd}).
\item Lower bound: unifiability in logics with frames weakly subreducing to $(\R+\nr n)^\R$ (Thm.~\ref{thm:lb-sig2exp}).
\end{itemize}
\item $\NEXP$
\begin{itemize}
\item Upper bounds:
inadmissibility in clx logics with bounded cluster size or bounded number of parameters (Thm.~\ref{thm:ub-clx}),
inadmissibility in tabular logics (Thm.~\ref{thm:ub-tab}).
\item Lower bounds: unifiability in nonlinear logics (Thm.~\ref{thm:lb-nexp}),
unifiability with $O(1)$ parameters in certain logics (Thms.~\ref{thm:nexp-1par}, \ref{thm:nexp-2par}, \ref{thm:nexp-0par}, Cor.~\ref{cor:nexp-1par-dp}),
inadmissibility without parameters in certain logics (Thm.~\ref{thm:nexp-0par-adm}).
\end{itemize}
\item $\coNEXP$
\begin{itemize}
\item Upper bounds: inadmissibility in linear clx logics (Thm.~\ref{thm:ub-lin-clx})
and in linear logics of bounded depth (Thm.~\ref{thm:ub-lin-bddp}).
\item Lower bound: unifiability in logics of unbounded cluster size (Thm.~\ref{thm:lb-conexp}).
\end{itemize}
\item $\psp$
\begin{itemize}
\item Upper bounds:
(non)derivability in clx logics (Thm.~\ref{thm:ub-der-clx}),
(in)admissibility in linear clx logics with bounded cluster size or with $O(1)$ parameters (Thm.~\ref{thm:ub-lin-clx}).
\item Lower bounds:
(non)derivability in logics subframe-universal for trees (Thm.~\ref{thm:lb-der-transl}, cf.\ Cor.~\ref{thm:dp-csf-univ}),
unifiability with $O(1)$ parameters in logics of unbounded depth (Thms.\ \ref{thm:lb-infdp}, \ref{thm:lb-infirrdp}).
\end{itemize}
\item $\Pi^p_{2d}$
\begin{itemize}
\item Upper bound: inadmissibility in linear tabular logics (Thm.~\ref{thm:ub-lin-bddp}, cf.\ Cor.~\ref{cor:ub-clx-br0-bdcl}).
\item Lower bound: unifiability in logics of depth $\ge d$ (Thm.~\ref{thm:lb-infdp}).
\end{itemize}
\item $\np$
\begin{itemize}
\item Upper bounds:
nonderivability in consistent linear clx logics (Thm.~\ref{thm:der-lin-clx}),
unifiability with $O(1)$ parameters in logics of bounded depth (Thm.~\ref{thm:ub-unif-bdpar-bddp}),
inadmissibility with $O(1)$ parameters in logics of bounded depth and width (Thm.~\ref{thm:ub-bdpar-bddp-bdwd}),
inadmissibility without parameters in consistent linear clx logics \cite[Thm.~2.6]{ej:admcomp}.
\item Lower bound: unifiability without parameters in consistent logics (Thm.~\ref{thm:lb-np}).
\end{itemize}
\end{itemize}

\section{Preliminaries}\label{sec:preliminaries}

This paper is a continuation of \citi{}, and we assume the reader has access to that paper. We generally
follow the same terminology and notational conventions as in~\citi{}, which we shall not repeat here, as it would
considerably add to the length of the paper. In particular, we assume the reader is familiar with the content of
\citi{\S2}, which lays out our basic concepts and terminology. Moreover, we will rely on the definition of clx
logics (Def.~4.1) and their main structural properties (\S4.3); the definition of tight predecessors and
extensible frames (Def.~5.1) along with the ensuing semantical characterization of admissibility in clx logics
(Thm.~5.18); and the characterization of admissibility in terms of pseudoextensible models (Def.~5.21,
Thm.~5.24).

We stress that unless stated otherwise, admissibility refers to admissibility with arbitrary parameters, and similarly
for unification and other related notions.

All logics in this paper are normal modal logics extending~$\kiv$, which we will not always state explicitly.

As a piece of a more obscure notation, we recall from~\citi{} that $\two=\{0,1\}$; if $P$ is a finite set of formulas
(in this paper, typically $P\sset\Par$), then $\two^P$ is the set of all assignments $t\colon P\to\two$, and for any
$t\in\two^P$, we put $P^t=\ET_{\fii\in P}\fii^{t(\fii)}$, where $\fii^1=\fii$, $\fii^0=\neg\fii$.

If $n\in\omega$ is a natural number (which we take to include~$0$), we will sometimes denote the set
$\{0,\dots,n-1\}=\{i:i<n\}$ as just~$n$. (This in fact agrees with the common von Neumann \emph{definition} of natural
numbers.) Given an indexed sequence $P=\{\fii_i:i<n\}$ and $t\in\two^n$, we extend the notation above so that
$P^t=\ET_{i<n}\fii_i^{t(i)}$. Finally, we will use this notation also for subsets $t\sset n$, that are identified in
this context with their characteristic functions: that is, $P^t=\ET_{i<n}\fii_i^{[i\in t]}$. Here, $[\dots]$ is the
\emph{Iverson bracket}: for any predicate $R$,
\[[R]=\begin{cases}1&\text{if $R$ is true,}\\0&\text{otherwise.}\end{cases}\]
The set of all subsets of~$n$ is denoted $\pw n$.

In accordance with~\citi{\S5.1}, we define a logic~$L$ to have the \emph{disjunction property} if it admits the rules
\begin{equation}
\tag{$\DP_n$}\LOR_{i<n}\Box x_i\Ru\{x_i:i<n\}
\end{equation}
for all $n\in\omega$, or equivalently, for $n=0,2$ (the admissibility of $\DP_0$ being equivalent to the
consistency of~$L$).

If $C$ is a cluster or a point in a model, a cluster type, or an extension condition, let $\refl(C)\in\{\I,\R\}$
denote its \emph{reflexivity}: $\refl(C)=\I$ if $C$ is irreflexive, and $\refl(C)=\R$ if it is reflexive.

Recall that a frame $\p{V,{\prec},B}$ is a \emph{generated subframe} of $\p{W,{<},A}$ (written as $V\sgen W$) if
$V\sset W$, ${\prec}={<}\cap(V\times W)$ (which implies $V\up\sset V$), and $B=\{X\cap V:X\in A\}$. The \emph{disjoint
sum} of frames $\p{W_i,<_i,A_i}$ ($i\in I$), denoted $\sum_{i\in I}W_i$, is the frame $\p{W,<,A}$ whose underlying set
$W$ is the disjoint union $\bigcupd_{i\in I}W_i$, ${<}=\bigcup_{i\in I}<_i$, and
$A=\{X\sset W:\forall i\in I\,(X\cap W_i\in A_i)\}$. A \emph{subreduction} of $\p{W,{<},A}$ to $\p{V,{\prec},B}$
(see~\cite[\S9.1]{cha-zax}) is a partial map $f\colon W\to V$ such that
\begin{enumerate}
\item\label{item:18} $Y\in B\implies f^{-1}[Y]\in A$,
\item\label{item:19} $w<w'\implies f(w)\prec f(w')$, and
\item\label{item:20} $f(w)\prec v\implies\exists w'>w\,f(w')=v$
\end{enumerate}
for all $w,w'\in W$, $v\in V$, and $Y\sset V$. If $V$ is a (Kripke) finite frame, condition~\ref{item:18} simplifies to
\begin{enumerate}
\item[(i${}'$)] $f^{-1}[v]\in A$ for all $v\in V$.
\end{enumerate}
A subreduction is \emph{cofinal} if $\dom(f)\up\sset\dom(f)\Down$. A subreduction of $W$ \emph{onto~$V$} is
is a subreduction whose image is all of~$V$. A frame $W$ \emph{(cofinally) subreduces} onto~$V$ if there exists a
(cofinal, resp.) subreduction of $W$ onto~$V$.

We will often encounter conditions concerning the occurrence of certain patterns as subframes that are oblivious to the
reflexivity of individual points. In order to facilitate their formulation, we define a \emph{weak subreduction} of $W$
to~$V$ to be a partial map $f\colon W\to V$ that satisfies \ref{item:18}, \ref{item:19}, and the modified condition
\begin{enumerate}
\item[(iii${}'$)] $f(w)\prec v\implies\exists w'\ge w\,f(w')=v$
\end{enumerate}
for all $w\in W$ and $v\in V$. We define other derived notions such as weak subreductions onto~$V$ and cofinal weak
subreductions similarly as above.

The \emph{reflexivization} of a frame $\p{W,{<},A}$, denoted $W_\R$, is the frame $\p{W,{\le},A}$. The following
is an immediate consequence of the definitions:
\begin{Obs}\label{obs:weaksubr}
Let $\p{W,{<},A}$ and $\p{V,{\prec},B}$ be frames, and $f$ a partial map from $W$ to~$V$.
\begin{itemize}
\item If $f$ is a weak subreduction of $W$ to~$V$, it is a subreduction of $W_\R$ to~$V_\R$.
\item If $f$ is a subreduction of $W_\R$ to~$V$, it is a weak subreduction of $W$ to~$V$.\qedhere
\end{itemize}
\end{Obs}

If $L\Sset\kiv$ is a logic with the finite model property, and $P$ a finite set of parameters, let $\unifr LP$ denote
the \emph{universal $L$-frame for parameters~$P$}: it is defined as the upper part of the universal $L$-frame of
rank~$\lh P$ (denoted $\mathcal F_L^{<\infty}(\lh P)$ in \cite[\S8.7]{cha-zax}), endowed with the canonical valuation
of parameters from~$P$ to make it a parametric frame. The double dual of $\unifr LP$ is the canonical frame
$C_L(P,\nul)$ (\cite[Cor.~8.89]{cha-zax}); in particular, if $\fii\in\Form(P,\nul)$ is unprovable in~$L$, then
$u\nmodel\fii$ for some $u\in\unifr LP$.

Explicitly, we may construct $\unifr\kiv P$ as the union
$\bigcup_{n\in\omega}W_n$ of the following inductively defined chain of finite parametric frames:
\begin{itemize}
\item We start with $W_0=\nul$.
\item The frame $W_{n+1}$ includes~$W_n$ as a generated subframe. Moreover, for every $X\sgen W_n$, $\RI\in\{\I,\R\}$,
and $\nul\ne E\sset\two^P$, where $\lh E=1$ if $\RI=\I$: if $X$ has no tight $\p{\RI,E}$-predecessor in~$W_n$, we
include one in~$W_{n+1}$.
\end{itemize}
Then $\unifr LP\sgen\unifr\kiv P$ consists of all points $u\in\unifr\kiv P$ such that $u\Up$ is an $L$-frame. We stress
that if $X\sgen W_n$ is a rooted subframe with a reflexive root cluster~$C$ such that all assignments from~$E$ are
realized in~$C$, then no tight $\p{\RI,E}$-predecessor of~$X$ is added to~$W_{n+1}$, as it already has
one (viz., a subset of~$C$).

If $L$ is a clx logic, we may also describe $\unifr LP$ in our terminology as the minimal locally finite
$\base(L)$-extensible parametric Kripke frame for the parameters~$P$.

It is well known that every point of~$\unifr LP$ is definable by a formula (see e.g.~\cite[Thm.~8.83]{cha-zax}). We
will need an explicit description of such formulas in the simplest case $P=\nul$. Notice that in this case, the
universal frame has no proper clusters.

For each $u\in\unifr\kiv\nul$, we define a formula $\beta_u$ by induction on the
depth of~$u$ as
\begin{equation}\label{eq:60}
\beta_u=\begin{cases}
\displaystyle\ET_{v>u}\dia\beta_v\land\Box\LOR_{v>u}\beta_v,&\refl(u)=\I,\\[1.5\bigskipamount]
\displaystyle\dia\Bigl(\ET_{v\gneq u}\dia\beta_v\Bigr)
     \land\Box\Bigl(\LOR_{v\gneq u}\beta_v\lor\dia\ET_{v\gneq u}\dia\beta_v\Bigr),&\refl(u)=\R.
\end{cases}
\end{equation}
\begin{Lem}\label{lem:unifr-flas}
Let $L\Sset\kiv$ have fmp, $u,u'\in\unifr L\nul$, and $\fii\in\Form(\nul,\nul)$.
\begin{alignat}{2}
\label{eq:58} u'&\model\beta_u&&\iff u'=u,\\
\label{eq:59} u&\model\fii&&\iff\vdash_L\beta_u\to\fii.
\end{alignat}
\end{Lem}
\begin{Pf}
We prove~\eqref{eq:58} by induction on the depth of~$u$. The right-to-left implication amounts to $u\model\beta_u$,
which is straightforward to check.

Assume that $u'\model\beta_u$. If $u$ is irreflexive, the definition of~$\beta_u$
and the induction hypothesis imply $u'\up=u\up$. Notice that there is no $v\in u\up$ such that $v\up=u\up$: this would
make $u\up$ rooted with a reflexive root~$v$, hence no additional tight predecessor~$u$ would be added to
$\unifr L\nul$. Thus, $u'\notin u\up$, hence $u'$ is irreflexive, i.e., it is the unique irreflexive tight predecessor
of~$u\up$: $u'=u$.

If $u$ is reflexive, we have $X:=u\up\bez\{u\}\sset u'\up$, which implies $u'\notin X$ as above. Assuming
$u'\ne u$ for contradiction, let $w\ge u'$ be maximal such that $w\notin u\up$. By maximality, $w\up\bez\{w\}\sset
u\up$. Since $u'\model\beta_u$, we have $w\model\dia\ET_{v\in X}\dia\beta_v$, i.e., there is $w'>w$ such that
$X\sset w'\up$. This again implies $w'\notin X$, hence $w'=u$ or $w'=w$ by maximality, i.e.,
$w<u$ or $w$ is reflexive. If $w<u$, then $w\up\bez\{w\}=u\up$, i.e., $w$ is a tight predecessor of~$u\up$ other
than $u$~itself, but no such tight predecessor was added into~$\unifr L\nul$. Otherwise $w\up\bez\{w\}=X$ and $w$ is
reflexive, i.e., $w$ is a reflexive tight predecessor of $X$ other than $u$, which is again
impossible.

\eqref{eq:59}: The right-to-left implication follows from $u\model\beta_u$. Left-to-right: assume that
$\nvdash_L\beta_u\to\fii$. Since $(\beta_u\to\fii)\in\Form(\nul,\nul)$, this implies $u'\nmodel\beta_u\to\fii$ for some
$u'\in\unifr L\nul$ by universality. We must have $u'=u$ by~\eqref{eq:58}, thus $u\nmodel\fii$.
\end{Pf}

\subsection{Complexity classes}\label{sec:complexity-classes}

Since the topic of this paper is computational complexity, we will assume some degree of familiarity with basic
computation models and complexity classes. We refer the reader to e.g.\ Arora and Barak~\cite{aro-bar} for general
background on complexity theory, but for convenience, we review the definitions of classes that appear in this paper.

For any function $t\colon\omega\to\omega$, let $\cxt{DTIME}(t)$ (or $\cxt{DTIME}\bigl(t(n)\bigr)$ for emphasis) denote
the class of all languages $L\sset\Sigma^*$ (where $\Sigma$ is a finite alphabet) computable by a \emph{deterministic
(multitape) Turing machine (DTM)} in time at most~$t(n)$, where $n$ is the length of input. If $F$ is a family of such
functions, such as the family of all polynomially bounded functions $\poly(n)=n^{O(1)}$, we put
$\cxt{DTIME}(F)=\bigcup_{t\in F}\cxt{DTIME}(t)$. The \emph{polynomial time} and \emph{exponential time} classes are
then defined as
\begin{align*}
\ptime&=\cxt{DTIME}\bigl(\poly(n)\bigr),\\
\EXP&=\cxt{DTIME}\bigl(2^{\poly(n)}\bigr).
\end{align*}
(This is exponential time with polynomial exponent; exponential classes with linear exponent, such as $\cxt
E=\cxt{DTIME}\bigl(2^{O(n)}\bigr)$, will not be used in this paper.) Likewise, $\cxt{DSPACE}\bigl(s(n)\bigr)$ is the
class of languages computable by a DTM using $s(n)$ cells of memory, and
\[\psp=\cxt{DSPACE}\bigl(\poly(n)\bigr).\]
For polynomial-space languages it does not matter, but recall that in general, space usage is defined so that it only
accounts for the content of work tapes, excluding the input tape (which is assumed read-only), and---if
we are computing a function rather than language membership---excluding the output tape (which is write-only).

In particular, our basic notion of reduction will be \emph{many-one logarithmic-space (logspace) reductions}. If $X$ is
a complexity class, a language $L_0$ is \emph{$X$-hard} if every language $L\in X$ is logspace-reducible to~$L_0$, and
$L_0$ is \emph{$X$-complete} if additionally $L_0\in X$.

\emph{Nondeterministic Turing machines (NTM)} may have a choice between multiple possible transitions in any
configuration; the machine is declared to accept a given input if there \emph{exists} a run of the machine that ends in
an accepting state. The class of languages accepted by a NTM in time~$t(n)$ is denoted $\cxt{NTIME}(t)$, and we put
\begin{align*}
\np&=\cxt{NTIME}\bigl(\poly(n)\bigr),\\
\NEXP&=\cxt{NTIME}\bigl(2^{\poly(n)}\bigr).
\end{align*}
An equivalent definition of $\np$ is that it consists of languages $L$ such that membership in~$L$ can be
\emph{witnessed} by a polynomial-size certificate whose validity can be checked in~$\ptime$.

We could also define nondeterministic space classes, but the only example we are interested in is
$\cxt{NPSPACE}$, which equals $\psp$ by Savitch's theorem.

For any class $X$ of languages $L\sset\Sigma^*$, $\cxt{co}X$ denotes the dual class $\{\Sigma^*\bez L:L\in X\}$.

The deterministic and nondeterministic time classes above can be relativized: for any language~$A$, a \emph{Turing
machine with oracle~$A$} may query membership of words in~$A$ (by writing them on a dedicated oracle query tape) at
unit time cost. Then for any class~$X$, $\np^X$ denotes the set of languages computable in polynomial-time by a
NTM with oracle $A\in X$, and similarly for other classes. The \emph{polynomial} and
\emph{exponential hierarchies} are defined by $\Sigma^p_0=\Delta^p_0=\Pi^p_0=\ptime$, and for $k>0$,
\begin{align*}
\Sigma^p_k     &=\np^{\Sigma^p_{k-1}},  &\Delta^p_k    &=\ptime^{\Sigma^p_{k-1}},&\Pi^p_k    &=\conp^{\Sigma^p_{k-1}},\\
\Sigma^{\exp}_k&=\NEXP^{\Sigma^p_{k-1}},&\Delta^{\exp}_k&=\EXP^{\Sigma^p_{k-1}},  &\Pi^{\exp}_k&=\coNEXP^{\Sigma^p_{k-1}}.
\end{align*}
(Notice that an $\cxt{(N)EXP}$ machine may supply exponentially long queries to the oracle, hence the
$\Sigma^p_{k-1}$~oracle in the definition of $\Sigma^{\exp}_k$ should be thought of actually having the power of
$\Sigma^{\exp}_{k-1}$.) In particular, $\Sigma^p_1=\np$, and $\Sigma^{\exp}_1=\NEXP$.

Many of the classes above can be equivalently defined using \emph{alternating Turing machines (ATM)}, which is a view
we will favour especially when proving upper bounds (Sections~\ref{sec:derivability}--\ref{sec:upper-bounds}). An ATM
is similar to a NTM in that multiple possible transitions may be defined for any given state. However, the definition of
acceptance is different. Each non-final state of an ATM is labelled as either \emph{existential} or \emph{universal},
and we define inductively the set of accepting configurations of the machine as the smallest set satisfying the
following conditions:
\begin{itemize}
\item A configuration in an accepting final state%
\footnote{In fact, we could dispense with final states altogether: an accepting (rejecting) final state is equivalent
to a universal (existential, resp.) state with no possible transitions out.}
is accepting.
\item A configuration in an existential state is accepting if there exists a transition to an accepting configuration.
\item A configuration in a universal state is accepting if all transitions lead to accepting configurations.
\end{itemize}
(States with exactly one possible transition can be thought of as deterministic; it makes no difference whether they
are labelled as existential or universal.)

If $\cxt{AP}$ denotes the class of languages computable by an ATM in polynomial time, we have
\begin{equation}\label{eq:64}
\cxt{AP}=\psp.
\end{equation}
(On a related note, $\cxt{APSPACE}=\EXP$, but we will not need this.)

We are especially interested in classes with bounded alternation. Here, an ATM is said to make an \emph{alternation}
when it transitions from an existential state to a universal state, or vice versa. For any $k>0$ and
$t\colon\omega\to\omega$, $\Sigma_k\text-\cxt{TIME}(t)$ denotes the class of languages computable in time $t(n)$ by an
ATM that starts in an existential state, and then makes at most $k-1$ alternations. The class
$\Pi_k\text-\cxt{TIME}(t)$ is defined similarly, but starting from a universal state. In particular,
$\Sigma_1\text-\cxt{TIME}(t)=\cxt{NTIME}(t)$, and $\Pi_1\text-\cxt{TIME}(t)=\cxt{coNTIME}(t)$. We have the following
characterization for any $k>0$:
\begin{align*}
\Sigma^p_k&=\Sigma_k\text-\cxt{TIME}\bigl(\poly(n)\bigr),&
\Pi^p_k&=\Pi_k\text-\cxt{TIME}\bigl(\poly(n)\bigr),\\
\Sigma^{\exp}_k&=\Sigma_k\text-\cxt{TIME}\bigl(2^{\poly(n)}\bigr),&
\Pi^{\exp}_k&=\Pi_k\text-\cxt{TIME}\bigl(2^{\poly(n)}\bigr).
\end{align*}

We will also need convenient complete languages for our classes. The set $\task{TAUT}$ of tautologies of $\CPC$ is the
canonical $\conp$-complete language, and the dual language $\task{SAT}$ of satisfiable classical propositional formulas
is $\np$-complete. The standard $\psp$-complete language%
\footnote{It is more transparent to think of it as an $\cxt{AP}$-complete language: the existential and universal
quantifiers almost directly correspond to existential and universal states of an ATM. The usual textbook proof of
$\psp$-completeness of $\task{QSAT}$ is for the most part actually a proof of the equality~\eqref{eq:64}.} 
is $\task{QSAT}$: the language of all true quantified Boolean sentences
\begin{equation}\label{eq:65}
Q_0x_0\,Q_1x_1\,\dots\,Q_{n-1}x_{n-1}\,\fii(\vec x),
\end{equation}
where $\fii$ is a propositional formula, and each $Q_i\in\{\exists,\forall\}$ quantifies over a truth value
$x_i\in\two$. 

Let $k>0$. A quantified Boolean sentence~\eqref{eq:65} is in $\Sigma^q_k$ if the quantifier prefix $Q_0\dots Q_{n-1}$
may be written as at most~$k$ alternating blocks, the first block consisting of existential quantifiers, the
second of universal quantifiers, and so on; $\Pi^q_k$ is defined dually (i.e., starting with a universal block). Then
the language $\Sigma^q_k\text-\task{SAT}$ consisting of all true $\Sigma^q_k$~sentences is $\Sigma^p_k$-complete, and
the dual language $\Pi^q_k\text-\task{SAT}$ is $\Pi^p_k$-complete. Notice that $\Sigma^q_1\text-\task{SAT}$ is just a
notational variant of~$\task{SAT}$: a propositional formula is satisfiable iff the corresponding existentially
quantified sentence is true.

Finally, we need complete languages for~$\Sigma^{\exp}_k$ (in particular, for $k=2$). Recall that in descriptive
complexity, we encode words $w=w_0\dots w_{n-1}\in\Sigma^*$ by models $M_w$ with domain~$n=\{0,\dots,n-1\}$
endowed with the order relation~$<$ (and possibly other arithmetical predicates, which we will not need here), and
unary predicates $P_a$ for each symbol $a\in\Sigma$, such that
\[M_w\model P_a(i)\iff w_i=a.\]
By Fagin's theorem, a language~$L$ is in~$\np$ iff there is a $\Sigma^1_1$ (i.e., existential second-order)
sentence~$\Phi$ such that
\begin{equation}\label{eq:66}
w\in L\iff M_w\model\Phi.
\end{equation}
More generally, $\Sigma^p_k$~languages are exactly those that are $\Sigma^1_k$-definable. This correspondence can be
generalized to the exponential hierarchy, using \emph{third-order} sentences: a language is in $\Sigma^{\exp}_k$ iff it
is $\Sigma^2_k$-definable. (See Ko\l odziejczyk \cite[Prop.~2.6]{kol:hol}, which also includes a brief historical
discussion. A similar statement in Hella and Turull-Torres \cite[Thm.~7]{hel-tuto:hol} suffers from an off-by-one
error.)

Since we only need complete problems rather than exact descriptions of the languages, we may simplify the
$\Sigma^2_k$~sentences to a convenient form. This was already done in \cite[L.~3.1]{ej:admcomp} for the
special case $k=1$ (i.e., $\NEXP$); here we generalize it to higher levels of the exponential hierarchy (with a more
detailed proof).
\begin{Thm}\label{thm:exphprob}
Let~$k\ge1$. Put $Q=\exists$ for $k$ odd, and $Q=\forall$ for $k$ even, and let $\ob Q$ be its dual. Then the set of
true $\Sigma^2_k$~sentences of the form
\begin{equation}\label{eq:1}
\exists X_1\sset\pw n\:\forall X_2\sset\pw n\,\dots\,QX_k\sset\pw n\:\ob Qt_0,\dots,t_{m-1}\sset n\:\fii
\end{equation}
is a $\Sigma^{\exp}_k$-complete language, where $n$ is given in unary, and $\fii$ is a Boolean combination of atomic
formulas of the form $i\in t_\alpha$ or~$t_\alpha\in X_j$ for $\alpha<m$, $i<n$, and $j=1,\dots,k$.
\end{Thm}
\begin{Pf}
For ease of notation, we will assume that $k$ is odd, so that $Q=\exists$. The case of $k$~even is dual. We will denote
third-order variables by capital letters $X,Y,\dots$ (with indices etc.), second-order variables by lower-case letters
$t,s,\dots$, and first-order variables with Greek letters $\xi,\eta,\dots$.

First, any $\Sigma^{\exp}_k$~language reduces to a language $L\in\Sigma_k\text-\cxt{TIME}(2^n)$ by a
simple padding argument: if $L_0\in\Sigma_k\text-\cxt{TIME}(2^{n^c})$, then $L=\{0^{n^c}1w:w\in
L_0,\lh w=n\}$ is computable in $\Sigma_k\text-\cxt{TIME}(2^n)$, and $w\mapsto0^{n^c}1w$ is a logspace reduction of
$L_0$ to~$L$. Thus, let us fix a $\Sigma^{\exp}_k$-complete language $L\in\Sigma_k\text-\cxt{TIME}(2^n)$.

By Ko\l odziejczyk~\cite{kol:hol} (Thm.~5.6 and a comment below Def.~5.9), there is a $s[\Sigma^2_k]^{\le1}$
sentence~$\Phi$ that defines~$L$ as in~\eqref{eq:66}: that is, $\Phi$ has the form
\[\exists\vec X_1\:\forall\vec X_2\,\dots\,\exists\vec X_k\:\Psi(\vec X_1,\dots,\vec X_k),\]
where $\Psi$ is a second-order formula, and all the second-order and third-order variables are unary: i.e., for a
model with domain~$n$, the second-order variables range over $\pw n$, and third-order variables over $\pw{\pw n}$.
(Second-order variables will remain unary for the rest of the proof, but we will introduce third-order variables of
higher arity during subsequent manipulations.) We may assume $\Psi$ only uses $=$ for first-order objects.

We may write $\Psi$ in prenex normal form, and moreover, we may assume that all second-order quantifiers precede all
first-order quantifiers: this is easily accomplished by exploiting the equivalences
\begin{align*}
\exists\xi\:Q_0t_0\,\dots\,Q_lt_l\:\psi
&\EQ\exists s\:\bigl(s\ne\nul\land Q_0t_0\,\dots\,Q_lt_l\:\forall\xi\,(s(\xi)\to\psi)\bigr)\\
&\EQ\exists s\:Q_0t_0\,\dots\,Q_lt_l\:\bigl(s\ne\nul\land\forall\xi\,(s(\xi)\to\psi)\bigr),
\end{align*}
where $s\ne\nul$ stands for $\exists \xi\,s(\xi)$; similarly for universal quantifiers. Thus, we may write
\[\Phi=\exists\vec X_1\:\forall\vec X_2\,\dots\,\exists\vec X_k\:
  \forall t_0\:\exists s_1\:\forall t_1\,\dots\,\forall t_{l-1}\:\exists s_l\:\psi(\vec X_1,\dots,\vec X_k,\vec t,\vec s),\]
where $\psi$ is first-order. Next, we get rid of existential second-order quantifiers by introducing Skolem functions:
$\Phi$ is equivalent to
\begin{multline*}
\exists\vec X_1\:\forall\vec X_2\,\dots\,\exists\vec X_k\:
  \exists Y_1\,\dots\exists Y_l\:\forall t_0\,\dots\,\forall t_{l-1}\\
\psi\bigl(\vec X_1,\dots,\vec X_k,\vec t,\{\xi:Y_1(t_0,\xi)\},\dots,\{\xi:Y_l(t_0,\dots,t_{l-1},\xi)\}\bigr),
\end{multline*}
where each $Y_i$ is a third-order relation variable with $i$~second-order and one first-order argument (i.e., it
ranges over $\pw{\pw n^i\times n}$.) Eliminating the comprehension symbols, $\Phi$ is equivalent to
\[\exists\vec X_1\:\forall\vec X_2\,\dots\,\exists\vec X_k\:
     \exists Y_1\,\dots\exists Y_l\:\forall t_0\,\dots\,\forall t_{2l-1}\:
          \bigl(\gamma(\vec Y,\vec t)\to\psi(\vec X_1,\dots,\vec X_k,\vec t)\bigr),\]
where $\gamma$ denotes the first-order formula
\begin{equation}\label{eq:68}
\ET_{i<l}\forall\xi\,\bigl(t_{l+i}(\xi)\eq Y_{i+1}(t_0,\dots,t_i,\xi)\bigr).
\end{equation}
By increasing~$l$ or adding dummy quantifiers if necessary, we may assume that all the tuples $\vec X_j$ also have
length~$l$.

In order to simplify each third-order quantifier block to a single unary variable, we pass from $M_w$ to a larger
model~$M^*_w$, defined as follows. The domain of $M^*_w$ is $n^*=n\times(l+1)^2$ (i.e., $(l+1)^2$ disjoint copies of
the domain of~$M_w$). For each $i<(l+1)^2$, $M^*_w$ includes a unary predicate $C_i$ that selects the $i$th copy
of~$M_w$ (i.e., $C_i$ is satisfied by elements of $n\times\{i\}$), and a binary predicate~$E$ that defines the
equivalence relation
\[\{\p{\p{\xi,i},\p{\xi,j}}:\xi<n\text{ and }i,j<(l+1)^2\}\]
(i.e., the kernel of the projection $M^*_w\to M_w$). The original relations of~$M_w$ are included on the $0$th copy
$n\times\{0\}$. Clearly, the mapping $w\mapsto M^*_w$ is still logspace computable. We will translate $\Phi$ to a
formula $\Phi^*$ such that
\begin{equation}\label{eq:67}
w\in L\iff M_w\model\Phi\iff M^*_w\model\Phi^*.
\end{equation}
We may represent elements $\xi\in M_w$ by elements of~$M^*_w$ that satisfy~$C_0$, and subsets $t\sset M_w$ by subsets
of~$C_0$; however, we will actually need to quantify over the copies of $t$ in each $C_i$ as well. We represent an
$l$-tuple $\vec X=\p{X_i:i<l}$ of third-order objects $X_i\sset\pw n$ by a single third-order object $X^*\sset\pw{n^*}$
defined as
\[X^*=\bigcup_{i<l}\bigl\{t\times\{i\}:t\in X_i\bigr\}.\]
Moreover, if we are in addition to $\vec X$ given a tuple $Y_1,\dots,Y_l$, where $Y_i\sset\pw n^i\times n$, we
represent $\vec X,\vec Y$ together by
\begin{multline*}
\bigcup_{i<l}\bigl\{t\times\{i\}:t\in X_i\bigr\}\\
\cup\bigcup_{i=1}^l\Bigl\{\bigcup_{j<i}\bigl(t_j\times\{i(l+1)+j\}\bigr)\cup\bigl(\{\xi\}\times\{i(l+1)+l\}\bigr):
\p{t_0,\dots,t_{i-1},\xi}\in Y_i\Bigr\}.
\end{multline*}
With this representation in mind, $\Phi^*$ will have the form
\[\exists X^*_1\:\forall X^*_2\,\dots\,\exists X^*_k\:\forall\vec t_0\,\dots\,\forall\vec t_{2l-1}\:
  \bigl(\gamma^*(X^*_k,\vec t_0,\dots,\vec t_{2l-1})\to\psi^*(X^*_1,\dots,X^*_k,\vec t_0,\dots,\vec t_{2l-1})\bigr),\]
where each $\vec t_i$ is an $(l+1)^2$-tuple $\p{t_{i,j}:j<(l+1)^2}$, the formula $\gamma^*$ is a conjunction of
\eqref{eq:69} and~\eqref{eq:70} below, and~$\psi^*$ is constructed from~$\psi$ as follows. We replace first-order
quantifiers $\exists\xi\,\dots$ and $\forall\xi\,\dots$ by $\exists\xi\,(C_0(\xi)\land\dots)$ and
$\forall\xi\,(C_0(\xi)\to\dots)$, respectively. Atomic subformulas of~$\psi$ that only mention first-order objects are
left intact, atomic formulas $t_i(\xi)$ are replaced with $t_{i,0}(\xi)$, and atomic formulas $X_{j,p}(t_i)$ are
replaced with $X_j^*(t_{i,p})$.

The first conjunct of~$\gamma^*$,
\begin{equation}\label{eq:69}
\ET_{\substack{i<2l\\\hbox to0pt{\hss$\scriptstyle j,j'<(l+1)^2$\hss}}}\forall\xi\,\bigl(t_{i,j'}(\xi)\eq
     C_{j'}(\xi)\land\exists\eta\,\bigl(C_j(\eta)\land E(\xi,\eta)\land t_{i,j}(\eta)\bigr)\bigr),
\end{equation}
ensures that the sets $\vec t_i$ are correctly formed: i.e., $t_{i,j}\sset C_j$, and for fixed $i$, all the
$t_{i,j}$ are copies of the same set $t_i\sset n$. The second conjunct is a translation of~\eqref{eq:68}, which can be
written as
\[\ET_{i=1}^l\forall\xi\:\Bigl(t_{l+i-1,i(l+1)+l}(\xi)\eq X^*_k\Bigl(\bigcup_{j<i}t_{j,i(l+1)+j}\cup\{\xi\}\Bigr)\Bigr).\]
Expanding the definition of~$\cup$, we arrive at
\begin{equation}\label{eq:70}
\ET_{i=1}^l\forall\xi\:\exists s\:\Bigl(\forall\eta\:\Bigl(s(\eta)\eq\eta=\xi\lor\LOR_{j<i}t_{j,i(l+1)+j}(\eta)\Bigr)
  \land\bigl(t_{j+i-1,i(l+1)+l}(\xi)\eq X^*_k(s)\bigr)\Bigr).
\end{equation}

By construction, \eqref{eq:67} holds. Notice that $\gamma^*$, specifically~\eqref{eq:70},
contains a second-order quantifier. When we bring $\Phi^*$ to prenex normal form, we obtain a sentence of the form
(dropping the ${}^*$ decoration from variables)
\[\exists X_1\:\forall X_2\,\dots\,\exists X_k\:\forall\vec t\:\exists\xi\:\forall\vec s\:\fii(X_1,\dots,X_k,\vec t,\vec s,\xi),\]
where $\fii$ is first-order. (This is still a single, constant-size sentence that only depends on~$L$, not on~$w$.)

As a final step, we transform $\Phi^*$ for any given word~$w$ to a sentence $\Phi_w$ that embeds the structure
of~$M^*_w$ as follows. Using constants for elements $i\in M^*_w$ (i.e., $i<n^*$), we expand each first-order
quantifier $\exists\xi\,\dots$ to a disjunction $\LOR_{i<n^*}\dots$, and $\forall\xi\,\dots$ to $\ET_{i<n^*}\dots$. Then we
evaluate each atomic formula that does not involve higher-order variables, and replace it with $\top$ or~$\bot$
according to its truth-value.

The resulting formula has size polynomial in~$n$ (the exponent being roughly the nesting depth of first-order
quantifiers in~$\Phi^*$), and it is easy to see that it is logspace computable. It has the form~\eqref{eq:1}, and we
have
\[w\in L\iff M^*_w\model\Phi^*\iff n^*\model\Phi_w\]
by construction.
\end{Pf}

\section{Derivability}\label{sec:derivability}

Before we embark on our main quest for the complexity of admissibility and unifiability in clx logics, let us first
settle a more basic question: what is the complexity of \emph{tautologicity} or \emph{derivability} in these logics.

\begin{Rem}\label{rem:clx-frames-polytime}
Notice that (single-conclusion) derivability has the same complexity as tautologicity for any transitive logic~$L$, as
\[\Gamma\vdash_L\fii\iff\vdash_L\boxdot\ET\Gamma\to\fii\]
gives a logspace reduction.

Recall that any clx logic~$L$ is $\forall\exists$-definable on finite frames \citi{Thm.~4.29}, and as a
consequence, finite $L$-frames are recognizable in polynomial time (in fact, in $\cxt{\Pi_2\text-TIME}(O(\log n))$, a
subclass of uniform~$\Ac$).
\end{Rem}

\begin{Thm}\label{thm:der-lin-clx}
For any consistent linear clx logic~$L$, derivability in~$L$ is $\conp$-complete.
\end{Thm}
\begin{Pf}
Since $L$ is a conservative extension of classical propositional logic, $\vdash_L$ is $\conp$-hard.

On the other hand, if $\fii$ is a formula of size $n=\lh\fii$ such that $\nvdash_L$, then there exists an $L$-model
$F\nmodel\fii$ of depth and cluster size at most~$n$, hence size at most~$n^2$, by \citi{Thm.~4.38}. (In fact, it is
easy to show that size~$n$ is enough.) Since
finite $L$-frames are polynomial-time recognizable by Remark~\ref{rem:clx-frames-polytime}, this shows that $\nvdash_L$ is
in~$\np$.
\end{Pf}

The complexity of nonlinear clx logics is a bit more difficult to establish (although it is just a variant of standard
$\psp$-completeness results for modal logics starting with Ladner~\cite{ladner}). For this reason, we state the upper
and lower bounds separately. We begin with the former.

Recall from \citi{Def.~4.20} that $\base(L)$ is a finite set of extension conditions that determines the shape of
$L$-frames.
\begin{Thm}\label{thm:ub-der-clx}
For any clx logic~$L$, $L$-derivability is in~$\psp$.
\end{Thm}
\begin{Pf}
Recall from \citi{Thm.~4.38} that if $\nvdash_L\fii$, then $\fii$ is falsified in a rooted $L$-model which is a tree of
depth $\le n$ of clusters of size $\le n$. Similarly to the standard case of $\kiv$ and other common logics, we may
search for such a tree in polynomial space---or as we prefer to think about it, in alternating polynomial time---by
exploring one branch at a time.

In more detail, let $\fii$ be a formula whose provability in~$L$ we want to determine. Since variables and parameters
work the same way with respect to derivability, we may assume $\fii$ contains no parameters. Put
$\Sigma=\Sub(\Box\fii)$, $B=\{\psi:\Box\psi\in\Sigma\}$, $V=\Sigma\cap\Var$, and $n=\lh\Sigma$. For any $v\in\two^V$
and $X\sset B$, let $A(v,X)$ denote the Boolean assignment to modal formulas that agrees with~$v$ on variables, and
that makes $\Box\psi$ true for $\psi\in X$, and false for $\psi\in B\bez X$. If $W$ is an $L$-model, let
$B^+(W)=\{\psi\in B:W\model\psi\}$.

We intend to construct an algorithm to compute the predicate
\[S(X)\iff\exists W\in\Mod_L\:B^+(W)=X\]
for $X\sset B$. (I.e., $S$ determines what assignments to the set~$\{\boxdot\psi:\psi\in B\}$ are consistent with~$L$;
recall that $\Mod_L$ denotes the class of rooted finite $L$-models.) We may then express (un)provability of~$\fii$ as
\begin{equation}\label{eq:12}
\nvdash_L\fii\iff\exists X\sset B\bez\{\fii\}\:S(X).
\end{equation}
We will present a recursive description of~$S(X)$ expressing how to build a model witnessing $S(X)$ by attaching a new
root cluster to a disjoint sum of rooted models witnessing $S(Z)$ for some sets $Z\sSset X$.

To this end, we consider auxiliary predicates $H_C(X,Y)$ and $U_m(X,Y)$ for $X\sset Y\sset B$, where $\p{C,m}\in\ECI$.
Their intended meaning is that $H_C(X,Y)$ holds if, given a (not necessarily rooted) model $W'$ such that $B^+(W')=Y$,
we can construct a model~$W$ such that $B^+(W)=X$ by attaching to~$W'$ a root cluster of type~$C$ with a suitable
valuation; $U_m(X,Y)$ holds if there are rooted models $\{W_i:i<m\}$ such that $B^+\bigl(\sum_{i<m}W_i\bigr)=Y$ and
$B^+(W_i)\sSset X$ for each~$i$. We interpret this for infinite extension conditions such that if $C=\nr\infty$, the
root cluster can be reflexive of arbitrary size, and if $m=\infty$, any positive number of $W_i$ is allowed.

Formally, we define $H_C$ and~$U_m$ as follows:
{\allowdisplaybreaks[3]
\begin{align*}
H_\I(X,Y)&\iff\exists v\in\two^V\:A(v,Y)\model\ET_{\psi\in X}\psi\land\ET_{\chi\in Y\bez X}\neg\chi,\\
H_{\nr k}(X,Y)&\iff\begin{aligned}[t]
      \exists\{v_i:i<k\}\sset\two^V\,\Bigl(&\forall i<k\:A(v_i,X)\model\ET_{\psi\in X}\psi\\
               &\quad\et\forall\chi\in Y\bez X\:\exists i<k\:A(v_i,X)\model\neg\chi\Bigr),
  \end{aligned}\\
H_{\nr\infty}(X,Y)&\iff\forall\chi\in(Y\bez X)\cup\{\bot\}\:\exists v\in\two^V\:
                           A(v,X)\model\ET_{\psi\in X}\psi\land\neg\chi,\\
U_m(X,Y)&\iff\begin{aligned}[t]
      \exists\{Z_j:j<m\}\,\Bigl(&\forall j<m\:\bigl(X\ssset Z_j\et S(Z_j)\et Y\sset Z_j\sset B\bigr)\\
               &\quad\et\forall\chi\in B\bez Y\:\exists j<m\:\chi\notin Z_j\Bigr),
  \end{aligned}\\
U_\infty(X,Y)&\iff
\begin{cases}
      \forall\chi\in B\bez Y\:\exists Z\,\bigl(X\ssset Z\et S(Z)\et Y\sset Z\sset B\bez\{\chi\}\bigr),&Y\ssset B,\\
      X\ssset B\et S(B),&Y=B.
\end{cases}
\end{align*}}
for $1\le k<\omega$ and $m<\omega$.
\begin{Cl}\label{cl:der-clx-rec}
For any $X\sset B$,
\begin{equation}\label{eq:11}
S(X)\iff\exists\p{C,m}\in\base(L)\,\exists Y\,\bigl(X\sset Y\sset B\et H_C(X,Y)\et U_m(X,Y)\bigr).
\end{equation}
\end{Cl}
\begin{Pf*}
Left-to-right: Fix a finite rooted $L$-model $W$ such that $B^+(W)=X$, and such that $B^+(w\Up)\sSset X$ for all $w\in
W\bez\rcl(W)$. Assume that $W$ is of type~$t$, and fix $\p{C,m}\in\base(L)$ such that $t\preceq\p{C,m}$. Put
$Y=B^+\bigl(W\bez\rcl(W)\bigr)$.

We claim that $H_C(X,Y)$: if $C=\I$, let $v\in\two^V$ be the assignment of variables in the (unique) root~$r$, i.e.,
$v=\sat_V(r)$. Since $Y=\{\psi\in B:r\model\Box\psi\}$, we have $\sat_\Sigma(r)=A(v,Y)$, and it is readily seen that
\[r\model\ET_{\psi\in X}\psi\land\ET_{\chi\in Y\bez X}\neg\chi,\] thus $v$ witnesses that $H_\I(X,Y)$ holds. If $C$ is
reflexive, then $X=\{\psi\in B:r\model\Box\psi\}$, thus for any $w\in\rcl(W)$, $\sat_\Sigma(w)=A(v,X)$, where
$v=\sat_V(w)$. If $C=\nr k$, fix a (not necessarily injective) enumeration $\rcl(W)=\{w_i:i<k\}$; then $H_{\nr k}(X,Y)$
is witnessed by $v_i=\sat_V(w_i)$. Likewise, if $C=\nr\infty$, we may take $v=\sat_V(w)$ for any $w\in\rcl(W)$ such
that $w\model\neg\chi$.

We also claim that $U_m(X,Y)$. If $m<\infty$, let $\{C_j:j<m\}$ be a (not necessarily injective) enumeration of
immediate successor clusters of $\rcl(W)$, and put $Z_j=B^+(C_j\Up)$. Then $S(Z_j)$ by definition, and
$Y=\bigcap_{j<m}Z_j$, which implies that $Y\sset Z_j\sset B$, and for each $\chi\in B\bez Y$, $\chi\notin Z_j$ for
some~$j$. Moreover, our assumptions on~$W$ ensure that $X\ssset Z_j$. Thus, $\{Z_j:j<m\}$
witness that $U_m(X,Y)$ holds. If $m=\infty$, then for any $\chi\in B\bez Y$, there exists a point $x\in W\bez\rcl(W)$
such that $x\nmodel\boxdot\chi$; then $Z=B^+(x\Up)$ has the required properties. If $Y=B$, we just do
the same for an arbitrary $x\in W\bez\rcl(W)$.

Right-to-left: Fix $\p{C,m}\in\base(L)$ and $X\sset Y\sset B$ such that $H_C(X,Y)$ and $U_m(X,Y)$. We will construct a
finite $L$-model $W$ whose root witnesses~$S(X)$. First, we build the root cluster. If $C=\I$, $\rcl(W)$ will be an
irreflexive point with valuation of variables $v$ chosen as a witness for~$H_\I(X,Y)$. Likewise, if $C=\nr k$, let
$\{v_i:i<k\}$ be witnesses for $H_{\nr k}(X,Y)$; then $\rcl(W)$ is a $k$-element reflexive cluster $\{w_i:i<k\}$ with
$\sat_V(w_i)=v_i$. If $C=\nr\infty$, then for each $\chi\in Y\bez X$, fix $v_\chi\in\two^V$ such that
$A(v_\chi,X)\model\ET_X\psi\land\neg\chi$; then $\rcl(W)$ is a reflexive cluster consisting of points $\{w_\chi:\chi\in
Y\bez X\}$ such that $\sat_V(w_\chi)=v_\chi$. This does not work if $X=Y$; in this case, we let $\rcl(W)$ be a
reflexive singleton satisfying $v$ where $A(v,X)\model\ET_{\psi\in X}\psi$.

Next, we construct the rest of the model. If $m<\infty$, let $\{Z_j:j<m\}$ be witnesses for $U_m(X,Y)$. Since $S(Z_j)$
for each $j<m$, we may fix a finite rooted $L$-model $W_j$ such that $B^+(W_j)=Z_j$. We let $W\bez\rcl(W)=\sum_{j<m}W_j$.
Similarly, if $m=\infty$, then for each $\chi\in B\bez Y$, we fix $Z_\chi$ satisfying the condition from the definition
of~$U_{\nr\infty}$, and we find a rooted $L$-model $W_\chi$ such that $B^+(W_\chi)=Z_\chi$.
We define $W\bez\rcl(W)$ as the disjoint union of $W_\chi$ for all $\chi\in B\bez Y$, as long as $Y\ssset B$. If $Y=B$,
let $W\bez\rcl(W)$ be a rooted $L$-model satisfying $\ET_{\psi\in B}\boxdot\psi$.

Since $Y=\bigcap_{j<m}Z_j$ or $Y=\bigcap_{\chi\in B\bez Y}Z_\chi$ (respectively), we see that 
\[W\bez\rcl(W)\model\ET_{\psi\in Y}\boxdot\psi,\]
and
\[\rcl(W)\model\ET_{\chi\in B\bez Y}\neg\Box\chi.\]
Then it is easy to check that $\rcl(W)$ witnesses~$S(X)$.
\end{Pf*}

Notice that if we expand the occurrences of $H_C$ and~$U_m$ in~\eqref{eq:11} using their definitions, we obtain a
recursive expression for~$S(X)$ in terms of $S(Z)$ or $S(Z_j)$, where $X\ssset Z,Z_j$. We may read it as a recipe for
an algorithm computing $S(X)$ (and $\vdash_L\fii$, in view of~\eqref{eq:12}) on an alternating Turing machine, where we
interpret the existential and universal quantifiers (and disjunctions and conjunctions) in the expression as
nondeterministic choices in existential or universal states (respectively). As just noted, the argument strictly
increases on each recursive call to~$S(X)$, hence the recursion depth is at most~$n$. Moreover, each quantifier takes
$O(n)$ bits, and the conditions in-between (e.g., the truth of some formulas from~$\Sigma$ under $A(v,X)$) can be
checked deterministically in polynomial time. Thus, the algorithm works in alternating polynomial time, i.e., in $\psp$
by~\eqref{eq:64}.
\end{Pf}
\begin{Rem}\label{rem:deriv-multi}
The bounds in Theorems \ref{thm:der-lin-clx} and~\ref{thm:ub-der-clx} apply not just to derivability of single-conclusion rules as in
Remark~\ref{rem:clx-frames-polytime}, but also to derivability of multiple-conclusion rules:
its definition in terms of single-conclusion derivability
\[\Gamma\vdash_L^m\Delta\iff\exists\fii\in\Delta\:\Gamma\vdash_L^1\fii\]
shows that $\vdash_L^m$ is in $\conp$ or~$\psp$ whenever $\vdash_L^1$ is. (In general, the definition gives a
logspace disjunctive truth-table (dtt) reduction of $\vdash_L^m$ to~$\vdash_L^1$; both $\psp$ and $\conp$ are closed
under dtt-reductions.)
\end{Rem}

We now turn to the lower bound. There is more than one way to approach it. One possibility is to extend Ladner's proof
to show that all normal modal logics with the disjunction property are $\psp$-hard. (Even though this is
straightforward to carry out, and seems to be a fundamental result, the author is not aware of any published
reference. The corresponding result for si logics was proved by Chagrov~\cite{chagl}, cf.~\cite[Thm.~18.30]{cha-zax}.)
We will use another method, namely by reduction from intuitionistic logic ($\IPC$) using a series of translations. This
route is more useful for our purposes, because the resulting statement is relatively more general in
the context of transitive modal logics (it applies to all transitive logics with the disjunction property, and it also
applies to various extensions of $\lgc{K4.2}$, which will be relevant in the sequel).

\begin{Def}\label{def:transl}
Let $\T$ denote the \emph{G\"odel--McKinsey--Tarski translation} of $\IPC$ (formulated using connectives
$\{{\to},{\land},{\lor},{\bot}\}$) in~$\lgc{S4}$: $\T(\fii)=\Box\fii$ if $\fii$ is an atom, $\T$ commutes with
$\land$, $\lor$, and $\bot$, and
\[\T(\fii\to\psi)=\Box\bigl(\T(\fii)\to\T(\psi)\bigr).\]

The \emph{relativization translation} $\reltr$ of $\kiv$ in itself is defined as follows. Given a formula $\fii$,
we fix an atom~$r$ (a parameter, in contexts where the distinction between variables and parameters becomes
relevant) that does not occur in~$\fii$. We define an auxiliary translation $\psi^r$ such that it preserves atoms,
commutes with Boolean connectives, and
\[(\Box\psi)^r=\Box(r\to\psi^r).\]
Then we define $\reltr(\fii)$ as $r\to\fii^r$.

Let $\bdtr$ denote the \emph{boxdot translation} of $\lgc{S4}$ in~$\kiv$, which preserves atoms, commutes with
Boolean connectives, and satisfies
\[\bdtr(\Box\fii)=\boxdot\bdtr(\fii).\]
Notice that when expanded out, this formula contains two occurrences of $\bdtr(\fii)$, and as a consequence, the boxdot
translation may exponentially enlarge formulas with deeply nested boxes. For this reason, we define the
\emph{efficient boxdot translation} $\bdtr'(\fii)$ as follows. We introduce a new variable $z_\psi$ for each
formula $\psi$ such that $\Box\psi\sset\fii$. For any $\chi\sset\fii$, let $\chi'$ denote the formula
resulting from~$\chi$ by replacing all topmost occurrences of boxed subformulas $\Box\psi$ with the
corresponding~$z_\psi$. Then
\[\bdtr'(\fii):=\ET_{\Box\psi\sset\fii}\boxdot(z_\psi\eq\boxdot\psi')\to\fii'.\]
\end{Def}
\begin{Lem}\label{lem:eff-boxdot}
For any formula~$\fii$, the formulas $\bdtr(\fii)$ and~$\bdtr'(\fii)$ are equiderivable.
Specifically, we have 
\begin{align}
\label{eq:9}&\vdash_\kiv\bdtr(\fii)\to\bdtr'(\fii),\\
\label{eq:10}&\vdash_\kiv\sigma\bigl(\bdtr'(\fii)\bigr)\to\bdtr(\fii),
\end{align}
where $\sigma$ is the substitution such that $\sigma(z_\psi)=\bdtr(\Box\psi)$.
\end{Lem}
\begin{Pf}

\eqref{eq:9}: We can prove
\[\vdash_\kiv\ET_{\Box\psi\sset\fii}\boxdot(z_\psi\eq\boxdot\psi')\to\bigl(\chi'\eq\bdtr(\chi)\bigr)\]
for each $\chi\sset\fii$ by induction on the complexity of~$\chi$.

\eqref{eq:10}: Since $\bdtr$ and~$\sigma$ both commute with Boolean connectives, we have $\sigma(\chi')=\bdtr(\chi)$
for each $\chi\sset\fii$, thus $\sigma(\bdtr'(\fii))$ is
\[\ET_{\Box\psi\sset\fii}\boxdot\bigl(\boxdot\bdtr(\psi)\eq\boxdot\bdtr(\psi)\bigr)\to\bdtr(\fii),\]
which is equivalent to just~$\bdtr(\fii)$.
\end{Pf}

Recall that for any frame $W$, $W_\R$ denotes its reflexivization. Following~\cite{cha-zax}, if $W$ is reflexive, let
$\roo W$ denote the \emph{skeleton} of~$W$, i.e., the quotient of $W$ by the cluster equivalence relation~$\sim$,
considered as an intuitionistic frame. The following is standard (see e.g.~\cite[L.~8.28]{cha-zax}):
\begin{Lem}\label{lem:godel}
For any reflexive frame~$W$, and an intuitionistic formula~$\fii$, we have
\[\roo W\model\fii\iff W\model\T(\fii).\qedhere\]
\end{Lem}

We leave the straightforward proof of the corresponding property of the boxdot translation to the reader:
\begin{Lem}\label{lem:boxdot}
For any frame $W$, and a formula~$\fii$, we have
\[W_\R\model\fii\iff W\model\bdtr(\fii).\qedhere\]
\end{Lem}

The $\reltr(\fii)$ translation is equivalent to $\fii^{sf}$ from Chagrov and Zakharyaschev \cite[Exer.~9.7]{cha-zax}.
\begin{Lem}\label{lem:relat}
Let $\fii$ be a formula, $r$ an atom, $W$ a frame, $W_0$ its subframe, and $\model$ a valuation in~$W$ such that
\[W,w\model r\iff w\in W_0.\]
Then
\[W,w\model\fii^r\iff W_0,w\model\fii\]
for all $w\in W_0$.
\end{Lem}
\begin{Pf}
By induction on the complexity of~$\fii$.
\end{Pf}
\begin{Cor}\label{cor:relat}
A frame~$W$ validates $\reltr(\fii)$ iff $\fii$ holds in all subframes of~$W$ (and thus in all frames that $W$
subreduces onto).

Consequently, if $L$ is a subframe logic and $\vdash_L\fii$, then $\vdash_L\reltr(\fii)$.
\noproof\end{Cor}

\begin{Def}\label{def:tree-universal}
A logic~$L\Sset\kiv$ is \emph{(cofinally) subframe-universal for trees} if for each finite tree~$T$, considered as a reflexive
frame, there exists an $L$-frame $W$ that (cofinally, resp.) weakly subreduces onto~$T$.
\end{Def}

Notice that in order to verify subframe universality for trees, it is enough to check \emph{binary} trees; on the other
hand, if a logic is subframe-universal for trees, it is also subframe-universal for all finite reflexive rooted frames
without proper clusters. This follows from the fact that every such frame is a p-morphic image of a binary tree. The
same considerations apply, mutatis mutandis, to cofinal subframe universality.

\begin{Thm}\label{thm:lb-der-transl}
Let a logic $L\Sset\kiv$ be subframe-universal for trees. Then $L$-derivability is $\psp$-hard.
\end{Thm}
\begin{Pf}
Derivability in~$\IPC$ is $\psp$-complete by Statman~\cite{stat}, hence it suffices to provide a reduction of~$\IPC$
to~$L$. Now, the translations $\T$, $\reltr$, and $\bdtr'$ increase the size of the formula only linearly, and are
readily seen to be logspace-computable, thus we only need to prove
\[\vdash_\IPC\fii\iff\vdash_L\bdtr'(\reltr(\T(\fii))).\]
By Lemma~\ref{lem:eff-boxdot}, this is equivalent to
\[\vdash_\IPC\fii\iff\vdash_L\bdtr(\reltr(\T(\fii))).\]
For the left-to-right implication, it is well known that $\T$ is an interpretation of $\IPC$ in~$\lgc{S4}$, and $\bdtr$
is an interpretation of $\lgc{S4}$ in~$\kiv$. By Corollary~\ref{cor:relat}, $\reltr$ is a self-interpretation of $\lgc{S4}$.

For the right-to-left implication, assume that $\nvdash_\IPC\fii$, whence there exists a finite tree~$T$ such that
$\fii$ is not valid in $\roo T$ (the intuitionistic version of~$T$). By assumption, there exists a weak subreduction
from an $L$-frame~$W$ onto~$T$, hence a subreduction from the reflexivization $W_\R$ to~$T$ by Observation~\ref{obs:weaksubr}.
Then
\[\roo T\nmodel\fii\implies T\nmodel\T(\fii)\implies W_\R\nmodel\reltr(\T(\fii))\implies W\nmodel\bdtr(\reltr(\T(\fii)))\]
by Lemma~\ref{lem:godel}, Corollary~\ref{cor:relat}, and Lemma~\ref{lem:boxdot}.
\end{Pf}

The previous theorem applies in particular to all logics with the disjunction property, as they are subframe universal
for trees. We will even prove that logics with the disjunction property are \emph{cofinally} subframe-universal for
trees, which fact we will need later. We note that the special case for logics $L\Sset\lgc{S4}$ follows from
\cite[Prop.~15.13]{cha-zax}; indeed, we could alternatively prove the general result by reducing it to the reflexive
case, but the proof below is more elementary (it avoids the machinery of canonical formulas).
\begin{Thm}\label{thm:dp-csf-univ}
Every logic~$L\Sset\kiv$ satisfying the disjunction property is cofinally subframe-universal for trees. Consequently, $L$-derivability
is $\psp$-hard.
\end{Thm}
\begin{Pf}
We will show by induction on the depth of~$T$ that for every finite tree~$T$, there exists a descriptive $L$-frame $W$
cofinally weakly subreducing to~$T$. For the base case, any nonempty $L$-frame (which exists as $L$ is consistent)
weakly subreduces to the one-element tree.

For the induction step, let $r$ be the root of~$T$, let $r_i$, $i<n$, be the immediate successors of~$r$, and for each
$i<n$, let $T_i$ be the subtree of~$T$ rooted at~$r_i$. We may assume $n\ge2$ by duplicating $T_0$ if necessary. By the
induction hypothesis, we may fix a descriptive $L$-frame $W_i$, and a cofinal weak subreduction $f_i$ from $W_i$
onto~$T_i$ for each $i<n$. We may assume that $W_i$ is rooted, and its root~$w_i$ is mapped to~$r_i$ by $f_i$.

By \cite[Thm.~15.1]{cha-zax}, there exists a rooted descriptive $L$-frame $\p{W,{<},A}$ such that the disjoint sum $\sum_{i<n}W_i$ is a
generated subframe of~$W$. Notice that $f_i$ is not necessarily a weak subreduction from $W$ to~$T_i$, as $W_i$ may not
be an admissible subset of~$W$: it is in general only an intersection of admissible subsets. Nevertheless, assume for now
that $\bigcup_{i<n}f_i$ extends to a cofinal weak subreduction $g$ from $W$ to $\bigcup_{i<n}T_i\sset T$. We may further
extend it to a subreduction $h\Sset g$ from $W$ onto~$T$ by putting
\[h(v)=r\iff v\in\bigcap_{i<n}g^{-1}[r_i]\down.\]
(In particular, the root of~$W$ is mapped to~$r$.) The subreduction $h$ is not necessarily cofinal. However, if $v\in W$ sees no
point of $\dom(h)$ (i.e., violates cofinality), then points below~$v$ cannot be mapped by~$h$ to anything else
than~$r$, as $g$ \emph{is} cofinal. Thus, if we fix an arbitrary leaf node $t\in T$, the partial mapping $f\colon W\to
T$ that extends $h$ by putting
\[f(v)=t\quad\text{if}\quad v\notin\dom(h)\Down\]
is still a weak subreduction from $W$ onto~$T$, and it is clearly cofinal.

Now, it remains to construct~$g$. Let us first fix $i<n$ and~$u\in W_i$. Since $W$ is refined, we may find for each
$j<n$, $j\ne i$, a set $X_{u,j}\in A$ such that $w_j\in\boxdot X_{u,j}$, and $u\notin X_{u,j}$. Putting
$X_u=W\bez\bigcup_{j\ne i}X_{u,j}$, we have $u\in X_u\in A$, while $X_u\Down\cap\bigcup_{j\ne i}W_j=\nul$. The sets
$X_u\cap W_i$ are admissible in~$W_i$; using the compactness of~$W_i$, there exists a finite set $\{u_k:k<m\}\sset W_i$
such that $W_i$ is covered by $Y_i=\bigcup_{k<m}X_{u_k}\in A$.

Let $\model$ be an admissible valuation on~$W$ such that
\[v\model y_i\iff v\in Y_i\]
for each $i<n$ and $v\in W$, and
\[v\model x_t\iff f_i(v)=t\]
for each $i<n$, $t\in T_i$, and $v\in W_i$ (the valuation of $x_t$ may be arbitrary outside~$W_i$, which allows
$\model$ to be admissible in~$W$). We define a partial function $g\colon W\to T$ such that for any $i<n$, $t\in T_i$,
and $v\in W$,
\[\begin{split}
  g(v)=t\iff v\model x_t&\land\boxdot y_i\land\boxdot\LOR_{s\in T_i}\diadot x_s
   \land\ET_{\substack{s,s'\in T_i\\s\ne s'}}\boxdot\neg(x_s\land x_{s'})\\
   &\land\ET_{\substack{s,s'\in T_i\\s\nleq s'}}\boxdot\bigl(x_s\to\Box\neg x_{s'}\bigr)
   \land\ET_{\substack{s,s'\in T_i\\s\lneq s'}}\boxdot\bigl(x_s\to\dia x_{s'}\bigr).
\end{split}\]
The $\boxdot y_i$ conjunct ensures that the purported preimages of distinct points are disjoint, hence $g$ is indeed a
well-defined partial mapping. It is then easy to check that $g\Sset f_i$ for each $i<n$, and that $g$ is a cofinal
weak subreduction from $W$ onto~$\bigcup_{i<n}T_i$.
\end{Pf}

\section{Upper bounds}\label{sec:upper-bounds}

\subsection{Nonlinear cluster-extensible logics}\label{sec:nonl-clust-extens}

In order to prove our basic estimate of the complexity of admissibility in clx logics, we will use the description of
admissibility in terms of pseudoextensible models from \citi{Thm.~5.24}, thus we need to know what it takes to check
pseudoextensibility.
\begin{Lem}\label{lem:psext}
Let us fix an extension condition $\p{C,m}\in\ECI$.

Given a finite set of formulas $\Sigma$ which is closed under subformulas, and a finite model $F$, we can check if $F$
is $\p{C,m}$-pseudoextensible wrt~$\Sigma$ in $\cxt{coNTIME}\bigl((N+2^{n^2})^{O(1)}\bigr)$, where $n=\lh\Sigma$, and
$N=\lh F$.

If moreover $C$ is finite, we can even check it in $\cxt{DTIME}\bigl((N+2^n)^{O(1)}\bigr)$.
\end{Lem}
\begin{Pf}
First, observe that we can compute the valuation of all formulas from~$\Sigma$ in~$F$ in about $O(nN^2)$ steps.

Let $P=\Par\cap\Sigma$, $B=\{\fii:\Box\fii\in\Sigma\}$, and  $\RI=\refl(C)$. (Recall that $\RI\in\{\I,\R\}$ denotes the
reflexivity of~$C$.)

Given $X\sset F$, $\nul\ne E\sset\two^P$ (with $\lh E=1$ if $\RI=\I$), and $u=\{u_e:e\in E\}\sset F$, we can test in
$\cxt{DTIME}\bigl((N+2^n)^{O(1)}\bigr)$ if $u$ is a tight $\p{\RI,E}$-pseudopredecessor of~$X$ wrt~$\Sigma$ just using
the definition: this amounts to checking $u_e\model P^e$ for all $e\in E$, and for each $\fii\in B$, to look at
the valuation of $\fii$ and~$\Box\fii$ in every $u_e$ and~$w\in X$.

Better yet, given $E$ and~$X$, we can check the \emph{existence} of a $\p{\RI,E}$-tpp of~$X$ wrt~$\Sigma$ in
$\cxt{DTIME}\bigl((N+2^{n^2})^{O(1)}\bigr)$: we compute $B^+=\bigcap_{w\in X}\{\fii\in B:w\model\boxdot\fii\}$,
and~$B^-=B\bez B^+$. If $\RI=\I$, we try every $u\in F$ to see if
\[u\model P^e\land\ET_{\fii\in B^+}\Box\fii\land\ET_{\fii\in B^-}\neg\Box\fii.\]
If $\RI=\R$, we try every $D\sset B^+$ and $f\colon D\to E$ to see if any satisfy the condition
\[\forall e\in E\,\exists u\in F\,u\model P^e\land\ET_{\fii\in B^+\bez D}\boxdot\fii\land\ET_{\fii\in B^-\cup
D}\neg\Box\fii\land\ET_{\substack{\fii\in D\\f(\fii)=e}}\neg\fii\]
(cf. \citi{Def.~5.21 and L.~5.22}). This condition in turn can be checked by trying all possible $e$ and~$u$. Notice
that there are at most $2^n$ choices for~$D$, and at most $\lh E^n\le2^{n^2}$ choices for~$f$; if $\lh E$ is bounded by
a constant, the latter bound is $2^{O(n)}$.

If $m$ is finite, then $F$ is $\p{C,m}$-pseudoextensible wrt~$\Sigma$ iff for all $E\sset\two^P$ of size $\lh
E\le_0\lh C$, and for all $X=\{w_i:i<m\}\sset F$, there exists a $\p{\RI,E}$-tpp of~$X$ wrt~$\Sigma$. We check this by
co-nondeterministically choosing $E$, and then deterministically trying all $N^m$ possibilities for~$X$, resulting in a
$\cxt{coNTIME}\bigl((N+2^{n^2})^{O(1)}\bigr)$ algorithm. If moreover $C$ is finite with $\lh C=k$, we only need to
check $E\sset\two^P$ of size $\lh E\le k$; we may do this deterministically by trying all $2^{kn}$ possibilities,
resulting in a $\cxt{DTIME}\bigl((N+2^n)^{O(1)}\bigr)$ algorithm.

If $m=\infty$, we could choose $X\sset F$ co-nondeterministically as well. However, in order to get a
deterministic algorithm for $C$ finite, we cannot afford to try all $2^N$ possibilities for~$X$. We observe that the
existence of a $\p{\RI,E}$-tpp of~$X$ wrt~$\Sigma$ does not depend on $X$ as such, but only on the set $B^+\sset B$
as defined above. Thus, instead of checking all~$X\sset F$, we only check all (at most $2^n$) possibilities
for~$B^+\sset B$. Given $B^+$, we can test if there exists a corresponding $\p{\RI,E}$-tpp by the same method as above;
moreover, we can test if there exists a set $X\sset F$ such that $B^+=\bigcap_{w\in X}\{\fii\in B:w\model\boxdot\fii\}$
as this is equivalent to the condition
\[\forall\fii\in B\bez B^+\,\exists w\in F\,w\model\neg\boxdot\fii\land\ET_{\psi\in B^+}\boxdot\psi,\]
easily verifiable by going through all $\fii$ and~$w$.
\end{Pf}

\begin{Thm}\label{thm:ub-clx}
For any clx logic~$L$, $L$-admissibility is computable in~$\Pi^{\exp}_2$, and $L$-unifiability is computable
in~$\Sigma^{\exp}_2$.

If $L$ has bounded cluster size, or if the number of parameters is bounded by a constant, $L$-admissibility is
computable in $\coNEXP$, and $L$-unifiability in $\NEXP$.
\end{Thm}
\begin{Pf}
Using \citi{Thm.~5.24}, a rule $\Gamma\ru\Delta$ is not $L$-admissible if and only if it fails
in some $L$-model $F$ of size at most~$4^n$ that is $\base(L)$-pseudoextensible wrt~$\Sigma$, where
$\Sigma=\Sub(\Gamma\cup\Delta)$, and $n=\lh\Sigma$.

We can check this by nondeterministically choosing such a model $F$ of size $N\le4^n$ equipped with an assignment
${\model}\colon\Sigma\to\two$, and verifying it is indeed a counterexample for admissibility of $\Gamma\ru\Delta$. Given
$F$, we can check in time $N^{O(1)}=2^{O(n)}$ whether $F$ is an $L$-frame (using \citi{Thm.~4.29}), whether $\model$ is an honest valuation respecting the connectives, and whether $\Gamma\ru\Delta$ is satisfied
in the model $\p{F,{\model}}$. By Lemma~\ref{lem:psext}, we can check that $F$ is $\p{C,m}$-pseudoextensible wrt~$\Sigma$
for each $\p{C,m}\in\base(L)$ by a $\cxt{coNTIME}(2^{O(n^2)})$ algorithm. Overall, this gives a
$\Sigma^{\exp}_2$~algorithm for checking inadmissibility.

If $L$ has bounded cluster size, all conditions $\p{C,m}\in\base(L)$ have $C$~finite, thus we can check
$\base(L)$-pseudoextensibility of~$F$ in $\cxt{DTIME}(2^{O(n)})$, and inadmissibility in $\NEXP$ (in fact, in
$\cxt{NE}$). The same bound also holds for arbitrary $L$ in case the number of parameters is bounded by a constant~$k$,
as we may reduce the size of each~$C$ to at most~$2^k$.
\end{Pf}

\subsection{Linear cluster-extensible logics}\label{sec:line-clust-extens}

Observe that linear clx logics are cofinal-subframe logics.

\begin{Thm}\label{thm:ub-lin-clx}
Let $L$ be a linear clx logic. Then $L$-admissibility is computable in $\NEXP$, and $L$-unifiability in $\coNEXP$.

If $L$ has bounded cluster size, or if the number of parameters is bounded by a constant, then $L$-admissibility and
$L$-unifiability are in $\psp$.
\end{Thm}
\begin{Pf}
Put $\base^i(L)=\{\p{C,m}\in\base(L):m=i\}$ and $\base_i(L)=\{C:\p{C,i}\in\base(L)\}$ for $i=0,1$.
Let us fix a rule $\Gamma\ru\Delta$, and put $\Sigma=\Sub(\Gamma\cup\Delta)$, $P=\Sigma\cap\Par$,
$B=\{\fii:\Box\fii\in\Sigma\}$, and $n=\lh\Sigma$.

If $W$ is a finite rooted $L$-model, we put $B^+(W)=\{\fii\in B:W\model\fii\}$, and we
define predicates $G_i(W)$ for $i\ge\lh{B^+(W)}$ inductively by
\begin{align}\label{eq:3}
G_i(W)\iff{}&\forall C\in\base_1(L)\:\forall E\sset\two^P,\lh E\le_0\lh C:\\
\nonumber&\begin{aligned}
    \exists W'\in\Mod_L\,\Bigl(W\sgen W'\model\Gamma&\et W'\bez W\text{ is a $\p{\refl(C),E}$-tp of $W$}\\
    &\et\bigl(B^+(W')\ssset B^+(W)\TO G_{i-1}(W')\bigr)\Bigr).
\end{aligned}
\end{align}
Notice that the constraint $B^+(W')\ssset B^+(W)$ ensures that the condition $i\ge\lh{B^+(W)}$ is preserved; in
particular, $G_0(W)$ never refers to the nonexistent $G_{-1}$, and it is well defined. Finally, we define $G(W)$ for
arbitrary finite $L$-models $W$ by
\begin{equation}\label{eq:7}
G(W)\iff W\model\Gamma\et\forall u\in W\:G_{\lh{B^+(W_u)}}(W_u).
\end{equation}
\begin{Cl}\label{cl:good-equi}
If $W_0$ and~$W_1$ are rooted $L$-models of~$\Gamma$ such that $B^+(W_0)=B^+(W_1)$, and $i\ge\lh{B^+(W_0)}$, then
$G_i(W_0)$ iff $G_i(W_1)$.
\end{Cl}
\begin{Pf*}
By induction on~$i$. Fix $C$ and~$E$ as in~\eqref{eq:3}, and let $W'_0$ be a witness for~$G_i(W_0)$, so that
$U=W'_0\bez W_0$ is a $\p{\refl(C),E}$-tp of~$W_0$. Put $W'_1=W_1\cup U$ (with $U<W_1$ so that $W_1\sgen W'_1$).
Then $W'_1$ is an $L$-model, and $W'_1\bez W_1$ is a $\p{\refl(C),E}$-tp of~$W_1$ in~$W'_1$. We have $W'_1\model\Gamma$:
$\Gamma$ holds in $W_1\sgen W'_1$ by assumption; it holds in~$U$ since it holds in~$W'_0$, and satisfaction of
formulas from~$\Sigma$ in~$U$ only depends (apart from valuation of atoms in~$U$, which did not change) on $B^+(W_1)$,
equal to $B^+(W_0)$ by assumption. For the same reason, we obtain $B^+(W'_1)=B^+(W'_0)$, therefore, in case
$B^+(W'_1)\ssset B^+(W_1)$, we have $G_{i-1}(W'_1)$ by the induction hypothesis.
\end{Pf*}
\begin{Cl}\label{cl:good-model}
$G(W)$ holds if and only if $W$ is a generated submodel of an $L$-model $F\model\Gamma$ such that $F$ is
$\base^1(L)$-pseudoextensible wrt~$\Sigma$.
\end{Cl}
\begin{Pf*}

Left-to-right: Start with $W$, and by unwinding \eqref{eq:7} and~\eqref{eq:3}, attach all the tight predecessors that
are asserted to exist. Call the resulting model~$F$. It is clear that $F$ is a finite $L$-model, $W\sgen F$, and
$F\model\Gamma$. In order to see that $F$ is $\base^1(L)$-pseudoextensible wrt~$\Sigma$, let $C\in\base_1(L)$,
and $E\sset\two^P$ with $\lh E\le_0\lh C$. By construction, all rooted submodels $W'\sgen F$ satisfy $G_i(W')$ for an
appropriate~$i$, which guarantees the existence of a $\p{\refl(C),E}$-tp (and a fortiori $\p{\refl(C),E}$-tpp
wrt~$\Sigma$), with the following exception: no tp's of~$W'$ were added when $W'$ resulted by attaching the cluster
$\rcl(W')$ below the model $W''=W'\bez\rcl(W')$ which satisfied $G_i(W'')$ for some~$i$, and such that $B^+(W')=B^+(W'')$. But then
$F$ includes a $\p{\refl(C),E}$-tp of~$W''$, and since $B^+(W')=B^+(W'')$, this is also a $\p{\refl(C),E}$-tpp of~$W'$
wrt~$\Sigma$.

Right-to-left: It suffices to show that $G_i(W)$ if furthermore $W$ is rooted, and $i\ge\lh{B^+(W)}$. Let us proceed by
induction on~$i$ as in the definition of~$G_i(W)$. For any $C\in\base_1(L)$ and $E\sset\two^P$ such that $\lh E\le_0\lh
C$, we may fix a $\p{\refl(C),E}$-tpp $\{u_e:e\in E\}\sset F$ of~$W$ wrt~$\Sigma$. We construct $W'$ by attaching to
$W$ a new cluster $\{v_e:e\in E\}$, reflexive or not according to $\refl(C)$, such that $v_e\model P^e$ satisfies the
same variables as~$u_e$. Clearly, $W\sgen W'$, $W'$ is an $L$-model, and $W'\bez W$ is a $\p{\refl(C),E}$-tp of~$W$.
Using the fact that $\{u_e:e\in E\}$ is a $\p{\refl(C),E}$-tpp wrt~$\Sigma$, we can prove $u_e\model\psi\EQ
v_e\model\psi$ for all $\psi\in\Sigma$ and $e\in E$ by induction on the complexity of~$\psi$. This implies
$W'\model\Gamma$, and $B^+(W')=B^+(F_{u_e})$ for any $e\in E$. Thus, either $B^+(W')=B^+(W)$, or $i>0$ and
$G_{i-1}(F_{u_e})$ by the induction hypothesis, which implies $G_{i-1}(W')$ by Claim~\ref{cl:good-equi}.
\end{Pf*}
\begin{Cl}\label{cl:good-adm}
$\Gamma\nadm_L\Delta$ if and only if
\begin{align}\label{eq:8}
\forall C\in\base_0(L)\:\forall E\sset\two^P,\lh E\le_0\lh C\:
&\exists W\in\Mod_L\:\bigl(G(W)\et W\text{ is a $\p{\refl(C),E}$-tp of $\nul$}\bigr)\\
\nonumber\et\forall\fii\in\Delta\:&\exists W\in\Mod_L\:\bigl(G(W)\et W\nmodel\fii\et\lh W\le n+1\bigr).
\end{align}
\end{Cl}
\begin{Pf*}

Right-to-left: For each $C\in\base_0(L)$ and $E\sset\two^P$ such that $\lh E\le_0\lh C$, there exists an $L$-model
$F_{C,E}\model\Gamma$ which is $\base^1(L)$-pseudoextensible wrt~$\Sigma$ by Claim~\ref{cl:good-model}. Likewise, for each
$\fii\in\Delta$, there exists a $\base^1(L)$-pseudoextensible wrt~$\Sigma$ $L$-model $F_\fii\model\Gamma$ such that
$F_\fii\nmodel\fii$. Let $F$ be the disjoint union of all the $F_{C,E}$'s and $F_\fii$'s. Then $F$ is an $L$-model
refuting $\Gamma\ru\Delta$ that is $\base(L)$-pseudoextensible wrt~$\Sigma$, hence $\Gamma\nadm_L\Delta$ by
\citi{Thm.~5.24}.

Left-to-right: Using \citi{Thm.~5.24} again, there exists a locally finite $\base(L)$-extensible $L$-model $F$ such
that $F\model\Gamma$, and $F\nmodel\fii$ for each $\fii\in\Delta$. For any $C$ and~$E$ as in~\eqref{eq:8}, let $W$ be a $\p{\refl(C),E}$-tp of~$\nul$ inside~$F$; then $G(W)$ by
Claim~\ref{cl:good-model}, as $F$ is in particular $\base^1(L)$-pseudoextensible wrt~$\Sigma$.

A similar argument shows that for any $\fii\in\Delta$, there is a finite rooted submodel $W\sgen F$ such that $G(W)$,
and $W\nmodel\fii$. Let $W_0\sset W$ be a subset
of size at most $n+1$ that contains a point in a final cluster of~$W$, and for each $\psi\in(B\bez
B^+(W))\cup\{\fii\}$, a point $u_\psi\nmodel\psi$ such that
$\psi$ holds in all clusters strictly above $\cls(u_\psi)$. Then $W_0$ is an $L$-model, and since $W$ is a chain of
clusters, it is easy to see that $W_0$ agrees with $W$ on the truth of all formulas from~$\Sigma$; in particular,
$W_0\model\Gamma$, and $W_0,u_\fii\nmodel\fii$. Moreover, we have~$G(W_0)$ using Claim~\ref{cl:good-equi}: for any $u\in
W_0$, $B^+((W_0)_u)=B^+(W_u)$.
\end{Pf*}

We will now use Claim~\ref{cl:good-adm} and the definition of~$G(W)$ to estimate the complexity of~$\nadm_L$. Assume first
that $L$ has cluster size bounded by~$k$, or that the number of parameters is bounded, in which case we put $k=2^{\lh
P}$. Then all the sets~$E$ referred to in~\eqref{eq:3} and~\eqref{eq:8} have size at most~$k$.
We may directly ``execute'' \eqref{eq:8}, \eqref{eq:7}, and~\eqref{eq:3} by an alternating Turing machine, with existential
and universal quantifiers implemented by nondeterministic choices in existential or universal states (respectively).

Since $\lh{B^+(W)}<n$, there are $O(n)$ alternations. For the universal quantifiers, we need $O(1)$ bits to specify~$C$,
$O(kn)$ bits to specify~$E$, and $O(\log n)$ bits to specify $\fii\in\Delta$. For the existential quantifiers, we need
$n^{O(1)}$ bits to specify the model $W$ of size $\le n+1$ on the second line of~\eqref{eq:8}; the remaining
existential quantifiers quantify over a single cluster of size~$\le k$ each, hence they can be specified with $O(kn)$
bits. The other conditions in the definition (e.g., truth of~$\Gamma$) may be tested deterministically in polynomial
time from the data. Thus, we obtain an alternating polynomial-time algorithm for checking $\Gamma\nadm_L\Delta$,
showing that it is in $\cxt{AP}=\psp$.

In the general case where both the cluster size and the number of parameters are unbounded, we need $O(2^n)$ bits to
specify the universally quantified~$E$ and the existentially quantified~$W'$. The former is expected as we are
shooting for $\coNEXP$, but the latter is a problem. Similarly to the proof of Lemma~\ref{lem:psext}, we solve it by
existentially quantifying only over the sets $B^+(W')\sset B$ instead of~$W'$ proper, which makes sense because of
Claim~\ref{cl:good-equi}.

In more detail, let $G'_i(B^+)$ (with $B^+\sset B$ and~$i\ge\lh{B^+}$) stand for $G_i(W)$ with
input $B^+(W)$ in place of~$W$. Put $V=\Sigma\cap\Var$. For any $v\in\two^V$, $e\in\two^P$, and $B^+\sset B$, let $A(v,e,B^+)$ denote the Boolean assignment that agrees with
$v$ and~$e$ on variables and parameters (respectively), and that makes $\Box\fii$ true for $\fii\in B^+$, and false for
$\fii\in B\bez B^+$. Then we can express $G'_i(B^+)$ as the conjunction of
\begin{multline*}
\forall E\sset\two^P\bez\{\nul\}\:\exists D\sset B^+\,\exists f\colon D\to E\\
\Bigl(\forall e\in E\:\exists v\in\two^V\:
   A(v,e,B^+\bez D)\model\ET\Gamma\land\ET_{\fii\in B^+\bez D}\fii\land\ET_{\substack{\fii\in D\\f(\fii)=e}}\neg\fii\\
  \et\bigl(D\ne\nul\TO G'_{i-1}(B^+\bez D)\bigr)\Bigr)
\end{multline*}
if $\nr\infty\in\base_1(L)$, a similar conjunct restricted to $\lh E\le k$ if $\nr k\in\base_1(L)$, and
the conjunct
\begin{multline*}
\forall e\in\two^P\:\exists v\in\two^V\,\exists D\sset B^+\:\Bigl(D=\{\fii\in B^+:A(v,e,B^+)\nmodel\fii\}\\
    \et A(v,e,B^+)\model\ET\Gamma\et\bigl(D\ne\nul\TO G'_{i-1}(B^+\bez D)\bigr)\Bigr)
\end{multline*}
if $\I\in\base_1(L)$. We may rewrite the first conjunct in~\eqref{eq:8} analogously.

In this way, we get an alternating algorithm for $\nadm_L$ using time $2^{n^{O(1)}}$, out of
which time $t_\exists(n)=n^{O(1)}$ is spent in existential states. We can simulate the existential states by
deterministically trying all possible choices; this will multiply the overall running time by a factor of
$2^{t_\exists(n)}$, which is still $2^{n^{O(1)}}$, hence we obtain a $\coNEXP$ algorithm.
\end{Pf}

\begin{Rem}\label{rem:ub-lin-clx-parfree}
Admissibility in linear clx logics with no parameters at all is even easier than $\psp$, viz.~$\conp$, by
\cite[Thm.~2.6]{ej:admcomp}.
\end{Rem}

\subsection{Logics of bounded depth}\label{sec:logics-bounded-depth}

Our basic goal is to completely determine the complexity of admissibility and unification for clx logics. As we
will see later, the only cases remaining where we can improve upon the upper bounds proved so far is for clx logics
of branching~$0$ (i.e., depth~$1$) when, moreover, the cluster size or the number of parameters is bounded. This is a
tiny class of problems, most of which are not really interesting. For this reason, we broaden the scope of this section
beyond clx logics, and we will look at larger classes of logics of bounded depth. We are primarily interested in cases
where we can get the complexity inside the polynomial hierarchy, but we will also include for completeness some results
of higher complexity, analogous to clx logics.

The added generality comes at a cost: the analysis of admissible rules from \citi{} does not apply to non-clx logics,
hence we need to come up with a relevant theory of admissible rules for logics of bounded depth. Fortunately, this is
fairly easy to do. The key factor is that logics of bounded depth are \emph{locally tabular}:
\begin{Thm}[{{\cite[Thm.~12.21]{cha-zax}}}]\label{thm:bddp-loc-tab}
Every logic~$L$ of finite depth is locally tabular, i.e., there are only finitely many nonequivalent formulas over any
finite set of atoms, or equivalently, the canonical frames $C_L(P,V)$ are finite for any finite $P\sset\Par$ and
$V\sset\Var$ (and more generally, all finitely generated refined $L$-frames are finite).

Consequently, all logics~$L$ of finite depth, as well as atomic multi-conclusion consequence relations that extend them
(such as $\adm_L$), have the finite model property.
\noproof\end{Thm}

The finite model property with respect to admissibility avoids most of the trouble we had to go through in \citi{}.

Let us start with a simple result on unifiability in \emph{all} logics of bounded depth as a teaser.
\begin{Thm}\label{thm:ub-unif-bdpar-bddp}
If $L$ is a logic of bounded depth, then $L$-unifiability with a constant number of parameters is in~$\np$.

In particular, this holds if $L$ is a clx logic of branching~$0$.
\end{Thm}
\begin{Pf}
A formula~$\fii$ with parameters from a fixed finite set~$P$ is $L$-unifiable iff it has a variable-free unifier iff
there is a (definable) valuation in the canonical frame $C_L(P,\nul)$ that makes $\fii$ true in all points of the
model. By Theorem~\ref{thm:bddp-loc-tab}, $C_L(P,\nul)$ is a fixed finite frame, which makes the condition above checkable
in~$\np$.
\end{Pf}

Subsequent complexity results will only apply under further
restrictions, not to all logics of bounded depth. When the number of parameters is unlimited, we will see in the next
section that $L$-unifiability is $\NEXP$-hard or $\coNEXP$-hard whenever $L$ has width $\ge2$ or unbounded cluster
size (Theorems \ref{thm:lb-nexp} and~\ref{thm:lb-conexp}). Thus, we may hope to get better complexity only for linear logics of bounded cluster size, which in view of
bounded depth means that $L$ is tabular.

Contrary to Theorem~\ref{thm:ub-unif-bdpar-bddp}, we need some restrictions even if the number of parameters is finite when
we discuss admissibility rather than unifiability: we know from~\cite{ej:admcomp} that logics such as $\lgc{S4BD_3}$
have a $\coNEXP$-hard admissibility problem already with no parameters at all. Even if we just wanted results
analogous to the previous sections, placing admissibility on a low level of the exponential hierarchy, this does not
seem possible for general logics of bounded depth: cf.\ the construction of logics of depth~$d$
in~\cite[\S18.4]{cha-zax} such that some nonderivable formulas require countermodels whose size is a tower of
exponentials of height~$d-2$. With these examples in mind, it seems we
can only expect reasonably general results if we restrict attention to logics of bounded width (which is equivalent to
bounded branching for logics of bounded depth).

But first, we need a rudimentary theory of admissibility in logics of bounded depth.
\begin{Def}\label{def:bddp-ext}
Let $L$ be a logic of bounded depth, and assume that $\Par$ is finite. A finite parametric $L$-frame~$F$ has a
\emph{loosely separated root} if it is rooted, and points of $\rcl(F)$ have pairwise distinct valuations of parameters.
The frame~$F$ has a \emph{separated root} if, moreover, it is \emph{not} the case that $F\bez\rcl(F)$ is a rooted frame
with reflexive root cluster~$C$, and for each $w\in\rcl(F)$ there is $x\in C$ such that $w\equiv_\Par x$.

A finite parametric $L$-frame~$W$ is \emph{$L$-extensible} (or \emph{strongly $L$-extensible}) if for every parametric
$L$-frame $F$ with a separated root (a loosely separated root, resp.) such that $F\bez\rcl(F)\sgen W$, $\id_{F\bez\rcl(F)}$ extends to an isomorphism of $F$ onto a generated subframe of~$W$.
\end{Def}

While the concept of strong $L$-extensibility looks simpler, and it is easier to work with, $L$-extensibility is a
structurally more fundamental property due to the following characterization: $L$-extensible frames are exactly the
finitely generated \emph{injective} $L$-frames, or duals of finitely generated projective $L$-algebras.
\begin{Lem}\label{lem:bddp-ext}
Let $L$ be a logic of bounded depth, and assume that $\Par$ is finite. The following are equivalent for any finite
parametric $L$-frame $W$.
\begin{enumerate}
\item\label{item:6}
$W$ is $L$-extensible.
\item\label{item:7} For every pair of finite parametric $L$-frames $F_0\sgen F$, and a p-morphism $f_0\colon F_0\to W$,
there exists a p-morphism $f\colon F\to W$ such that $f\Sset f_0$.
\end{enumerate}
\end{Lem}
\begin{Pf}

\ref{item:6}\txto\ref{item:7}:
We may extend the p-morphism to one cluster after another in a top-down fashion, hence we may assume that $F$ is
rooted, and $F_0=F\bez\rcl(F)$. Let $F'$ be $f_0[F_0]\sgen W$ extended with a copy of~$\rcl(F)$ as a new root cluster: then
$f_0$ extends to a p-morphism of $F$ onto~$F'$, hence replacing $F$ with~$F'$, we may assume without loss of generality
that $F_0\sgen W$, and $f_0=\id_{F_0}$. We may shrink the root cluster of~$F$ by a p-morphism identifying all points
that satisfy the same parameters, hence we may assume that points of $\rcl(F)$ have pairwise distinct valuations of
parameters. If $F_0$ has a reflexive root cluster, and for each $w\in\rcl(F)$ there exists $x_w\in\rcl(F_0)$ such that
$w\equiv_\Par x_w$, then $\id_{F_0}$ extended with the mapping $w\mapsto x_w$ is a p-morphism $F\to F_0$. Otherwise $F$
has a separated root, hence $f_0=\id_{F_0}$ extends to a p-morphism $F\to W$ using~\ref{item:6}.

\ref{item:7}\txto\ref{item:6}:
Assume that $F$ has a separated root, and $F_0=F\bez\rcl(F)\sgen W$. Using~\ref{item:7}, $\id_{F_0}$ extends to a
p-morphism $f\colon F\to W$. Since $F$ has a separated root, $f$ cannot identify distinct points of~$\rcl(F)$, and it
cannot map any point of~$\rcl(F)$ to~$F_0$, thus it is an isomorphic embedding in~$W$.
\end{Pf}
\begin{Lem}\label{lem:bddp-can-ext}
Let $L$ be a logic of bounded depth, and assume that $\Par$ is finite.
\begin{enumerate}
\item\label{item:8}
For any finite $V\sset\Var$, the canonical parametric frame $C_L(\Par,V)$ is $L$-extensible.
\item\label{item:9}
If $W$ is an $L$-extensible finite parametric $L$-frame, there exists a finite $V\sset\Var$ such that $W$ is a
p-morphic image of~$C_L(\Par,V)$.
\end{enumerate}
\end{Lem}
\begin{Pf}

\ref{item:8}:
The frame $C_L(\Par,V)$ is finite by Theorem~\ref{thm:bddp-loc-tab}. If $F$ is a finite $L$-frame with a separated root such
that $F_0=F\bez\rcl(F)\sgen C_L(\Par,V)$, let $\model$ be a valuation in~$F$ that agrees in~$F_0$ with the canonical
valuation of $C_L(\Par,V)$, and is arbitrary in~$\rcl(F)$. Define a mapping $f\colon F\to C_L(\Par,V)$ by
$f(x)=\sat_{\Form(\Par,V)}(F,x)$. Then $f$ is a p-morphism of $F$ to~$C_L(\Par,V)$ extending $\id_{F_0}$. As in
Lemma~\ref{lem:bddp-ext}, $f$ is an isomorphic embedding because $F$ has separated root.

\ref{item:9}:
By \citi{Lem.~2.3}, $W$ is isomorphic to a generated subframe $W'\sgen C_L(\Par,V)$ for sufficiently large finite
$V\sset\Var$. The isomorphism $W'\to W$ extends to a p-morphism $C_L(\Par,V)\to W$ by Lemma~\ref{lem:bddp-ext}.
\end{Pf}
\begin{Rem}\label{exm:bddp-ext-warn}
In fact, $L$-extensible frames are exactly the \emph{retracts} of the finitely generated canonical frames $C_L(\Par,V)$.

It is not generally true that p-morphic images of $C_L(\Par,V)$ are $L$-extensible. For example, let $F_0$ be the
disjoint union of the two-element cluster and the two-element reflexive chain, and let $L$ be the logic of~$F_0$. If
$V\ne\nul$, then one of the connected components of $C_L(\Par,V)$ is a two-element cluster where both elements satisfy
the same parameters. We may contract this cluster to a single point by a p-morphism, but then the resulting frame is
not $L$-extensible (if $\Par\ne\nul$). For much the same reason, there exist inadmissible rules $\Gamma\nadm_L\Delta$
with $\Gamma\cup\Delta\sset\Form(\Par,V)$ such that there is no projective formula $\theta\in\Form(\Par,V)$ such that
$\theta\vdash_L\ET\Gamma$, and $\theta\nvdash_L\fii$ for each $\fii\in\Delta$. An explicit example is the rule
$p\lor\dia x\to\Box x\ru x$ for $\Par=\{p\}$ and $V=\{x\}$.

It is easy to show (e.g.\ using the frame rules from~\cite{ej:canrules}) that p-morphic images of finitely generated
canonical frames are exactly the finite $L$-frames that validate all $L$-admissible rules.

The canonical frames $C_L(\Par,V)$ are in general not strongly $L$-extensible. For example, let
$L=\lgc{S4.3}\oplus\lgc{BD_2}$. Then $C_L(\Par,V)$ contains a cluster~$C$ such that every possible valuation of
$\Par\cup V$ is realized by a point of~$C$. There cannot be any cluster below~$C$, as any such could be p-morphically
contracted into~$C$ while preserving the canonical valuation. Thus, $C_L(\Par,V)$ is not strongly $L$-extensible.
\end{Rem}
\begin{Lem}\label{lem:bddp-emd-strong-ext}
Let $L$ be a logic of bounded depth, and assume that $\Par$ is finite. Then any finite parametric $L$-frame is a
generated subframe of a strongly $L$-extensible finite parametric $L$-frame.
\end{Lem}
\begin{Pf}
Let $W_0$ be the given frame, and $d$ be the depth of~$L$. We define a sequence of parametric $L$-frames $W_0\sgen W_1\sgen\dots\sgen W_d$ by
induction as follows: for every parametric $L$-frame~$F$ with a loosely separated root such that $F\bez\rcl(F)\sgen
W_i$, but $\id_{F\bez\rcl(F)}$ does not extend to an isomorphism of~$F$ to a generated subframe of~$W_i$, we add a copy
of $\rcl(F)$ to~$W_{i+1}$. (We do not distinguish among isomorphic frames~$F$: they all get a single common copy
of~$\rcl(F)$.)

Since all the clusters being added have size at most~$2^{\lh\Par}$, there are only finitely many choices for~$F$ up to
isomorphism, hence all the frames~$W_i$ are finite. Moreover, we see by induction on~$i$ that all frames $F$ as above
with $F\bez\rcl(F)\sgen W_i$ such that the depth of~$F$ is less than~$i$ already have a copy in~$W_i$, i.e., only
clusters of depth $\ge i+1$ are added to~$W_{i+1}$. It follows that $W_d$ is strongly $L$-extensible.
\end{Pf}

Each of the classes of finitely generated canonical frames, p-morphic images of canonical frames, $L$-extensible
frames, or strongly $L$-extensible frames, provides an adequate semantics for admissible rules. We will mostly work with
strongly $L$-extensible frames, as the other possibilities are not suitable for our purposes for various reasons:
canonical frames are astronomically large, and are not flexible enough so that they could be constructed from small
pieces as needed; p-morphic images of canonical frames are difficult to algorithmically recognize. There is not much
difference between $L$-extensible and strongly $L$-extensible frames, however strongly $L$-extensible frames have a
somewhat simpler definition, and are easier to manipulate.
\begin{Thm}\label{thm:bddp-adm-ext}
Let $L$ be a logic of bounded depth, and assume that $\Par$ is finite. The following are equivalent for any rule
$\Gamma\ru\Delta$:
\begin{enumerate}
\item\label{item:10}
$\Gamma\adm_L\Delta$.
\item\label{item:11}
$\Gamma\ru\Delta$ holds in all $L$-extensible finite parametric $L$-frames.
\item\label{item:12}
$\Gamma\ru\Delta$ holds in all strongly $L$-extensible finite parametric $L$-frames.
\end{enumerate}
\end{Thm}
\begin{Pf}

\ref{item:10}${}\eq{}$\ref{item:11}:
By \citi{Lem.~2.2}, $\Gamma\ru\Delta$ is $L$-admissible iff it holds in $C_L(\Par,V)$ for all
finite sets of variables~$V$. Thus, on the one hand, if $\Gamma\adm_L\Delta$, then $\Gamma\ru\Delta$ holds in all
$L$-extensible frames by Lemma~\ref{lem:bddp-can-ext}, as the validity of rules is preserved by p-morphic images. On the
other hand, if $\Gamma\ru\Delta$ holds in $L$-extensible frames, it holds in particular in every $C_L(\Par,V)$ by
Lemma~\ref{lem:bddp-can-ext}, hence $\Gamma\adm_L\Delta$.

\ref{item:11}\txto\ref{item:12} is trivial. \ref{item:12}\txto\ref{item:11}: Assume that $\Gamma\ru\Delta$ holds in all
strongly $L$-extensible frames, and let $W$ be an $L$-extensible frame. By Lemma~\ref{lem:bddp-emd-strong-ext},
there exists a strongly $L$-extensible frame~$W'$ such that $W\sgen W'$. By Lemma~\ref{lem:bddp-ext}, there exists a
p-morphism $f\colon W'\to W$ extending~$\id_W$. Thus, $W'\model\Gamma\ru\Delta$ implies $W\model\Gamma\ru\Delta$.
\end{Pf}
\begin{Rem}\label{rem:bddp-adm-infpar}
We can, and will, apply the characterization from Theorem~\ref{thm:bddp-adm-ext} even if $\Par$ is infinite, by considering
frames with valuation of a finite set of parameters $P\sset\Par$, which will be formally accomplished by shrinking
$\Par$ to a finite subset for the duration of an argument. We formulated the criteria the way we did for notational
convenience.
\end{Rem}

\begin{Lem}\label{lem:bddp-frame-flas}
Any logic~$L$ of finite depth~$d$ is axiomatizable over $\lgc{K4BD}_d$ by a set of frame formulas
$\{\alpha^\sharp(F_i,\bot):i\in I\}$.
\end{Lem}
\begin{Pf}
Let $\{F_i:i\in I\}$ be an enumeration of finite rooted frames of depth~$d$ such that $F_i\nmodel L$, and put
$L'=\lgc{K4BD}_d\oplus\{\alpha^\sharp(F_i,\bot):i\in I\}$. On the one hand, $F_i\nmodel L$ implies
$\vdash_L\alpha^\sharp(F_i,\bot)$, hence $L'\sset L$. Assume for contradiction that $\vdash_L\fii$, but
$\nvdash_{L'}\fii$. Being a logic of bounded depth, $L'$ has the finite model property, hence there exists a finite
rooted $L'$-frame $F\nmodel\fii$ of depth~$d$. Since $\vdash_L\fii$, $F$ is not an $L$-frame. But then
$\alpha^\sharp(F,\bot)$ is an axiom of~$L'$, which implies that $F$ is not an $L'$-frame after all, a contradiction.
\end{Pf}

We now focus on logics of bounded depth and width. All such logics are finitely axiomatizable (see
\cite[Thm.~5.16]{ej:sfef}), and therefore their finite frames are $\conp$-recognizable. In fact, we can improve this
to~$\ptime$:
\begin{Lem}\label{lem:bddp-bdwd-ptime}
Let $L$ be a logic of bounded depth and width. Then finite $L$-frames are recognizable in polynomial time.
\end{Lem}
\begin{Pf}
By \cite[Thm.~5.16]{ej:sfef} and Lemma~\ref{lem:bddp-frame-flas}, $L$ is axiomatizable by $\lgc{K4BD}_k\lgc{BW}_k$ for some
constant~$k$, plus finitely many frame formulas $\alpha^\sharp(F_i,\bot)$. We can compute the depth and width
of a finite frame in polynomial time. Moreover, given a rooted frame~$W$ of width and depth at most~$k$, we can
recognize in polynomial time if $W$ reduces to~$F_i$: if $F_i$ has no proper clusters, then for any p-morphism $f\colon
W\onto F_i$, preimages of all points are unions of clusters of~$W$. Thus, there are only $O(1)$ choices for~$f$, which
are easy to check. In the general case, let $F'_i$ be $F_i$ with all clusters reduced to size~$1$. Then a p-morphism
$W\to F_i$ induces a p-morphism $W\to F'_i$; conversely, a p-morphism $W\to F'_i$ is induced by a p-morphism $W\to F_i$
if and only if the maximal clusters in the preimage of each point of~$F'_i$ have size at least the size of the
corresponding cluster of~$F_i$. This is again easy to check in polynomial time.
\end{Pf}
\begin{Cor}\label{cor:bddp-bdwd-cls-exp}
For any logic $L$ of bounded depth and width, there exists a constant~$c$ with the following property: whenever $W$ is
a Kripke $L$-frame, and $C$ is a cluster of~$W$ of size $\lh C\ge c$, then the frames obtained from $W$ by enlarging
$C$ to arbitrary cardinality are also $L$-frames.
\end{Cor}
\begin{Pf}
Let $c$ be the maximum of cluster sizes of the frames $F_i$ in the proof of Lemma~\ref{lem:bddp-bdwd-ptime}.
\end{Pf}

Incidentally, since logics of bounded depth and width have the polynomial-size model property (see
e.g.~\cite[L.~5.13]{ej:sfef}), Lemma~\ref{lem:bddp-bdwd-ptime} implies that all such logics are in~$\conp$. (This will also
follow from Theorem~\ref{thm:ub-bdpar-bddp-bdwd}.)
\begin{Cor}\label{cor:ub-der-bddp-bdwd}
If $L$ is a logic of bounded depth and width, then $L$-derivability is in $\conp$.
\noproof\end{Cor}
\begin{Lem}\label{lem:bddp-bdwd-bdpar-polysize}
Let $L$ be a logic of bounded depth and width, and assume that $\Par$ is finite. There is a polynomial $p(x)$ such that
whenever $\Gamma\ru\Delta$ is a rule of length~$n$, and $\Gamma\nadm_L\Delta$, then $\Gamma\ru\Delta$ is falsified in
a strongly $L$-extensible parametric $L$-frame of size at most~$p(n)$.
\end{Lem}
\begin{Pf}
Put $\Sigma=\Sub(\Gamma\cup\Delta)$, $B=\{\fii:\Box\fii\in\Sigma\}$, and $n=\lh\Sigma$. Let $\model$ be a valuation
falsifying $\Gamma\ru\Delta$ in a strongly $L$-extensible parametric finite $L$-frame~$W$, using
Theorem~\ref{thm:bddp-adm-ext}. For each $\fii\in\Delta$, fix $x_\fii\in W$ such that $x_\fii\nmodel\fii$, and put
$W_0=\bigcup_{\fii\in\Delta}x_\fii\Up$.

Starting with~$W_0$, we define a sequence $W_0\sset W_1\sset\cdots\sset W_d$ of generated subframes of~$W$ as follows:
$W_{i+1}\Sset W_i$, and whenever $F\sgen W$ has a loosely separated root, $F\bez\rcl(F)\sset W_i$, but
$\id_{F\bez\rcl(F)}$ does not extend to a p-morphic embedding $F\to W_i$, we add a copy of $\rcl(F)$ to~$W_{i+1}$ using
strong $L$-extensibility of~$W$. As in Lemma~\ref{lem:bddp-emd-strong-ext}, we do not distinguish among isomorphic~$F$.

By the same argument as in Lemma~\ref{lem:bddp-emd-strong-ext}, the frame $W_d$ is strongly $L$-extensible. Let $s_i$ be
the number of clusters in~$W_i$. Since $L$ has bounded depth and width, any rooted $L$-frame has only $O(1)$ clusters,
thus $s_0=O(\lh\Delta)=O(n)$. When constructing $W_{i+1}$ from~$W_i$, each $F\bez\rcl(F)$ is generated by an antichain
in~$W_i$; since $L$ has bounded width~$w$, there are at most $s_i^{w+1}$ choices for $F\bez\rcl(F)$. For each of them,
there are at most $2^{2^{\lh\Par}}+2^{\lh\Par}=O(1)$ choices for $\rcl(F)$ up to isomorphism. Thus,
$s_{i+1}=s_i^{O(1)}$. Since a constant number of iterations of a polynomial is still polynomial, $s_d=n^{O(1)}$.

We construct a submodel $W'\sset W_d$ as follows:
\begin{itemize}
\item The points $x_\fii$, $\fii\in\Delta$, are included in~$W'$.
\item For each cluster $C$ of~$W_d$, and each assignment $e\in\two^\Par$ that is realized in~$C$, we include in~$W'$ at
least one point from~$C$ satisfying~$e$.
\item For each cluster $C$ of~$W_d$, and each formula $\fii\in B$ that is falsified in some point of~$C$,
we include at least one such point in~$W'$.
\item All clusters of~$W_d$ of size $\le c$ are fully included in~$W'$, where $c$ is the constant from
Corollary~\ref{cor:bddp-bdwd-cls-exp}.
\item For each cluster $C$ of~$W_d$ of size $\ge c$, we make sure $W'$ includes at least $c$ points of~$C$.
\end{itemize}
We may choose $W'$ satisfying these requirements so that it has cluster size at most $\lh
B+2^{\lh\Par}+\lh\Delta+c=O(n)$, thus the overall size of~$W'$ is polynomial in~$n$. Notice that $W'$ has the same
skeleton as~$W_d$: we merely shrunk large clusters to control their size.

The choice of~$W'$ ensures that it is an $L$-model, and that it agrees with~$W$ on the satisfaction of formulas
from~$\Sigma$; in particular, $\Gamma\ru\Delta$ fails in~$W'$. We claim that $W'$ is strongly $L$-extensible. Let $F'$
be a parametric $L$-frame with a loosely separated root such that $F'\bez\rcl(F')\sgen W'$. Let $F$ denote
$(F'\bez\rcl(F'))\Up\sgen W_d$ with $\rcl(F')$ attached below (i.e., we expand the clusters of~$W'$ included in
$F'\bez\rcl(F')$ back to what they were in~$W_d$). By Corollary~\ref{cor:bddp-bdwd-cls-exp} and the choice
of~$W'$, $F$ is still an $L$-frame with a loosely separated root. Since $W_d$ is strongly $L$-extensible,
$\id_{F\bez\rcl(F)}$ extends to a p-morphic embedding $f\colon F\to W_d$. Since all points of $\rcl(F)=\rcl(F')$ are
distinguished by valuation of parameters, all of~$\rcl(F)$ is included in~$W'$, thus $f\colon F'\to W'$.
\end{Pf}

\begin{Thm}\label{thm:ub-bdpar-bddp-bdwd}
Let $L$ be a logic of bounded depth and width. Then $L$-admissibility with a constant number of parameters is
in~$\conp$.

In particular, this holds if $L$ is a clx logic of branching~$0$.
\end{Thm}
\begin{Pf}
By Lemma~\ref{lem:bddp-bdwd-bdpar-polysize}, we have $\Gamma\nadm_L\Delta$ iff $\Gamma\ru\Delta$ is falsified in a
strongly $L$-extensible $L$-model~$W$ of polynomial size. Clearly, we can check in polynomial time that
$W\nmodel\Gamma\ru\Delta$, and by Lemma~\ref{lem:bddp-bdwd-ptime}, we can check that $W$ is an $L$-model.

We can also check in polynomial time that $W$ is strongly $L$-extensible: by the same argument as in the proof of
Lemma~\ref{lem:bddp-bdwd-bdpar-polysize}, there are only polynomially many choices for a parametric $L$-frame $F$ with a
loosely separated root such that $F\bez\rcl(F)\sgen W$. We can generate each such frame in polynomial time, check that
it is indeed an $L$-frame with a loosely separated root, and check if we can find a suitable copy of $\rcl(F)$
inside~$W$.
\end{Pf}

The situation for admissibility with unrestricted number of parameters is more complicated. As we already mentioned,
cluster size and linearity will come into play. Notice that logics of bounded depth, width, and cluster size are
tabular.

\begin{Thm}\label{thm:ub-tab}
Let $L$ be a tabular logic. Then $L$-admissibility is decidable in $\coNEXP$, and $L$-unifiability in $\NEXP$.
\end{Thm}
\begin{Pf}
Let us fix a rule $\Gamma\ru\Delta$, and put $\Sigma=\Sub(\Gamma\cup\Delta)$ and $n=\lh\Sigma$. In order to apply the
criterion from Theorem~\ref{thm:bddp-adm-ext}, we will proceed as if $\Par$ were finite, consisting of only the parameters
from~$\Sigma$ (cf.\ Remark~\ref{rem:bddp-adm-infpar}). However, we bear in mind that $\lh\Par$ is bounded by~$n$, not by a constant, for complexity estimates.

We claim that if $\Gamma\nadm_L\Delta$, there exists a strongly $L$-extensible $L$-model~$W$ falsifying
$\Gamma\ru\Delta$ of size $2^{O(n)}$. This can be shown by constructing a sequence of models $W_0\sgen
W_1\sgen\dots\sgen W_d$ as in the proof of Lemma~\ref{lem:bddp-bdwd-bdpar-polysize}. Since $L$ has bounded cluster size,
there are only $2^{O(n)}$ possible root clusters in frames with loosely separated roots; with this in mind, it is easy
to compute that each $W_i$ has size $2^{O(n)}$. (We do not need to select $W'\sset W_d$, as $W_d$ already has bounded
cluster size.)

Given a model $W$ of size $2^{O(n)}$, it is easy to check in deterministic exponential time that $W$ is an $L$-frame,
and that $W\nmodel\Gamma\ru\Delta$. We may also check in exponential time if $W$ is strongly $L$-extensible:
by the same counting argument as above, there are $2^{O(n)}$ possible frames~$F$ that we need to test, and we can
generate and check them in exponential time.
\end{Pf}

\begin{Thm}\label{thm:ub-lin-bddp}
Let $L$ be a linear logic of depth~$d$. Then $L$-admissibility is decidable in $\NEXP$, and $L$-unifiability in
$\coNEXP$.

If $L$ has bounded cluster size (i.e., it is tabular), then $L$-admissibility is in~$\Sigma^p_{2d}$, and
$L$-unifiability in~$\Pi^p_{2d}$.
\end{Thm}
\begin{Pf}
We use a variant of the algorithm presented in Theorem~\ref{thm:ub-lin-clx}. We fix $\Gamma\ru\Delta$, and put
$\Sigma=\Sub(\Gamma\cup\Delta)$, $B=\{\fii:\Box\fii\in\Sigma\}$, and $n=\lh\Sigma$. We will again proceed as if $\Par$
were finite of size $\le n$.

Given a finite parametric $L$-frame~$F$, let $\Ext_L(F)$ denote the set of single-cluster parametric frames~$C$ such
that $F^C$ is an $L$-frame with a loosely separated root, where $F^C$ denotes $F$ with $C$ added as a new root cluster. Using
Lemma~\ref{lem:bddp-bdwd-ptime}, it is polynomial-time decidable if $C\in\Ext_L(F)$. If $W$ is an $L$-model, let $\frx(W)$
denote its underlying parametric frame. By abuse of notation, we will write $\Ext_L(W)$ instead of $\Ext_L(\frx(W))$.

Similarly to the proof of Theorem~\ref{thm:ub-lin-clx}, but avoiding the complicated set-up using $B^+(W)$ to guard the
search, we define predicates $G_i(W)$ for $W$ a finite rooted $L$-model, and $i<d$, such that $G_0(W)$ is always
true, and
\begin{equation}\label{eq:14}
G_{i+1}(W)\iff
  \forall C\in\Ext_L(W)\:\exists W'\,\bigl(W\sgen W'\model\Gamma\et\frx(W')=\frx(W)^C\et G_i(W')\bigr).
\end{equation}
We define $G(W)$ by
\begin{equation}\label{eq:15}
G(W)\iff W\model\Gamma\et\forall u\in W\:G_{d-1}(W_u).
\end{equation}
\begin{Cl}\label{cl:ub-lin-bddp-char}
$\Gamma\nadm_L\Delta$ if and only if
\begin{align}\label{eq:13}
\forall C\in\Ext_L(\nul)\:&\exists W\in\Mod_L\:\bigl(G(W)\et\frx(W)=C\bigr)\\
\nonumber\et\forall\fii\in\Delta\:&\exists W\in\Mod_L\:\bigl(G(W)\et W\nmodel\fii\bigr).
\end{align}
Moreover, we may restrict the size of $W$ on the second line of~\eqref{eq:13} to $n+O(1)$.
\end{Cl}
\begin{Pf*}

Left-to-right: Let $U$ be a finite strongly $L$-extensible $L$-model falsifying $\Gamma\ru\Delta$, using
Theorem~\ref{thm:bddp-adm-ext}. First, we show by induction on~$i$ that if $u\in U$ is of depth at least~$d-i$, then
$G_i(U_u)$ holds. This is trivial for~$i=0$. For the induction step, if $i>0$, and $C\in\Ext_L(U_u)$, then by
strong $L$-extensibility of~$U$, we can find a copy of $C$ in~$U$ so that the model $W=C\Up$  satisfies $U_u\sgen W$
and $\frx(W)=\frx(U_u)^C$. Since $W\sgen U$, $W\model\Gamma$. We have $G_{i-1}(W)$ by the induction
hypothesis as $\rcl(W)$ has depth at least $d-i+1$.

By taking $i=1$, it follows that $G(U_u)$ for all $u\in U$. Then the first line of~\eqref{eq:13} holds by
$L$-extensibility of~$U$, and the second one as each $\fii\in\Delta$ is falsified in~$U$.

Concerning the bound on~$\lh W$, let $W=U_u$ for some $u\nmodel\fii$. If $L$ is tabular, $\lh W=O(1)$ and we are done.
In general, $W$ is a chain of $j\le d$ clusters $C_j<C_{j-1}<\dots<C_1$. We first modify the model~$U$ to a model~$U'$
so that for each $1\le i<j$, we put a new copy of the submodel $C_{i+1}\Down$ below~$C_i$, and likewise, we include a
copy of~$C_1\Down$ as a new connected component of~$U'$. Then we shrink each original cluster $C_i$ to a subcluster $C'_i\sset C_i$ such
that:
\begin{itemize}
\item $\lh{C'_i}\ge c$, where $c$ is the constant from Corollary~\ref{cor:bddp-bdwd-cls-exp}.
\item $u\in C'_j$.
\item For each formula $\psi\in B$ such that $W\nmodel\psi$, let $i$ be minimal such that $C_i\Up\nmodel\psi$. Then
$C'_i$ includes a point falsifying~$\psi$.
\end{itemize}
Let $W''=\bigcup_iC'_i\sset W$, and $U''=(U'\bez W)\cup W''\sset U'$. Clearly, $U''$ is an $L$-frame, $W''\sgen U''$, and the
construction ensures that $\sat_\Sigma(U',x)=\sat_\Sigma(U'',x)$ for each $x\in U''$; in particular, $U''\model\Gamma$,
and $W''\nmodel\fii$. We have $\lh{W''}\le cd+n$.

We claim that $U''$ is strongly $L$-extensible, which implies $G(U''_x)$ for all $x\in U''$ by the first part of
the proof, and, in particular, $G(W'')$. So, let $F'_0\sgen U''$, and $C\in\Ext_L(F'_0)$. Let $F_0\Sset F'_0$ be the
frame obtained by replacing each $C'_i$ with~$C_i$ (if it is included in $F'_0$). By Corollary~\ref{cor:bddp-bdwd-cls-exp},
$C\in\Ext_L(F_0)$. If $F_0\sset U$, strong 
$L$-extensibility of $U$ ensures there is an isomorphism~$f$ of $F_0^C$ to a generated subframe of~$U$ extending $F_0$.
If $f[C]=C_i$ for some~$i$, we may modify $f$ so that $f[C]$ is instead the new copy of~$C_i$ included
in~$U'$. Thus, we may assume that $f[C]$ is disjoint from~$W$, hence $f[C]\sset U''$. But then $f$ maps $F_0^{\prime C}$
to a generated subframe of~$U''$ extending $F'_0$.

If $F_0\nsset U$, it is a rooted frame whose root cluster is inside one of the new copies of $C_i\Down$ in~$U'$. Then
strong $L$-extensibility of~$U$, applied in the original copy of $C_i\Down$ and transferred back, ensures that
$\id_{F_0}$ extends to an isomorphism $f$ of $F_0^C$ to a generated subframe of~$U'$. Since $f[C]$ is also in the
new copy of $C_i\Down$, it is included in~$U''$, hence $f$ restricts to an isomorphism of $F_0^{\prime C}$ to a generated
subframe of~$U''$.

Right-to-left: For each $C\in\Ext_L(\nul)$, we fix a model $W_C$ based on~$C$ such that $G(W_C)$, and for each
$\fii\in\Delta$, we fix a model $W_\fii\nmodel\fii$ such that $G(W_\fii)$. For each $W_C$ and~$W_\fii$, we unwind the
definition of~$G$ to extend the models downwards with all the ``tight
predecessors'' to an upside-down tree of clusters. We let $U$ be the disjoint sum of the models we obtain. Clearly, $U$
is a finite $L$-model such that $U\model\Gamma$, and $U\nmodel\fii$ for each~$\fii\in\Delta$. By induction on~$i$, we
see that any point $u\in U$ of depth $i$ satisfies $G_j(U_u)$ for some $j\ge d-i$, which was unwinded in the course of
the construction of~$U$. It follows that $U$ is strongly $L$-extensible: let $F_0\sgen U$, and $C\in\Ext_L(F_0)$. If
$F_0=\nul$, then a copy of $C$ was added to~$U$ as~$W_C$. Otherwise $F_0$ is a rooted $L$-model of depth at most $d-1$,
hence $G_j(F_0)$ for some $j\ge1$, hence $F_0$ was extended with a copy of~$C$ while unwinding it.
\end{Pf*}

As in the proof of Theorem~\ref{thm:ub-lin-clx}, \eqref{eq:14}--\eqref{eq:13} give an algorithm deciding
$\Gamma\nadm_L\Delta$ on an alternating Turing machine. If $L$ is tabular, the clusters~$C$ and models $W$ or~$W'$
appearing in \eqref{eq:14}--\eqref{eq:13} have bounded size, hence they can be specified using $O(n)$ bits (encoding
the relevant valuations of parameters or variables). Thus, the algorithm works in alternating polynomial time. Since it
consists of $d$~pairs of universal quantifiers followed by existential quantifiers, it is a $\Pi^p_{2d}$~algorithm.

If $L$ has unbounded cluster size, the quantifiers over $C$, $W$, and~$W'$ in \eqref{eq:14}--\eqref{eq:13} have
exponential size, hence as such, we only get a $\Pi^{\exp}_{2d}$ algorithm. We keep the $\forall C$ quantifiers,
but we will need something more efficient for the existential quantifiers. We can do it in the same way as in
Lemma~\ref{lem:psext} or Theorem~\ref{thm:ub-lin-clx}: instead of specifying the full models $W$ or~$W'$, we only specify the
set of formulas $\boxdot\psi$, $\psi\in B$, that they satisfy, while checking separately that this choice of the subset
of~$B$ is satisfiable in the frame, and validates~$\Gamma$. Of course, we also need to know the underlying frame of the
model, but in \eqref{eq:14} and on the first line of~\eqref{eq:13}, this is already fully specified by~$C$. The models
$W$ on the second line of~\eqref{eq:13} may be written down explicitly, as they have polynomial size.

In this way, the existential quantifiers in the algorithm become polynomial-size, hence they can be simulated in
deterministic exponential time by trying all possibilities. The universal quantifiers remain, hence we obtain a
$\coNEXP$ algorithm for $\Gamma\nadm_L\Delta$.
\end{Pf}
\begin{Cor}\label{cor:ub-clx-br0-bdcl}
If $L$ is a clx logic of branching~$0$ and bounded cluster size, then $L$-admissibility is in~$\Sigma^p_2$, and
$L$-unifiability in~$\Pi^p_2$.
\noproof\end{Cor}

For completeness, we will also include an algorithm for admissibility with unlimited parameters in general logics of
bounded depth and width with no bound on cluster size. The argument in 
Lemma~\ref{lem:bddp-bdwd-bdpar-polysize} and Theorem~\ref{thm:ub-tab} only shows that an inadmissible rule is refuted in a
strongly extensible model of size $2^{O(2^n)}$, and this bound cannot be improved. This shows that $L$-admissibility is
decidable in $\cxt{coNTIME}\bigl(2^{O(2^n)}\bigr)$, i.e., in~$\cxt{coNEE}$. We need a little more elaborate
argument to obtain complexity that matches lower bounds from the next section (Theorem~\ref{thm:lb-sig2exp}).
\begin{Def}\label{def:prune}
Assume that $\Par$ is finite. Let $L$ be a logic of finite width and depth,
and $\Sigma$ a finite set of formulas closed under subformulas.

A finite $L$-model~$W$ is a \emph{$\Sigma$-pruned $L$-extensible model} if for every parametric
$L$-frame~$F$ with a loosely separated root such that $F\bez\rcl(F)$ is a generated subframe of (the underlying
parametric frame of) $W$, there exists a subframe $F\bez\rcl(F)\ssset F_1\sset F$ such that $\id_{F\bez\rcl(F)}$
extends to an isomorphism $f$ of~$F_1$ to a generated parametric subframe of~$W$, and the valuation on~$F_1$ lifted
from~$W$ via~$f$ extends to a valuation on~$F$ such that
\[F\model\fii\iff F_1\model\fii\]
for all $\fii\in\Sigma$.
\end{Def}
\begin{Lem}\label{lem:prune}
Assume that $\Par$ is finite. Let $L$ be a logic of finite width and depth, $\Sigma$ a finite set of formulas closed
under subformulas of size $n=\lh\Sigma$, and $\Gamma\cup\Delta\sset\Sigma$.
\begin{enumerate}
\item\label{item:14}
If $\Gamma\nadm_L\Delta$, then $\Gamma\ru\Delta$ fails in some $\Sigma$-pruned $L$-extensible model of size $2^{O(n^2)}$.
\item\label{item:13}
If $\Gamma\adm_L\Delta$, then $\Gamma\ru\Delta$ holds in all $\Sigma$-pruned $L$-extensible models.
\end{enumerate}
\end{Lem}
\begin{Pf}

\ref{item:14}: Let $c$ be the constant from Corollary~\ref{cor:bddp-bdwd-cls-exp}. We may assume $n\ge c$.
Since $\Gamma\ru\Delta$ is not admissible, we may fix a 
strongly $L$-extensible model~$W$ where it fails.

We first construct a submodel $W'\sset W$ by reducing each
cluster $C\sset W$ as follows:
\begin{itemize}
\item If $\lh C\le c$, then $C\sset W'$. If $\lh C\ge c$, we make sure $\lh{C\cap W'}\ge c$.
\item For each formula $\fii\in\Sigma$ falsified in some point of~$C$, we include at least one such point in~$W'$.
\end{itemize}
Clearly $W'$ is an $L$-model, and it has cluster size at most~$n$. The satisfaction of all formulas
from~$\Sigma$ is preserved in all points of~$W'$; in particular, $W'$ refutes $\Gamma\ru\Delta$. We claim that $W'$ is
a $\Sigma$-pruned $L$-extensible model, by a similar argument as in the proof of Lemma~\ref{lem:bddp-bdwd-bdpar-polysize}:
if $F'$ is a parametric $L$-frame with a loosely separated root such that $F'\bez\rcl(F')\sgen W'$, let $F$ be
$\rcl(F')$ attached below $(F'\bez\rcl(F'))\Up\sgen W$. By Corollary~\ref{cor:bddp-bdwd-cls-exp}, $F$ is still an $L$-frame,
hence $\id_{F\bez\rcl(F)}$ extends to an isomorphism $f$ of $F$ to a generated subframe of~$W$. Then
$F_1=f^{-1}[f[F]\cap W']$ has the property required by Definition~\ref{def:prune}.

Next, if $C$ and~$C'$ are two clusters of~$W'$ such that $G:=C\Up\bez C=C'\Up\bez C'$, and $\id_G$ extends to an
isomorphism of $C\Up$ to~$C'\Up$ (including valuation), let $W''$ be the model obtained from $W'$ by deleting $C'$ and
everything below it. Then $W''$ is an $L$-model refuting $\Gamma\ru\Delta$, and it is still a $\Sigma$-pruned
$L$-extensible model: for any $F$ as in the definition with $F\bez\rcl(F)\sgen W''$, let us find $F_1\sset F$ and
$f\colon F_1\to W'$ using $\Sigma$-pruned $L$-extensibility of~$W'$ so that the conclusion of the definition holds. If
$\rcl(f[F_1])=C'$, we modify it so that $\rcl(f[F_1])=C$. Then $f\colon F_1\to W''$, and it still has the required
property.

By continuing in this fashion, we obtain a $\Sigma$-pruned $L$-extensible model $U$ of cluster size $\le n$ such that
no two clusters have the same successors and valuations of parameters and variables. We claim that this implies $\lh
U=2^{O(n^2)}$. For $i\le d$, let $U_i$ denote the set of points of depth at most~$i$ in~$U$. We will prove by induction
that $\lh{U_i}=2^{O(n^2)}$. We have $U_0=\nul$. For the induction step, each cluster~$C$ of $U_{i+1}\bez U_i$ has
$C\Up\bez C$ generated by an antichain in~$U_i$, for which there are $\lh{U_i}^{O(1)}=2^{O(n^2)}$ choices. Apart from
this, $C$ is uniquely determined by its reflexivity, and its multiset of $\le n$ valuations to $\Par\cup V$, of which
there are $\le\bigl(2^{\lh{\Par\cup V}}\bigr)^{n+1}=2^{O(n^2)}$ choices. Thus, $\lh{U_{i+1}}=2^{O(n^2)}$.

\ref{item:13}:
Assume that $W_0$ is a $\Sigma$-pruned $L$-extensible model refuting $\Gamma\ru\Delta$. Put $f_0=\id_{W_0}$.
We will construct a sequence of models
$W_0\sgen W_1\sgen\dots\sgen W_d$, and maps $f_i\colon W_i\to W_0$ such that $f_0\sset f_1\sset\dots\sset f_d$, by
induction on~$i$. Each $f_i$ will be a depth-preserving p-morphism of the underlying (non-parametric) frames. It will
not preserve the valuation as such, but it will preserve the satisfaction of formulas of the form $\boxdot\fii$ for
$\fii\in\Sigma$.

Assume that $W_i$ has been constructed. For each parametric $L$-frame $F$ with a loosely separated root such that
$F_0=F\bez\rcl(F)\sgen W_i$, let $F'$ denote the subframe $f_i[F_0]\sgen W_0$ with $\rcl(F)$ attached below. Since
$f_i$ is a frame p-morphism, $F'$ is still an $L$-frame with a loosely separated root, and we may apply
Definition~\ref{def:prune} to find $F'_1\sset F'$ with $f_i[F_0]\ssset F'_1$, and a parametric isomorphism
$f\Sset\id_{f_i[F_0]}$ of $F'_1$ to a generated subframe $f[F'_1]\sgen W_0$. Moreover, we can fix a valuation in
$\rcl(F')$ so that the same formulas from~$\Sigma$ hold in~$F'$ as in $f[F'_1]$. The same formulas from~$\Sigma$ also
hold in~$F$ under the valuation obtained using $\rcl(F)=\rcl(F')$, as $f_i$ preserves the truth of formulas
$\boxdot\fii$, $\fii\in\Sigma$, between $F_0$ and $f_i[F_0]$. We include $\rcl(F)$ with this valuation in~$W_{i+1}$,
and extend $f_i$ to~$f_{i+1}$ so that $\rcl(F)$ is mapped onto~$\rcl(f[F'_1])$. We carry out this construction for all
frames~$F$ simultaneously, so that $W_{i+1}$ is $W_i$ together with all the new clusters $\rcl(F)$.

The construction ensures that $W_d$ is a model based on a strongly $L$-extensible parametric frame. Since $f_d$
preserves the truth of $\boxdot\fii$ for $\fii\in\Gamma$, we have $W_d\model\Gamma$, while $W_d\nmodel\fii$ for any
$\fii\in\Delta$, as $W_0\sgen W_d$. Thus, $\Gamma\nadm_L\Delta$ by Theorem~\ref{thm:bddp-adm-ext}.
\end{Pf}
\begin{Thm}\label{thm:ub-bddp-bdwd}
Let $L$ be a logic of bounded depth and width. Then $L$-admissibility is decidable in $\Pi^{\exp}_2$, and
$L$-unifiability in~$\Sigma^{\exp}_2$.
\end{Thm}
\begin{Pf}
Let us fix a rule $\Gamma\ru\Delta$, and put $\Sigma=\Sub(\Gamma\cup\Delta)$.  As in Theorem~\ref{thm:ub-tab}, we will
proceed as if $\Par$ were finite of size $\le n$.

By Lemma~\ref{lem:prune}, we have $\Gamma\nadm_L\Delta$ iff there exists a $\Sigma$-pruned $L$-extensible model
$W\nmodel\Gamma\ru\Delta$ of size $\lh W=2^{O(n^2)}$. This will give a $\Sigma^{\exp}_2$~algorithm for inadmissibility if
we show how to check that $W$ is a $\Sigma$-pruned $L$-extensible model in $\cxt{coNTIME}(2^{O(n^2)})$. We do this by
co-nondeterministically choosing $F$ as in Definition~\ref{def:prune}, and then deterministically testing it satisfies the
required condition.

We specify $F$ by selecting one of the $\lh W^{O(1)}$ choices for $F\bez\rcl(F)\sgen W$, and by indicating the
reflexivity of~$\rcl(F)$ as well as the set of assignments $E\sset\two^\Par$ that are realized in~$\rcl(F)$. This can be
done with $O(2^n)$ bits, and we can make sure the choice is sound (i.e., that $F$ is an $L$-frame, and that $\lh E=1$
if $\rcl(F)$ is irreflexive).

In order to test a given~$F$, we go through the clusters $C$ of~$W$ such that $C\Up\bez C=F\bez\rcl(F)$,
$\refl(C)=\refl(\rcl(F))$, $C$ is loosely separated, and the set $E_1$ of assignment to parameters realized in~$C$ is a
subset of~$E$. Notice that $E_1\ssset E$ may only happen in the reflexive case. Then the condition on extension of
valuation amounts to
\[\forall e\in E\bez E_1\:\exists v\in\two^V\:A(v,e,X)\model\ET_{\fii\in X}\fii,\]
where $V=\Sigma\cap\Var$, $X=\{\fii\in\Sigma:C\model\boxdot\fii\}$, and $A(v,e,X)$ is as in the proof of
Theorem~\ref{thm:ub-lin-clx}.
\end{Pf}

\section{Lower bounds}\label{sec:lower-bounds}

Our lower bounds generally have the form that unifiability in all logics from a certain class (usually determined
by the presence of particular frames as subframes) is hard for some complexity class~$\mathcal C$, and we will in fact
prove more: the reduction showing $\mathcal C$-hardness, and even the unifiers of positive instances, will be
independent of the logic. We introduce terminology so that we can state this more concisely.
\begin{Def}
Let $\mathcal L$ be a class of logics, and $\mathcal C$ a complexity class. We say that \emph{unifiability
in~$\mathcal L$ is uniformly $\mathcal C$-hard,} if for every $A\in\mathcal C$, there exists a logspace-computable
function $w\mapsto\xi_w$ such that:
\begin{enumerate}
\item If $w\in A$, there is a substitution~$\sigma$ such that $\vdash_L\sigma(\xi_w)$ for every~$L\in\mathcal L$.
\item If $w\notin A$, then $\xi_w$ is not $L$-unifiable for any~$L\in\mathcal L$.
\end{enumerate}
\end{Def}
Let us illustrate the concept on a simple example:
\begin{Thm}\label{thm:lb-np}
Parameter-free unifiability in all consistent logics~$L$ is uniformly $\np$-hard.
\end{Thm}
\begin{Pf}
A propositional formula $\fii$ with no modal operators or parameters is classically satisfiable if and only if it is
$L$-unifiable: on the one hand, a satisfying assignment provides a $\{\bot,\top\}$-substitution that makes the formula
a classical tautology; on the other hand, if $\vdash_L\sigma(\fii)$, then the evaluation of $\sigma$ in any fixed
element of a fixed $L$-model gives a satisfying assignment for~$\fii$.
\end{Pf}

\subsection{Exponential hierarchy}\label{sec:expon-hier}

In this section, we prove hardness of unifiability for several levels of the exponential hierarchy ($\NEXP$, $\coNEXP$,
$\Sigma_2^{\exp}$) for logics admitting suitable patterns as subframes, complementing the matching upper bounds from
Section~\ref{sec:upper-bounds}. In all three cases, we proceed by reducing the truth of third-order formulas from
Theorem~\ref{thm:exphprob} to unifiability. Generally speaking, we use variables to encode existential quantifiers, and
parameters to encode universal quantifiers; while second-order objects $t\sset n$ may be encoded by $n$~atoms in a
straightforward way, third-order objects $X\sset\pw n$ require more complex machinery to ensure they are
specified consistently, and it is at this point that the frame patterns are needed.
\begin{Thm}\label{thm:lb-nexp}
Unifiability in nonlinear logics is uniformly $\NEXP$-hard.
\end{Thm}
\begin{Pf}
By Theorem~\ref{thm:exphprob}, it is enough to exhibit a logspace reduction from the problem of determining the truth of
sentences of the form
\begin{equation}\label{eq:22}
\Phi=\exists X\sset\pw n\:\forall t_0,\dots,t_{m-1}\sset n\:\fii\bigl(\dots,i\in t_\alpha,\dots,t_\alpha\in X,\dots\bigr),
\end{equation}
where $n$ is given in unary and $\fii$ is a Boolean formula. Given $\Phi$, we define a formula
$\xi_\Phi$ in variables $x,x_\alpha$, and parameters $p_i,p_{\alpha,i},q,r$ for $i<n$ and~$\alpha<m$, as the conjunction
of the formulas
\begin{multline}\label{eq:23}
\ET_{i<n}\left[\boxdot\bigl((q\to p_{\alpha,i})\land(r\to p_i)\bigr)
                \lor\boxdot\bigl((q\to\neg p_{\alpha,i})\land(r\to\neg p_i)\bigr)\right]\\
\to\boxdot\bigl((q\to x_\alpha)\land(r\to x)\bigr)\lor\boxdot\bigl((q\to\neg x_\alpha)\land(r\to\neg x)\bigr)
\end{multline}
for $\alpha<m$, and
\begin{equation}\label{eq:24}
\fii(\dots,p_{\alpha,i},\dots,x_\alpha,\dots).
\end{equation}
Clearly, $\lh{\xi_\Phi}=\poly(n,\lh\Phi)$, and the mapping $\Phi\mapsto\xi_\Phi$ is logspace-computable.
\begin{Cl}\label{cl:nexp-true}
If $\Phi$ is true, then $\xi_\Phi$ is $\kiv$-unifiable.
\end{Cl}
\begin{Pf*}
Let $W=\unifr[\big]\kiv{\{p_i,p_{\alpha,i},q,r:i<n,\alpha<m\}}$, and let us fix a witness $X$ for the existential quantifier in~\eqref{eq:22}. By \citi{Thm.~5.18}, it
suffices to construct a valuation of the variables in~$W$ that makes $\xi_\Phi$ true in~$W$. For any point $u\in W$, we
define $t_\alpha^u,t^u\sset n$ for $\alpha<m$ by
\begin{align*}
t_\alpha^u&=\{i<n:u\model p_{\alpha,i}\},\\
t^u&=\{i<n:u\model p_i\}.
\end{align*}
Then we define the valuation of variables in~$u$ by
\begin{align*}
u&\model x_\alpha\iff t_\alpha^u\in X,\\
u&\model x_{\phantom\alpha}\iff t^u\in X.
\end{align*}
The choice of~$X$ ensures that $\fii(\dots,t_\alpha^u(i),\dots,t_\alpha^u\in X,\dots)$ is true for any $u\in W$, hence
\eqref{eq:24} is true in~$u$ by the definition of $t_\alpha^u$ and of the valuation of~$x_\alpha$.

Concerning~\eqref{eq:23}, let $\alpha<m$, and assume that
\[u\model\boxdot\bigl((q\to p_{\alpha,i})\land(r\to p_i)\bigr)
             \lor\boxdot\bigl((q\to\neg p_{\alpha,i})\land(r\to\neg p_i)\bigr)\]
for all $i<n$. This means there exists $t\sset n$ such that $t_\alpha^v=t$ for all $v\ge u$ satisfying~$q$, and
$t^v=t$ for all $v\ge u$ satisfying~$r$. It follows that
\[u\model\boxdot\bigl((q\to x_\alpha)\land(r\to x)\bigr)\]
if $t\in X$, and
\[u\model\boxdot\bigl((q\to\neg x_\alpha)\land(r\to\neg x)\bigr)\]
if $t\notin X$.
\end{Pf*}
\begin{Cl}\label{cl:nexp-false}
If $L$ is a nonlinear logic, and $\xi_\Phi$ is $L$-unifiable, then $\Phi$ is true.
\end{Cl}
\begin{Pf*}
Let $\sigma$ be an $L$-unifier of~$\xi_\Phi$. We may assume that the target formulas of~$\sigma$ include no parameters
except those occurring in~$\xi_\Phi$, and no variables. By assumption, there is an $L$-frame $W$ that weakly subreduces
onto a two-prong fork; that is, we may assume that $W$ is rooted, and we may fix disjoint
nonempty admissible subsets $A,B\sset W$ such that $A\cap B\Down=B\cap A\Down=\nul$. Put $P=\{p_i:i<n\}$, and
$P_\alpha=\{p_{\alpha,i}:i<n\}$ for $\alpha<m$.

For any $t\sset n$, let $W_t$ denote the parametric frame based on~$W$ where we make $r\land P^t$ true in~$A$, and
all parameters false elsewhere. In the root of~$W_t$, the premise of~\eqref{eq:23} (for arbitrary $\alpha<m$) is true, hence
$W_t\model\sigma(\xi_\Phi)$ implies
\[W_t\model r\to\sigma(x)\quad\text{or}\quad W_t\model r\to\neg\sigma(x).\]
We define a set $X\sset\pw n$ by
\[X=\bigl\{t\sset n:W_t\model r\to\sigma(x)\bigr\}.\]
We claim that $X$ is a witness for the truth of~$\Phi$. Let $t_0,\dots,t_{m-1}\sset n$; we need to show that
\begin{equation}\label{eq:25}
\fii\bigl(\dots,i\in t_\alpha,\dots,t_\alpha\in X,\dots\bigr)
\end{equation}
is true.

Define a new parametric frame $W^{t_0,\dots,t_{m-1}}$ based on~$W$, where
$q\land\ET_{\alpha<m}P_\alpha^{t_\alpha}$ holds in~$B$, and all parameters are false elsewhere. Since
$W^{\vec t}\model\sigma(\xi_\Phi)$, we have
\[W^{\vec t}\model\fii(\dots,p_{\alpha,i},\dots,\sigma(x_\alpha),\dots),\]
and fixing $u\in B$,
\[W^{\vec t},u\model p_{\alpha,i}\iff i\in t_\alpha\]
by the definition of~$W^{\vec t}$, hence \eqref{eq:25} will follow if we establish
\begin{equation}\label{eq:26}
W^{\vec t},u\model\sigma(x_\alpha)\iff t_\alpha\in X
\end{equation}
for each $\alpha<m$.

Consider the mixed parametric frame $W^{\vec t}_{t_\alpha}$: that is, make $r\land P^{t_\alpha}$ true in~$A$, 
$q\land\ET_{\beta<m}P_\beta^{t_\beta}$ true in~$B$, and all parameters false elsewhere. On the one hand, $W^{\vec
t}_{t_\alpha}$ coincides with $W^{\vec t}$ everywhere except in~$A$, and $B\Up\cap A=\nul$, hence
\[W^{\vec t},u\model\sigma(x_\alpha)\iff W^{\vec t}_{t_\alpha},u\model\sigma(x_\alpha).\]
On the other hand, $W^{\vec t}_{t_\alpha}$ coincides with $W_{t_\alpha}$ everywhere except in~$B$, and $A\Up\cap
B=\nul$, hence 
\[W^{\vec t}_{t_\alpha},v\model\sigma(x)\iff W_{t_\alpha},v\model\sigma(x)\iff t_\alpha\in X\]
for any $v\in A$. To connect these, the premise of~\eqref{eq:23} for~$\alpha$ holds in the root of~$W^{\vec t}_{t_\alpha}$, hence
\[W^{\vec t}_{t_\alpha}\model\boxdot\bigl((q\to\sigma(x_\alpha))\land(r\to\sigma(x))\bigr)
                \lor\boxdot\bigl((q\to\neg\sigma(x_\alpha))\land(r\to\neg\sigma(x))\bigr),\]
which implies
\[W^{\vec t}_{t_\alpha},u\model\sigma(x_\alpha)\iff W^{\vec t}_{t_\alpha},v\model\sigma(x).\]
This shows~\eqref{eq:26}.
\end{Pf*}
The result now follows from Claims \ref{cl:nexp-true} and~\ref{cl:nexp-false}.
\end{Pf}
\begin{Cor}\label{cor:nexp-compl}
Let $L$ be a nonlinear clx logic of bounded cluster size, or a nonlinear tabular logic. Then $L$-unifiability is
$\NEXP$-complete, and $L$-admissibility is $\coNEXP$-complete.
\end{Cor}
\begin{Pf}
By Theorems \ref{thm:ub-clx}, \ref{thm:ub-tab}, and \ref{thm:lb-nexp}.
\end{Pf}
\begin{Exm}\label{exm:nexp-complete}
The previous corollary applies to the logics $\lgc{GL}$, $\lgc{S4Grz}$, and $\lgc{K4Grz}$, as well as to their
extensions by the $\lgc{BB}_k$ axioms for $k\ge2$.
\end{Exm}

\begin{Thm}\label{thm:lb-conexp}
Unifiability in logics of unbounded cluster size is uniformly $\coNEXP$-hard.
\end{Thm}
\begin{Pf}
Using Theorem~\ref{thm:exphprob} again, we will present a logspace reduction from the truth of
sentences of the form
\begin{equation}\label{eq:27}
\Phi=\forall Y\sset\pw n\:\exists t_0,\dots,t_{m-1}\sset n\:\fii\bigl(\dots,i\in t_\alpha,\dots,t_\alpha\in Y,\dots\bigr),
\end{equation}
where $n$ is given in unary and $\fii$ is a Boolean formula. Given $\Phi$, we define a formula
$\xi_\Phi$ in parameters $p_i,q$, and variables $z_{\alpha,i}$ for $i<n$ and~$\alpha<m$, as the conjunction
of the formulas
\begin{align}
\label{eq:29}
\dia q&\to\dia(q\land\gamma),\\
\label{eq:30}
q\land\gamma&\to\fii\bigl(\dots,z_{\alpha,i},\dots,\eta(z_{\alpha,0},\dots,z_{\alpha,n-1}),\dots\bigr),
\end{align}
where
\begin{align*}
\gamma&=\ET_{\substack{\alpha<m\\i<n}}
     \bigl(\boxdot(\diadot q\to z_{\alpha,i})\lor\boxdot(\diadot q\to\neg z_{\alpha,i})\bigr),\\
\eta(x_0,\dots,x_{n-1})&=\dia\Bigl(\neg q\land\dia q\land\ET_{i<n}(p_i\eq x_i)\Bigr).
\end{align*}     
The result follows from the next two claims.
\begin{Cl}\label{cl:conexp-true}
If $\Phi$ is true, then $\xi_\Phi$ is $\kiv$-unifiable.
\end{Cl}
\begin{Pf*}
Putting $W=\unifr\kiv{\vec p,q}$, we will construct a valuation in~$W$ that makes $\xi_\Phi$ true. Since $\Phi$ holds,
we may fix for every $Y\sset\pw n$ some sets $t_0^Y,\dots,t_{m-1}^Y\sset n$ that make $\fii$ true. Write
$P=\{p_i:i<n\}$. For each point $u\in W$, we put
\[Y(u)=\bigl\{t\sset n:u\model\dia(\neg q\land\dia q\land P^t)\bigr\},\]
and define
\[u\model z_{\alpha,i}\iff i\in t_\alpha^{Y(u)}.\]

In order to show that \eqref{eq:29} holds, assume $u\model\dia q$. We can find $u<v\model q$ which is $\lnsim$-maximal,
i.e., $\neg q$ holds in all points strictly above $\cls(v)$. We claim that this makes $v\model\gamma$: indeed, the only
points $v'\ge v$ that satisfy $\diadot q$ are $v'\sim v$, and these have $Y(v')=Y(v)$, hence they agree on the
satisfaction of all the variables~$z_{\alpha,i}$.

As for~\eqref{eq:30}, assume that $u\model q\land\gamma$. This means that for each $\alpha<m$, there is
$t_\alpha\sset n$ such that
\begin{equation}\label{eq:28}
u\model\boxdot(\diadot q\to z_{\alpha,i}^{[i\in t_\alpha]})
\end{equation}
for all $i<n$. Since $u\model
q$, we must have $t_\alpha=t_\alpha^{Y(u)}$, thus
\[\fii\bigl(\dots,i\in t_\alpha,\dots,t_\alpha\in Y(u),\dots\bigr)\]
is true. This will imply
\[u\model\fii\bigl(\dots,z_{\alpha,i},\dots,\eta(z_{\alpha,0},\dots,z_{\alpha,n-1}),\dots\bigr)\]
if we show
\[u\model\eta(z_{\alpha,0},\dots,z_{\alpha,n-1})\iff t_\alpha\in Y(u).\]
But \eqref{eq:28} ensures that
\begin{align*}
u\model\eta(z_{\alpha,0},\dots,z_{\alpha,n-1})&\iff u\model\eta\bigl([0\in t_\alpha],\dots,[n-1\in t_{\alpha}]\bigr)\\
&\iff u\model\dia(\neg q\land\dia q\land P^{t_\alpha}),
\end{align*}
which is the definition of $t_\alpha\in Y(u)$.
\end{Pf*}
\begin{Cl}\label{cl:conexp-false}
If $\xi_\Phi$ is unifiable in a logic $L$ of cluster size at least $2^n+1$, then $\Phi$ is true.
\end{Cl}
\begin{Pf*}
Let $\sigma$ be an $L$-unifier of~$\xi_\Phi$ such that the target formulas of~$\sigma$ include no parameters
except those occurring in~$\xi_\Phi$, and no variables. Fix $Y\sset\pw n$: we need to find
$t_0,\dots,t_{m-1}\sset n$ satisfying~\eqref{eq:27}.

Let $\{s_j:j<l\}$ be an enumeration of~$Y$ with $l=\lh Y\le2^n$.
By assumption, there exists an $L$-frame $W$ that subreduces onto the $(l+1)$-element cluster: i.e., we may fix disjoint
nonempty admissible subsets $A_j\sset W$, $j\le l$, such that $A_j\sset A_{j'}\down$ for all $j,j'\le l$. We may assume
w.l.o.g.\ that
\[A_l\down\sset\bigcup_{j\le l}A_j\]
by adding $A_l\down\bez\bigcup_{j\le l}A_j$ to $A_l$ if necessary. (Notice that $A_j\down=A_l\down$ for all $j\le l$.)
With this condition, $\{A_j:j\le l\}$ is in fact a partition of $A_l\down$.

We consider the parametric frame based on~$W$ such that $A_j$ satisfies~$P^{s_j}$ for $j<l$, $A_l$ satisfies~$q$, and no
parameters hold elsewhere. Since $\vdash_L\sigma(\xi_\Phi)$, there exists $u\in A_l$ such that
$u\model\sigma(\gamma)$ by~\eqref{eq:29}, thus there are $t_0,\dots,t_{m-1}\sset n$ such that
\[u\model\boxdot\bigl(\diadot q\to\sigma(z_{\alpha,i})^{[i\in t_\alpha]}\bigr)\]
for each $\alpha<m$ and~$i<n$. As in the previous claim, this implies
\[u\model\sigma\bigl(\eta(z_{\alpha,0},\dots,z_{\alpha,n-1})\bigr)\eq\dia\bigl(\neg q\land\dia q\land P^{t_\alpha}\bigr).\]
Now,
\[u\model\dia\bigl(\neg q\land\dia q\land P^{t_\alpha}\bigr)\iff t_\alpha\in Y\colon\]
on the one hand, for each $j<l$, $u$ sees a point $v\in A_j$, and we have $v\model\neg q\land\dia q\land P^{s_j}$; on
the other hand, all points satisfying $\neg q\land\dia q$ are of this form.

Thus,
\[\fii\bigl(\dots,i\in t_\alpha,\dots,t_\alpha\in Y,\dots\bigr)\]
follows from $u\model\sigma(\ref{eq:30})$.
\end{Pf*}
\end{Pf}
\begin{Cor}\label{cor:conexp-compl}
Let $L$ be a linear clx logic of unbounded cluster size, or a linear logic of bounded depth and unbounded cluster size.
Then $L$-unifiability is $\coNEXP$-complete, and $L$-admissibility is $\NEXP$-complete.
\end{Cor}
\begin{Pf}
By Theorems \ref{thm:ub-lin-clx}, \ref{thm:ub-lin-bddp}, and \ref{thm:lb-conexp}.
\end{Pf}
\begin{Exm}\label{exm:conexp-complete}
The previous corollary applies to~$\lgc{S5}$ and to the related logics $\lgc{K45}$, $\lgc{D45}$, and~$\lgc{K4B}$. It
also applies to the logics $\lgc{S4.3}$, $\lgc{K4.3}$, and $\lgc{D4.3}$, and their extensions by the $\lgc{BD}_k$
axioms for $k\ge1$.
\end{Exm}

For our most complex logics, we will prove a $\Sigma_2^{\exp}$ lower bound. The argument will combine ideas from both
Theorems \ref{thm:lb-nexp} and~\ref{thm:lb-conexp}. First, let us fix a notation for frames that appear in our $\Sigma_2^{\exp}$-hardness
criterion.
\begin{Def}\label{def:sig2exp-frames}
Let $n\ge1$. The symbol $(\R+\nr n)^\R$ denotes a finite reflexive Kripke frame consisting of a root and two successor
clusters, one of which is an $n$-cluster, and the other one a single point.
\end{Def}
\begin{Thm}\label{thm:lb-sig2exp}
Unifiability is uniformly $\Sigma_2^{\exp}$-hard in the class of logics $L\Sset\kiv$ such that for every $n\ge1$, there exists
an $L$-frame weakly subreducing onto $(\R+\nr n)^\R$.
\end{Thm}
\begin{Pf}
We use Theorem~\ref{thm:exphprob} once again. Given a sentence
\begin{equation}\label{eq:31}
\Phi=\exists X\sset\pw n\:\forall Y\sset\pw n\:\exists t_0,\dots,t_{m-1}\sset n\:\fii\bigl(\dots,i\in t_\alpha,\dots,t_\alpha\in X,\dots,t_\alpha\in Y,\dots\bigr),
\end{equation}
where $n$ is written in unary and $\fii$ is a Boolean formula, we define a formula
$\xi_\Phi$ in parameters $p_i,q,r$, and variables $x,x_\alpha,z_{\alpha,i}$ for $i<n$ and~$\alpha<m$, as the conjunction
of the formulas
\begin{align}
\label{eq:33}
\dia q&\to\dia(q\land\gamma),\\
\label{eq:34}
q\land\gamma&\to\fii\bigl(\dots,z_{\alpha,i},\dots,x_\alpha,\dots,\eta(z_{\alpha,0},\dots,z_{\alpha,n-1}),\dots\bigr),
\end{align}
and
\begin{multline}\label{eq:32}
\ET_{i<n}\left[\boxdot\bigl((q\to z_{\alpha,i})\land(r\to p_i)\bigr)
                \lor\boxdot\bigl((q\to\neg z_{\alpha,i})\land(r\to\neg p_i)\bigr)\right]\\
\to\boxdot\bigl((q\to x_\alpha)\land(r\to x)\bigr)\lor\boxdot\bigl((q\to\neg x_\alpha)\land(r\to\neg x)\bigr)
\end{multline}
for $\alpha<m$, where
\begin{align*}
\gamma&=\ET_{\substack{\alpha<m\\i<n}}
     \bigl(\boxdot(\diadot q\to z_{\alpha,i})\lor\boxdot(\diadot q\to\neg z_{\alpha,i})\bigr),\\
\eta(x_0,\dots,x_{n-1})&=\dia\Bigl(\neg q\land\dia q\land\ET_{i<n}(p_i\eq x_i)\Bigr).
\end{align*}     
\begin{Cl}\label{cl:sig2exp-true}
If $\Phi$ is true, then $\xi_\Phi$ is $\kiv$-unifiable.
\end{Cl}
\begin{Pf*}
Let $W=\unifr\kiv{\vec p,q,r}$; we will construct a valuation in~$W$ that makes $\xi_\Phi$ true. Let $X$ be a witness
for the truth of~$\Phi$, and for each $Y\sset\pw n$, let us fix sets $t_0^Y,\dots,t_{m-1}^Y\sset n$ such that
\begin{equation}\label{eq:35}
\fii\bigl(\dots,i\in t_\alpha^Y,\dots,t_\alpha^Y\in X,\dots,t_\alpha^Y\in Y,\dots\bigr)
\end{equation}
is true. Let $P=\{p_i:i<n\}$. For each point $u\in W$, we put
\[Y(u)=\bigl\{t\sset n:u\model\dia(\neg q\land\dia q\land P^t)\bigr\},\]
and we define
\begin{alignat*}{2}
u&\model x&&\iff\sat_P(u)\in X,\\
u&\model x_\alpha&&\iff t_\alpha^{Y(u)}\in X,\\
u&\model z_{\alpha,i}&&\iff i\in t_\alpha^{Y(u)}.
\end{alignat*}

As in the proof of Claim~\ref{cl:conexp-true}, if $u$ is a $\lnsim$-maximal point satisfying~$q$, then $u\model\gamma$,
which shows that $W$ satisfies~\eqref{eq:33}. Moreover, if $u$ is any point such that $u\model q\land\gamma$, then
\begin{align*}
u\model\eta(z_{\alpha,0},\dots,z_{\alpha,n-1})&\iff u\model\eta\bigl([0\in t_\alpha^{Y(u)}],\dots,[n-1\in t_{\alpha}^{Y(u)}]\bigr)\\
&\iff u\model\dia(\neg q\land\dia q\land P^{t_\alpha^{Y(u)}})\\
&\iff t_\alpha^{Y(u)}\in Y(u),
\end{align*}
hence \eqref{eq:35} (with $Y=Y(u)$) implies
\[u\model\fii\bigl(\dots,z_{\alpha,i},\dots,x_\alpha,\dots,\eta(z_{\alpha,0},\dots,z_{\alpha,n-1}),\dots\bigr),\]
establishing the truth of~\eqref{eq:34}.

It is also straightforward to check~\eqref{eq:32}: if its premise holds in~$u$, let $t\sset n$ be such that
\[u\model\boxdot\bigl((q\to z_{\alpha,i}^{[i\in t]})\land(r\to p_i^{[i\in t]})\bigr)\]
for all $i<n$. Then all points $v\ge u$ satisfying~$q$ have $t_\alpha^{Y(v)}=t$, thus
\[v\model x_\alpha\iff t\in X,\]
while all points $w\ge u$ satisfying~$r$ have $\sat_P(w)=t$, thus
\[w\model x\iff t\in X\]
again.
\end{Pf*}
\begin{Cl}\label{cl:sig2exp-false}
If $\xi_\Phi$ is $L$-unifiable, and some $L$-frame weakly subreduces onto $(\R+\nr l)^\R$ for $l>2^n$, then $\Phi$ is true.
\end{Cl}
\begin{Pf*}
Let $\sigma$ be an $L$-unifier of~$\xi_\Phi$ such that the target formulas of~$\sigma$ include no parameters
except those occurring in~$\xi_\Phi$, and no variables. By the assumption, we may fix a rooted $L$-frame $W$, and
disjoint nonempty admissible subsets $A_j,B\sset W$, $j\le2^n$, such that $B\cap A_j\Down=A_j\cap B\Down=\nul$, and
$A_j\sset A_{j'}\down$ for $j,j'\le2^n$. This implies $A_j\down=A_{j'}\down$ for all $j,j'\le2^n$. Moreover, we may
assume
\[A_j\down\cap A_j\up\sset A_j\down\bez B\down\sset\bigcup_{i\le2^n}A_i\]
by adding the excess $A_j\down\bez\bigl(B\down\cup\bigcup_{i\le2^n}A_i\bigr)$ to~$A_{2^n}$.

In order to verify that $\Phi$ is true, we first construct a purported witness $X\sset\pw n$ to the first existential
quantifier. For any $t\sset n$, let $W_t$ denote the parametric frame based on~$W$ such that $r\land P^t$ holds
in~$B$, and no parameters elsewhere. Then the root of~$W$ satisfies the premise of~$\sigma(\ref{eq:32})$ (for
arbitrary $\alpha<m$), hence
\[W_t\model r\to\sigma(x)\quad\text{or}\quad W_t\model r\to\neg\sigma(x).\]
We define
\[X=\bigl\{t\sset n:W_t\model r\to\sigma(x)\bigr\}.\]

Next, let an arbitrary $Y\sset\pw n$ be given; we need to find $t_0,\dots,t_{m-1}\sset n$ satisfying~$\fii$. Let
$l=\lh Y\le2^n$, and fix an enumeration $Y=\{s_j:j<l\}$. We define a parametric frame $W^Y$ based on~$W$ so that
$A_j\model P^{s_j}$ for $j<l$, $A_j\model q$ for $l\le j\le2^n$, and no parameters are true elsewhere. Since the root
satisfies $\dia q$, $W^Y\model\sigma(\ref{eq:33})$ ensures we can find $u\model q\land\sigma(\gamma)$, thus there are
$t_0,\dots,t_{m-1}\sset n$ such that
\begin{equation}\label{eq:36}
W^Y,u\model\boxdot\bigl(\diadot q\to\sigma(z_{\alpha,i})^{[i\in t_\alpha]}\bigr)
\end{equation}
for all $\alpha<m$ and~$i<n$. It remains to prove that
\[\fii\bigl(\dots,i\in t_\alpha,\dots,t_\alpha\in X,\dots,t_\alpha\in Y,\dots\bigr)\]
is true. We have
\[W^Y,u\model\fii\bigl(\dots,\sigma(z_{\alpha,i}),\dots,\sigma(x_\alpha),\dots,\eta(\sigma(z_{\alpha,0}),\dots,\sigma(z_{\alpha,n-1})),\dots\bigr)\]
by $W^Y\model\sigma(\ref{eq:34})$,
\[W^Y,u\model\sigma(z_{\alpha,i})\iff i\in t_\alpha\]
for all $\alpha<m$ and $i<n$ by~\eqref{eq:36}, and \eqref{eq:36} also implies
\begin{align*}
W^Y,u\model\eta\bigl(\sigma(z_{\alpha,0}),\dots,\sigma(z_{\alpha,n-1})\bigr)
&\iff W^Y,u\model\dia\bigl(\neg q\land\dia q\land P^{t_\alpha}\bigr)\\
&\iff t_\alpha\in Y
\end{align*}
for all $\alpha<m$ as in the proof of Claim~\ref{cl:conexp-false}. Thus, to complete the proof, we only need to show
\begin{equation}\label{eq:37}
W^Y,u\model\sigma(x_\alpha)\iff t_\alpha\in X
\end{equation}
for all $\alpha<m$.

We proceed by constructing a parametric frame combining $W^Y$ and~$W_{t_\alpha}$, exploiting
$\vdash_L\sigma(\ref{eq:32})$, but we will need to be more careful than in the proof of Claim~\ref{cl:nexp-false}, because
there may be other points satisfying $q\land\sigma(\gamma)$ in the model than~$u$, and they may disagree
on~\eqref{eq:36}. Thus, let us consider the parametric frame $W'$ based on~$W$ such that $r\land P^{t_\alpha}$ holds in~$B$
(as in~$W_{t_\alpha}$), $P^{s_j}$ holds in $A_j$ for $j<l$ (as in~$W^Y$), and $q$ holds in the set
\[A'=\Bigl\{v:W^Y,v\model q\land\ET_{\substack{\alpha<m\\i<n}}\boxdot\bigl(\diadot q\to\sigma(z_{\alpha,i})^{[i\in t_\alpha]}\bigr)\Bigr\}.\]
Notice that $u\in A'$, and in fact $W'$ agrees with~$W^Y$ in~$u\Up$ (using $u\Up\cap B=\nul$), thus
\[W^Y,u\model\sigma(x_\alpha)\iff W',u\model\sigma(x_\alpha).\]
The definition of~$W'$ ensures that the premise of $\sigma(\ref{eq:32})$ for~$\alpha$ is satisfied in the root, hence
\[W'\model\bigl(q\to\sigma(x_\alpha)\bigr)\land\bigl(r\to\sigma(x)\bigr)\quad\text{or}\quad
  W'\model\bigl(q\to\neg\sigma(x_\alpha)\bigr)\land\bigl(r\to\neg\sigma(x)\bigr).\]
But $W'$ agrees with~$W_{t_\alpha}$ in $B\Up$, hence
\[W'\model r\to\sigma(x)\iff t_\alpha\in X\]
by the definition of~$X$, which implies
\[W',u\model\sigma(x_\alpha)\iff t_\alpha\in X\]
in view of $W',u\model q$. Thus, \eqref{eq:37} holds.
\end{Pf*}
\end{Pf}
\begin{Cor}\label{cor:sig2exp-compl}
Let $L$ be a nonlinear clx logic of unbounded cluster size, or a nonlinear logic of bounded depth, bounded width, and
unbounded cluster size. Then $L$-unifiability is $\Sigma_2^{\exp}$-complete, and $L$-admissibility is $\Pi_2^{\exp}$-complete.
\end{Cor}
\begin{Pf}
By Theorems \ref{thm:ub-clx}, \ref{thm:ub-bddp-bdwd}, and \ref{thm:lb-sig2exp}.
\end{Pf}
\begin{Exm}\label{exm:sig2exp-complete}
The previous corollary applies to the basic transitive logics $\kiv$, $\lgc{D4}$, and $\lgc{S4}$, their
variants with the McKinsey axiom $\lgc{K4.1}$, $\lgc{D4.1}$, $\lgc{S4.1}$, and extensions of these logics with the
bounded branching axioms $\lgc{BB}_k$ for $k\ge2$, possibly combined with the bounded depth axioms $\lgc{BD}_l$ for
$l\ge2$.
\end{Exm}

\subsection{$\NEXP$ lower bounds with $O(1)$ parameters}\label{sec:nexp-const-param}

The lower bounds in the previous subsection relied on the supply of arbitrarily many parameters, and it is clear from
the upper bounds in Section~\ref{sec:upper-bounds} that Theorems \ref{thm:lb-conexp} and~\ref{thm:lb-sig2exp} intrinsically need them.
This leaves the possibility of an $\NEXP$ lower bound on unifiability or inadmissibility with a constant number of
parameters. However, Theorems \ref{thm:ub-unif-bdpar-bddp} and~\ref{thm:ub-bdpar-bddp-bdwd} show that we need considerably stronger
assumptions than in Theorem~\ref{thm:lb-nexp} for that to happen: in particular, no fixed finite frame pattern is enough,
even allowing for infinite clusters. It is unclear what should be the optimal condition that, on the one hand,
guarantees $\NEXP$-hardness of $L$-unifiability (or at least $L$-admissibility) with finitely many parameters, and on
the other hand applies to a class of logics as wide as possible.

One such lower bound already appeared in \cite{ej:admcomp}: inadmissibility with no parameters
at all is $\NEXP$-complete in logics satisfying a certain ad hoc extensibility condition that operates only with frames
of depth~$\le3$, but requires unbounded width at that depth. As a special case, it applies to all clx logics of
\emph{unbounded branching}. While, as we mentioned, it is not clear what is the optimal lower bound to expect, we will
present variants of this construction that apply to $\p{\I,2}$-extensible or $\p{\R,2}$-extensible logics,
including all \emph{nonlinear} clx logics.

We will actually present several lower bounds, trading the number of parameters for the strength of assumptions on the
logic. All the statements will be proved by varying the details of the same generic construction. As in
Theorem~\ref{thm:lb-nexp}, the overall strategy is to provide a reduction from the truth problem for sentences of the form
\begin{equation}\label{eq:38}
\Phi=\exists X\sset\pw n\:\forall t_0,\dots,t_{m-1}\sset n\:\fii\bigl(\dots,i\in t_\alpha,\dots,t_\alpha\in X,\dots\bigr).
\end{equation}
Assume we have an efficiently computable list of formulas
\begin{equation}\label{eq:42}
\beta_0,\dots,\beta_{n-1},\theta_0,\dots,\theta_{m-1},\theta
\end{equation}
that do not contain the variable~$x$. (In fact, they will not contain \emph{any} variables in most cases.) We will use
points satisfying~$\beta_i$ as a kind of labels for encoding of the sets
$t_\alpha\sset n$ in~\eqref{eq:38}: a valid encoding of $t_\alpha$ will be a point that satisfies~$\theta_\alpha$,
and sees $\beta_i$ for $i\in t_\alpha$. We will also make use of encoding of $t\sset n$ without
reference to~$\alpha<m$, which will employ $\theta$ instead of~$\theta_\alpha$. The formulas $\theta_\alpha$
and~$\theta$ in turn will be defined, on a case-by-case basis, in terms of seeing points satisfying certain other
formulas $\gamma_\alpha$ and~$\delta$. (The precise meaning of this vague description will be made clear in the
specific proofs.)

Given~\eqref{eq:42}, we define a formula~$\xi_\Phi$ as the conjunction of the two formulas
\begin{equation}\label{eq:40}
\ET_{i<n}\bigl(\Box(\theta\to\dia\beta_i)\lor\Box(\theta\to\neg\dia\beta_i)\bigr)
\to\Box(\theta\to x)\lor\Box(\theta\to\neg x)
\end{equation}
and
\begin{multline}\label{eq:41}
\ET_{\alpha<m}\Bigl(\dia\theta_\alpha\land
     \ET_{i<n}\bigl(\Box(\theta_\alpha\to\dia\beta_i)\lor\Box(\theta_\alpha\to\neg\dia\beta_i)\bigr)\Bigr)\\
\to\fii\bigl(\dots,\dia(\theta_\alpha\land\dia\beta_i),\dots,\dia(\theta_\alpha\land x),\dots\bigr).
\end{multline}
\begin{Lem}\label{lem:nexp-bdpar-true}
If $\Phi$ is true, then $\xi_\Phi$ is $\kiv$-unifiable.
\end{Lem}
\begin{Pf}
Let $W=\unifr[\big]\kiv{\Par\cap\Sub(\xi_\Phi)}$; we will construct a valuation in~$W$ that makes $\xi_\Phi$ true. If there
are any variables other than~$x$, we fix their valuation in an arbitrary way. In order to define the valuation of~$x$,
we fix a witness $X\sset\pw n$ for~$\Phi$. For any $u\in W$, we define
\[t^u=\{i<n:u\model\dia\beta_i\},\]
and we put
\[u\model x\iff t^u\in X.\]

It is easy to see that this makes \eqref{eq:40} true, as the value of~$x$ in any given point is completely determined
by the values of $\dia\beta_i$ for $i<n$.

As for~\eqref{eq:41}, if its premise holds in~$u$, let $t_0,\dots,t_{m-1}\sset n$ be such that
\[u\model\Box\bigl(\theta_\alpha\to(\dia\beta_i)^{[i\in t_\alpha]}\bigr)\]
for each $\alpha<m$ and~$i<n$. Since also $u\model\dia\theta_\alpha$, this makes
\[u\model\dia(\theta_\alpha\land\dia\beta_i)\iff i\in t_\alpha.\]
Moreover, $t^v=t_\alpha$ for any $v>u$ such that $v\model\theta_\alpha$, hence the definition of $v\model x$ implies
\[u\model\dia(\theta_\alpha\land x)\iff t_\alpha\in X.\]
Thus,
\[u\model\fii\bigl(\dots,\dia(\theta_\alpha\land\dia\beta_i),\dots,\dia(\theta_\alpha\land x),\dots\bigr)\]
follows from the truth of $\fii(\dots,i\in t_\alpha,\dots,t_\alpha\in X,\dots)$.
\end{Pf}
\begin{Rem}\label{rem:next-bdpar-true-expl}
We can extract from the proof of Lemma~\ref{lem:nexp-bdpar-true} a completely explicit unifier $\sigma$ of~$\xi_\Phi$, namely
\begin{equation}\label{eq:53}
\sigma(x)=\LOR_{t\in X}\ET_{i<n}(\dia\beta_i)^{[i\in t]},
\end{equation}
with $\sigma(y)=y$ for all variables $y$ other than~$x$.
\end{Rem}

We are ready for our first lower bound. Recall Definition~\ref{def:tree-universal}.
\begin{Thm}\label{thm:nexp-1par}
Unifiability of formulas with $1$~parameters and $1$~variable is uniformly $\NEXP$-hard in the class of logics
cofinally subframe-universal for trees.
\end{Thm}
\begin{Pf}
We need to define the formulas~\eqref{eq:42} in such a way that $\xi_\Phi$ is not $L$-unifiable if $\Phi$ is false.
The basic idea is that in order to construct a sufficiently large antichain of
formulas in one parameter, we identify suitable antichains in~$\unifr\kiv p$ (actually, in $\unifr{\lgc{S4GrzBB_2}}p$),
and describe their points by formulas in a uniform and algorithmically efficient way. This would be easy
enough, but there is an additional serious complication that will force us to tread very carefully: our formulas have
to work not only in the original subframe $F\sgen\unifr\kiv p$, but in an $L$-frame that just weakly cofinally
subreduces to~$F$.

\begin{figure}
\centering
\magicparoff
\unitlength=1pt
\begin{picture}(120,110)

\put(0,103){$\scriptscriptstyle0$}
\put(52.5,104){\circle*{4}}
\put(0,87){$\scriptscriptstyle1$}
\put(52.5,88){\circle{4}}
\put(52.5,90){\vector(0,1){12}}
\put(77.5,96){\oval(4,20)}

\put(0,73){$\scriptscriptstyle2$}
\put(40,76){\circle*{4}}
\put(40,76){\vector(1,1){10.7}}
\put(65,76){\circle*{4}}
\put(65,76){\vector(-1,1){10.7}}
\put(65,76){\vector(1,1){10.7}}
\put(90,76){\circle*{4}}
\put(90,76){\vector(-1,1){10.7}}

\put(0,58){$\scriptscriptstyle3$}
\put(40,60){\circle{4}}
\put(40,62){\vector(0,1){12}}
\put(65,60){\circle{4}}
\put(65,62){\vector(0,1){12}}
\put(90,60){\circle{4}}
\put(90,62){\vector(0,1){12}}

\put(0,40){$\scriptscriptstyle4$}
\put(21,41){\circle*{4}}
\put(21,41){\vector(1,1){17.5}}
\put(40,41){\circle*{4}}
\put(40,41){\vector(0,1){17}}
\put(40,41){\vector(4,3){23.5}}
\put(54,41){\circle*{4}}
\put(54,41){\vector(-3,4){13}}
\put(54,41){\vector(2,1){34.7}}
\put(74,41){\circle*{4}}
\put(74,41){\vector(-1,2){8.5}}
\put(90,41){\circle*{4}}
\put(90,41){\vector(-4,3){23.5}}
\put(90,41){\vector(0,1){17}}
\put(109,41){\circle*{4}}
\put(109,41){\vector(-1,1){17.5}}

\put(0,23){$\scriptscriptstyle5$}
\put(21,25){\circle{4}}
\put(21,27){\vector(0,1){12}}
\put(40,25){\circle{4}}
\put(40,27){\vector(0,1){12}}
\put(54,25){\circle{4}}
\put(54,27){\vector(0,1){12}}
\put(74,25){\circle{4}}
\put(74,27){\vector(0,1){12}}
\put(90,25){\circle{4}}
\put(90,27){\vector(0,1){12}}
\put(109,25){\circle{4}}
\put(109,27){\vector(0,1){12}}

\put(13,4){$\iddots$}
\put(65,4){$\vdots$}
\put(100,4){$\ddots$}
\end{picture}
\caption{The frame $F\sgen\unifr{\lgc{S4Grz}}p$ from the proof of Theorem~\ref{thm:nexp-1par}. All points are
reflexive; here, $\I$ denotes points satisfying~$p$, and $\R$ points satisfying~$\neg p$.}
\label{fig:nexp-1par}
\end{figure}
Formally, we will define a sequence of auxiliary formulas $\{\beta^d_i,\eta^d_i:i<n_d\}$ in parameter~$p$ by induction
on~$d$, and we will simultaneously build a finite parametric frame~$F\sgen\unifr{\lgc{S4}}p$ with
domain $\{a^d_i:d\in\omega,i<n_d\}$ such that $\beta^d_i$ holds in~$a^d_i$, and $\eta^d_i$ in~$a^d_i\Up$.

For $d=0$, we put $n_0=2$, and
\[\beta^0_i=\eta^0_i=\boxdot p^i\]
for $i<2$. The points $a^0_0,a^0_1\in F$ are incomparable, and $a^0_i\model p^i$.

Assume everything has been defined for~$2d$. We put $n_{2d+1}=n_{2d}$, and for each $i<n_{2d}$, let
\begin{align*}
\eta^{2d+1}_i&=\boxdot(p\lor\Box\neg p\to\eta^{2d}_i),\\
\beta^{2d+1}_i&=\eta^{2d+1}_i\land\neg p\land\diadot\beta^{2d}_i.
\end{align*}
We introduce new points $\{a^{2d+1}_i:i<n_{2d}\}$ in~$F$ such that $a^{2d+1}_i$ is an immediate predecessor
of~$a^{2d}_i$, and $a^{2d+1}\model\neg p$; however, as an exception, $a^1_0$ will not be a new point: we put
$a^1_0=a^0_0$. (Notice that $\beta^1_0$, $\eta^1_0$, $\beta^0_0$, and~$\eta^0_0$ are all equivalent.)

At the next level, we put $n_{2d+2}=\binom{n_{2d}+1}2$. We will identify $i<n_{2d+2}$ with sets $\{j,k\}$ for
$j,k<n_{2d}$ (where $j$ may be equal to~$k$) in some canonical way, and we define
\begin{align*}
\eta^{2d+2}_{\{j,k\}}
&=\boxdot\bigl(p\to\eta^{2d}_j\lor\eta^{2d}_k\lor(\dia\beta^{2d+1}_j\land\dia\beta^{2d+1}_k)\bigr)
\land\boxdot\bigl(\neg p\to\boxdot(p\lor\Box\neg p\to\eta^{2d}_j\lor\eta^{2d}_k)\bigr),\\
\beta^{2d+2}_{\{j,k\}}&=\eta^{2d+2}_{\{j,k\}}\land p\land\dia\beta^{2d+1}_j\land\dia\beta^{2d+1}_k.
\end{align*}
For each $j,k<n_{2d}$, the point $a^{2d+2}_{\{j,k\}}\in F$ is an immediate predecessor of
$a^{2d+1}_j$ and~$a^{2d+1}_k$, and $a^{2d+2}_{\{j,k\}}\model p$.

Thus, we have $a^{2d+e}_i\model p^e$ for $e=0,1$, except that $a^0_0=a^1_0\model\neg p$.

The topmost part of~$F$ is depicted in Fig.~\ref{fig:nexp-1par}.
\begin{Cl}\label{cl:nexp-1par-beta}
For each $d,d'\in\omega$ and $i<n_d$, $i'<n_{d'}$,
\begin{align}
\label{eq:39} &\vdash_\kiv\beta^d_i\to\eta^d_i,\\
\label{eq:43} &\vdash_\kiv\eta^d_i\to\Box\eta^d_i,\\
\label{eq:44} a^{d'}_{i'}\le a^d_i\implies&\vdash_\kiv\eta^d_i\to\eta^{d'}_{i'},\\
\label{eq:45} a^{d'}_{i'}\nleq a^d_i\implies&\vdash_\kiv\neg(\beta^d_i\land\eta^{d'}_{i'}).
\end{align}
\end{Cl}
\begin{Pf*}
\eqref{eq:39} and~\eqref{eq:43} are obvious from the definition. As for~\eqref{eq:44}, apart from the exceptional case
$a^0_0\le a^1_0$ (where it holds), it suffices to prove it when $a^d_i$ is an immediate successor of~$a^{d'}_{i'}$,
i.e., $d'=d+1$, and $i'=i$ for $d$ even, or $i'=\{i,j\}$ for $d$ odd. In both cases it follows immediately from the
definition of~$\eta^{d'}_{i'}$, using~\eqref{eq:43}.

We will prove~\eqref{eq:45} by induction on $d+d'$. We distinguish several cases.

Assume $d$ is odd. If $a^{d'}_{i'}\nleq a^{d-1}_i$, then $\vdash\beta^d_i\to\diadot\beta^{d-1}_i$ by definition, while
$\vdash\eta^{d'}_{i'}\to\boxdot\neg\beta^{d-1}_i$ by \eqref{eq:43} and the induction hypothesis. Otherwise
$a^{d'}_{i'}\le a^{d-1}_i$ and $a^{d'}_{i'}\nleq a^d_i$, which is only possible if $\p{d',i'}=\p{d-1,i}\ne\p{0,0}$. In
particular, $d'$ is even. If $d'=0$, we thus have $\eta^{d'}_{i'}=\boxdot p$, while $\beta^d_i$ implies $\neg p$.
If $d'>0$, let $i=\{j,k\}$. Since $\beta^d_i$ implies $\neg p$, the definition of $\eta^{d'}_{i'}$ ensures that
$\eta^{d'}_{i'}\land\beta^d_i$ implies $\boxdot(p\to\eta^{d'-2}_j\lor\eta^{d'-2}_k)$. Also,
$\vdash\beta^d_i\to\diadot\beta^{d'}_i$, where $\vdash\beta^{d'}_i\to p$, hence
\[\vdash\eta^{d'}_{i'}\land\beta^d_i\to\diadot\bigl(\beta^{d'}_i\land(\eta^{d'-2}_{j'}\lor\eta^{d'-2}_{k'})\bigr).\]
This implies a contradiction by the induction hypothesis, as $a^{d'-2}_j\nleq a^{d'}_i$ and $a^{d'-2}_k\nleq
a^{d'}_i$.

If $d$ is even and $d'$ is odd, then $\vdash\beta^d_i\to p\lor\Box\neg p$, hence the definition
of~$\eta^{d'}_{i'}$ ensures $\vdash\beta^d_i\land\eta^{d'}_{i'}\to\eta^{d'-1}_{i'}$, which leads to a contradiction by
the induction hypothesis as $a^{d'-1}_{i'}\nleq a^d_i$.

The remaining case is if both $d$ and~$d'$ are even. First, if one of the immediate successors $a^{d-1}_j$ of~$a^d_i$
satisfies $a^{d'}_{i'}\nleq a^{d-1}_j$, then $\vdash\beta^d_i\to\dia\beta^{d-1}_j$, while
$\vdash\eta^{d'}_{i'}\to\Box\neg\beta^{d-1}_j$ by \eqref{eq:43} and the induction hypothesis. Thus, we may assume that
all immediate successors of~$a^d_i$ are above~$a^{d'}_{i'}$.

If $d'=0$, this forces $d=0$ as well; since $i\ne i'$, $\beta^d_i\land\eta^{d'}_{i'}$ is $\boxdot p\land\boxdot\neg p$,
which is contradictory.

If $d'>0$, let $i'=\{j',k'\}$. If $\p{d,i}=\p{0,0}$, i.e., $\beta^d_i=\boxdot\neg p$, the definition
of~$\eta^{d'}_{i'}$ gives
\[\vdash\beta^d_i\land\eta^{d'}_{i'}\to\eta^{d'-2}_{j'}\lor\eta^{d'-2}_{k'},\]
and the disjuncts $\eta^{d'-2}_{j'}$ and $\eta^{d'-2}_{k'}$ lead to contradiction by the induction hypothesis, as
$a^{d'-2}_{j'}\nleq a^d_i$ and $a^{d'-2}_{k'}\nleq a^d_i$. Otherwise $\vdash\beta^d_i\to p$, hence the definition
of~$\eta^{d'}_{i'}$ gives
\[\vdash\beta^d_i\land\eta^{d'}_{i'}\to\eta^{d'-2}_{j'}\lor\eta^{d'-2}_{k'}\lor
    (\dia\beta^{d'-1}_{j'}\land\dia\beta^{d'-1}_{k'}),\]
where $\eta^{d'-2}_{j'}$ and $\eta^{d'-2}_{k'}$ lead to contradiction as above, hence
\begin{equation}\label{eq:46}
\vdash\beta^d_i\land\eta^{d'}_{i'}\to\dia\beta^{d'-1}_{j'}\land\dia\beta^{d'-1}_{k'}.
\end{equation}
Notice that $\vdash\beta^{d'-1}_{j'/k'}\to\neg p$. Thus, \eqref{eq:46} directly leads to contradiction if
$\p{d,i}=\p{0,1}$, as $\beta^d_i=\boxdot p$. If $d>0$ and $i=\{j,k\}$, then the definition of $\beta^d_i$ ensures
\[\vdash\beta^d_i\to\boxdot\bigl(\beta^{d'-1}_{j'/k'}\to\boxdot(p\lor\Box\neg p\to\eta^{d-2}_j\lor\eta^{d-2}_k)\bigr).\]
Since also $\vdash\beta^{d'-1}_{j'/k'}\to\diadot\beta^{d'-2}_{j'/k'}$, where $\beta^{d'-2}_{j'/k'}$ implies $p\lor\Box\neg
p$, \eqref{eq:46} yields
\begin{equation}\label{eq:47}
\vdash\beta^d_i\land\eta^{d'}_{i'}\to\dia\bigl(\beta^{d'-2}_{j'}\land(\eta^{d-2}_j\lor\eta^{d-2}_k)\bigr)
           \land\dia\bigl(\beta^{d'-2}_{k'}\land(\eta^{d-2}_j\lor\eta^{d-2}_k)\bigr).
\end{equation}
Recall that $a^{d'}_{i'}\le a^{d-1}_j$ and $a^{d'}_{i'}\le a^{d-1}_k$. This means
\begin{equation}\label{eq:48}
a^{d'-1}_{j'}\le a^{d-1}_j\quad\text{or}\quad a^{d'-1}_{k'}\le a^{d-1}_j,
\end{equation}
and similarly for $a^{d-1}_k$; in particular, $d'\ge d$. On the other hand, using the
induction hypothesis, \eqref{eq:47} gives the desired $\vdash\neg(\beta^d_i\land\eta^{d'}_{i'})$ unless
\begin{equation}\label{eq:49}
a^{d-2}_j\le a^{d'-2}_{j'}\quad\text{or}\quad a^{d-2}_k\le a^{d'-2}_{j'},
\end{equation}
and similarly for $a^{d'-2}_{k'}$. In particular, this implies $d\ge d'$, thus $d=d'$. Then \eqref{eq:48} actually
means $j\in\{j',k'\}$, and together with the analogous property of~$k$ we obtain $\{j,k\}\sset\{j',k'\}$, while
\eqref{eq:49} and its analogue for~$k'$ give $\{j',k'\}\sset\{j,k\}$. Thus $i=\{j,k\}=\{j',k'\}=i'$, a contradiction.
\end{Pf*}

Notice that $n_{2d}$ is strictly increasing. Since $n_d$ roughly squares every other step, it is
asymptotically doubly exponential in~$d$: $n_d=2^{2^{\Omega(d)}}$. On the other hand, the size of the formulas
$\beta^d_i$, $\eta^d_i$ is $2^{O(d)}$.

For our application, we take the least even~$D$ such that $n_D\ge n+m+2$, and we define the formulas~\eqref{eq:42} as
follows: $\beta_i=\beta^D_i$ for $i<n$, $\gamma_\alpha=\beta^D_{n+\alpha}$ for $\alpha<m$, $\delta=\beta^D_{n+m}$, and
\begin{align}
\label{eq:61}\theta_{\phantom\alpha}&=p\land\dia\delta\land\neg\delta,\\
\label{eq:62}\theta_\alpha&=p\land\dia\delta\land\dia\gamma_\alpha\land\ET_{\alpha'\ne\alpha}\neg\dia\gamma_{\alpha'}
\end{align}
for each $\alpha<m$. We have $D=O(\log\log(n+m))$, thus
$\lh{\beta^D_i}=(\log(n+m))^{O(1)}$. It is not difficult to check that the formulas are logspace-constructible.
Claim~\ref{cl:nexp-1par-beta} shows the key property
\begin{equation}\label{eq:50}
i\ne i'\implies{}\vdash_\kiv\beta^D_i\to\boxdot\neg\beta^D_{i'}.
\end{equation}
In particular, $\theta_\alpha$ implies~$\theta$.

We also need to construct some finite trees (in the form of parametric reflexive frames) related to the representation
of $\Phi$ by~$\xi_\Phi$. We use $+$ or $\sum$ to denote disjoint sum as usual, and $F^p$ (or~$F^{\neg p}$) denotes $F$
extended below with a with new root, satisfying~$p$ (or~$\neg p$, respectively).

First, for any $d\in\omega$ and $i<n_d$, let $T^d_i$ be $F_{x^d_i}$ unravelled into a tree: that is, $T^0_i=\nul^{p^i}$
(i.e., the trivial tree satisfying~$p^i$) for $i<2$; $T^{2d+1}_i$ is $(T^{2d}_i)^{\neg p}$ for $i<n_{2d}$,
$\p{d,i}\ne\p{0,0}$, while $T^1_0=T^0_0$; and $T^{2d+2}_{\{j,k\}}=(T^{2d+1}_j+T^{2d+1}_k)^p$ (it does not
matter if we simplify this to $(T^{2d+1}_j)^p$ in the case $j=k$ or not).

We put $T_{\beta_i}=T^D_i$ for $i<n$, $T_{\gamma_\alpha}=T^D_{n+\alpha}$, and $T_\delta=T^D_{n+m}$ in accordance with the
definition of the formulas. We also define a ``dummy'' tree $T_-=T^D_{n+m+1}$.

For every $t\sset n$, let $T_t=\bigl(T_-+T_-+\sum_{i\in t}T_{\beta_i}\bigr)^p$. Next, we put
$T_{t,\alpha}=(T_\delta+T_{\gamma_\alpha}+T_t)^p$ for each $\alpha<m$. Then, for every
$\vec t=\p{t_0,\dots,t_{c-1}}\in(\pw n)^m$, we define $T_{\vec t}=\bigl(\sum_{\alpha<m}T_{t_\alpha,\alpha}\bigr)^p$.
Finally, we pack all this into one huge tree $T=\bigl(\sum_{\vec t}T_{\vec t}\bigr)^p$.

If $T_{\dots}$ is any of the trees above, let $r_{\dots}$ denote its root; in particular, $r$ denotes the root of~$T$.

We now assume that $L$ is cofinally subframe-universal for trees, thus there exists an $L$-frame $W$ and a weak cofinal
subreduction $g$ from $W$ onto (the underlying frame of) $T$. We might assume that $W$ is rooted, and its root is
mapped to $r$ by~$f$. We adjust the subreduction by putting
\[f(u)=x\iff u\in g^{-1}[x]\Down\bez\bigcup_{\substack{y\in T\\x\nleq y}}g^{-1}[y]\Down.\]
for all $u\in W$ and $x\in T$. It is easy to see that the purported $f$-preimages of points of~$T$ are disjoint, hence
$f$ is a well-defined partial map $W\to T$. Moreover, $g^{-1}[x]\sset f^{-1}[x]$ and $f^{-1}[x]\Down=g^{-1}[x]\Down$
for each $x\in T$. This implies that $f$ is still a cofinal weak subreduction from $W$ to~$T$. Moreover, the preimages
$f^{-1}[x]$ are \emph{convex}: $f^{-1}[x]\Up\cap f^{-1}[x]\Down\sset f^{-1}[x]$. We now define a valuation of the
parameter~$p$ in~$W$ by
\[u\model p\iff u\in\dom(f)\land f(u)\model p.\]

Let $h$ be the p-morphism from $T$ to~$F$ mapping the root of each embedded copy of $T^d_i$ to~$x^d_i$. Since $\dom(h)$
is a cofinal generated subframe of~$T$, $h\circ f$ is a cofinal weak subreduction from $W$ to~$F$ with convex preimages.
\begin{Cl}\label{cl:nexp-1par-sat}
For any $u\in\dom(h\circ f)$,
\[h(f(u))=a^d_i\implies u\model\beta^d_i.\]
\end{Cl}
\begin{Pf*}
By induction on~$d$. If $d=0$, then $a^d_i$ is a leaf of~$F$, thus $(h\circ f)[u\Up]=\{a^d_i\}$. By cofinality and
convexity of~$h\circ f$, we have in fact $u\Up\sset(h\circ f)^{-1}[a^d_i]$. Thus, $u\model\boxdot p^i$.

Assume the result holds up to~$2d$, and let $h(f(u))=a^{2d+1}_i$. Clearly, $u\model\neg p$, and
$u\model\diadot\beta^{2d}_i$ by the induction hypothesis. In order to check $u\model\eta^{2d+1}_i$, let $u\le v\model
p\lor\Box\neg p$. If $v\model p$, we have $v\in\dom(f)$, thus $h(f(v))=a^{d'}_{i'}\ge a^{2d+1}_i$; since $h(f(v))\model
p$, actually $a^{d'}_{i'}\ge a^{2d}_i$. By the induction hypothesis, $v\model\beta^{d'}_{i'}$, hence
$v\model\eta_i^{2d}$ by Claim~\ref{cl:nexp-1par-beta}. Otherwise, $v\model\boxdot\neg p$; by cofinality, there is $w\ge v$
such that $w\in\dom(h\circ f)$, and we may choose it so that $h(f(w))$ is a leaf. By the $d=0$ case, we cannot have
$h(f(w))=a^0_1$, as $w\model\boxdot\neg p$. Thus, $h(f(w))=a^0_0$; since $h\circ f$ is a subreduction, $a^{2d}_i\le
a^0_0$, thus $\vdash\boxdot\neg p\to\eta^{2d}_i$ by Claim~\ref{cl:nexp-1par-beta}, thus $v\model\eta^{2d}_i$.

Now, let $h(f(u))=a^{2d+2}_{\{j,k\}}$. Again, $u\model p$ holds by definition, and
$u\model\dia\beta^{2d+1}_j\land\dia\beta^{2d+1}_k$ by the induction hypothesis, hence it suffices to show
$u\model\eta^{2d+2}_{\{j,k\}}$.

If $u\le v\model p$, then $v\in\dom(h\circ f)$, and $a^{d'}_{i'}:=h(f(v))$ satisfies
$a^{d'}_{i'}=a^{2d+2}_{\{j,k\}}$ or $a^{d'}_{i'}\ge a^{2d}_j$ or $a^{d'}_{i'}\ge a^{2d}_k$. In the first case, we have
$v\model\dia\beta^{2d+1}_j\land\dia\beta^{2d+1}_k$ by the induction hypothesis; otherwise, the induction hypothesis
gives $v\model\beta^{d'}_{i'}$, hence $v\model\eta^{2d}_j$ or $v\model\eta^{2d}_k$ by Claim~\ref{cl:nexp-1par-beta}.

Assume $u\le v\le w$, $v\model\neg p$, and $w\model p\lor\Box\neg p$. If $w\model p$, then again $w\in\dom(h\circ f)$,
and $a^{d'}_{i'}:=h(f(w))$ satisfies $a^{d'}_{i'}=a^{2d+2}_{\{j,k\}}$ or $a^{d'}_{i'}\ge j$ or $a^{d'}_{i'}\ge a^{2d}_k$. The
first case is in fact impossible: if $h(f(u))=h(f(w))=a^{2d+2}_{\{j,k\}}$, then $h(f(v))=a^{2d+2}_{\{j,k\}}$ as well by
convexity, contradicting $v\nmodel p$. Thus, we have $w\model\eta^{2d}_j$ or $w\model\eta^{2d}_k$ as above%
\footnote{It may happen here that $v\notin\dom(f)$, while $v\in(h\circ f)^{-1}[a^{2d}_j]$ and
$v\in(h\circ f)^{-1}[a^{2d}_k]$, even if $j\ne k$. Then $v$ satisfies neither $\eta^{2d+1}_j$ nor~$\eta^{2d+1}_k$. It is
precisely for this reason that we did not simplify the second conjunct of $\eta^{2d+2}_{\{j,k\}}$ to
$\boxdot(\neg p\to\eta^{2d+1}_j\lor\eta^{2d+1}_k)$, as one might be tempted to do just looking at~$F$.}%
. The
remaining case $w\model\boxdot\neg p$ is left to the reader.
\end{Pf*}

Claim~\ref{cl:nexp-1par-sat} and~\eqref{eq:50} imply that each of the formulas
$\beta_0,\dots,\beta_{n-1},\gamma_0,\dots,\gamma_{m-1},\delta$ holds in~$W$ only in points $u$ such that $f(u)$ is the
root of a copy of the corresponding tree $T_{\beta_0},\dots,T_\delta$. Consequently, $\theta_\alpha$ holds only in
preimages of copies of $r_{t,\alpha}$ for some~$t\sset n$.

We are getting ready to bring the proof of Theorem~\ref{thm:nexp-1par} to an end. Assume that $\xi_\Phi$ is $L$-unifiable,
and let $\sigma$ be its unifier. We may assume $\sigma(x)$ uses no parameters except for~$p$, and no variables.

We need to define a witness $X$ for~$\Phi$. For any $t\sset n$, we put $t\in X$ if a preimage of some copy of
$r_{t,\alpha}$ for some $\alpha<m$ satisfies $\sigma(x)$. Crucially, this definition does not depend on the choices:
\begin{Cl}\label{cl:nexp-1par-consX}
If $t\sset n$, and $u,v\in W$ are such that $f(u)$ is a copy of $r_{t,\alpha}$ and $f(v)$ a copy of $r_{t,\beta}$
for some $\alpha,\beta<m$, then
\[u\model\sigma(x)\iff v\model\sigma(x).\]
\end{Cl}
\begin{Pf*}
Let $T'$ be the frame~$T$ with valuation of $p$ modified so that $p$ is false everywhere outside copies of
$T_{t,\alpha}$ or~$T_{t,\beta}$.
Let $W'$ be the corresponding modification of~$W$:
\[W',w\model p\iff w\in\dom(f)\text{ and }T',f(w)\model p.\]
In $W'$, $\theta$ holds only in points~$w$ such that $f(w)$ is a copy of $r_{t,\alpha}$ or~$r_{t,\beta}$. All these
points satisfy the same formulas of the form $\dia\beta_i$, namely $w\model\dia\beta_i$ iff $i\in t$. Thus, the root $r$ of
$W'$ satisfies the premise of $\sigma(\ref{eq:40})$. This implies
\[W',r\model\Box\bigl(\theta\to\sigma(x)\bigr)\quad\text{or}\quad W',r\model\Box\bigl(\theta\to\neg\sigma(x)\bigr),\]
and in particular, $u\model\sigma(x)$ iff $v\model\sigma(x)$.
\end{Pf*}

Now, let $t_0,\dots,t_{m-1}\sset n$; we need to show
\begin{equation}\label{eq:51}
\fii(\dots,i\in t_\alpha,\dots,t_\alpha\in X,\dots).
\end{equation}
Let $u\in\dom(f)$ be such that $f(u)$ is a copy of $r_{\vec t}$. The only points above $u$ that satisfy~$\theta_\alpha$
are preimages of the embedded copy of $T_{t_\alpha,\alpha}$, and these satisfy the same formulas of the form
$\dia\beta_i$ (viz., such that $i\in t_\alpha$). In other words, $u$ satisfies the premise
of~$\sigma(\ref{eq:41})$, hence
\[u\model\fii\bigl(\dots,\dia(\theta_\alpha\land\dia\beta_i),\dots,\dia(\theta_\alpha\land\sigma(x)),\dots\bigr).\]
Moreover,
\begin{alignat*}{2}
u&\model\dia(\theta_\alpha\land\dia\beta_i)&&\iff i\in t_\alpha,\\
u&\model\dia(\theta_\alpha\land\sigma(x))&&\iff t_\alpha\in X,
\end{alignat*}
hence \eqref{eq:51} follows. This completes the proof of Theorem~\ref{thm:nexp-1par}.
\end{Pf}
\begin{Cor}\label{cor:nexp-1par-dp}
Unifiability of formulas with $1$~parameters and $1$~variable is $\NEXP$-hard for all logics
with the disjunction property.
\end{Cor}
\begin{Pf}
By Theorems \ref{thm:nexp-1par} and~\ref{thm:dp-csf-univ}.
\end{Pf}
\begin{Exm}\label{exm:s42grz}
The cofinality condition in the statement of Theorem~\ref{thm:nexp-1par} cannot be dropped. For example, the logic
$L=\lgc{S4.2Grz}$ is subframe-universal for trees, but $L$-unifiability of formulas with one parameter is
$\np$-complete.  The universal frame $\unifr Lp$ consists of two disjoint infinite descending chains alternating
between $p$ and~$\neg p$, one with $p$ on top, and the other with $\neg p$ on top. It is easy to show that a formula
$\fii\in\Form(\{p\},\Var)$ is $L$-unifiable iff there exists a valuation $\model$ in~$U_L(p)$ that makes
$U_L(p)\model\fii$ iff there exists a valuation that makes $\fii$ true in the top $2n$ levels of~$U_L(p)$, where
$n=\lh{\{\psi:\Box\psi\sset\fii\}}+1$. This condition can be checked in~$\np$.

Nevertheless, the result holds under somewhat weaker assumptions. In particular, if the single-atom fragment of~$L$ is
included in a logic cofinally subframe-universal for trees, then $L$-unifiability of
formulas in $1$~parameter and $1$~variable is $\NEXP$-hard. For example, this applies to bounded top-width logics
such as $L=\lgc{S4BTW_2}$: the single-atom fragment of~$L$ is included in~$\lgc{S4Grz}$ (which is cofinally
subframe-universal for trees), as $U_\lgc{S4Grz}(p)$ has top width~$2$.

Unification with $1$~parameter is also $\NEXP$-hard for logics that are subframe-universal in a way that respects
reflexivity of points, see Theorem~\ref{thm:nexp-0par}.

Moreover, we can easily generalize the lower bound to all subframe-universal logics at the expense of using one more
parameter:
\end{Exm}
\begin{Thm}\label{thm:nexp-2par}
Unifiability of formulas with $2$~parameters and $1$~variable is uniformly $\NEXP$-hard in the class of logics
subframe-universal for trees.
\end{Thm}
\begin{Pf}
We use almost the same construction as in the proof of Theorem~\ref{thm:nexp-1par}, except that we take another
parameter~$q$, and \emph{relativize} the formulas~\eqref{eq:42} (recall Definition~\ref{def:transl}): instead of $\beta_i$,
$\theta_\alpha$, and~$\theta$, we use $q\land\beta_i^q$, $q\land\theta_\alpha^q$, and~$q\land\theta^q$. The proof
proceeds as before, but the subreduction $f$ is not necessarily cofinal, and we put
\[W,u\model q\iff u\in\dom(f).\]
The only place where we used the cofinality of~$f$ in the proof of Theorem~\ref{thm:nexp-1par} was in
Claim~\ref{cl:nexp-1par-consX}, to ensure that $(h\circ f)^{-1}[a^0_1]\model\boxdot p$, and that certain points satisfying
$\boxdot\neg p$ are in $(h\circ f)^{-1}[a^0_0]\Down$. Relativization side-steps both issues: the former property turns
into $(h\circ f)^{-1}[a^0_1]\model\boxdot(q\to p)$, which follows immediately from $h\circ f$ being a subreduction and
the definition of satisfaction of~$q$; the latter property holds because the points in question will now satisfy~$q$,
hence they will be in the domain of~$h\circ f$.
\end{Pf}

Going in the opposite direction, we may ask if at least one parameter is needed for $\NEXP$-hardness of unification in,
say, nonlinear clx logics. In general, the answer is yes: parameter-free unifiability is in~$\np$ for a large class of
logics, including all extensions of~$\lgc{D4}$ or~$\lgc{GL}$. However, we can prove $\NEXP$-hardness for logics that
allow enough mixing of reflexive and irreflexive points. The theorem below gives a typical and reasonably useful
sufficient condition, but similar results also holds for some other logics---basically, what is needed is that we can
construct embeddings of arbitrarily large trees into $\unifr L\nul$ in a uniform and efficient manner.

Using the relativization technique, we can also obtain additional $\NEXP$-hardness results for unifiability with
\emph{one} parameter in this way. Specifically, it applies to logics that satisfy a variant of subframe-universality
for trees including both reflexive and irreflexive points. As with Definition~\ref{def:tree-universal}, we may equivalently
state the definition for trees, for binary trees, or for arbitrary frames without proper clusters; this time we chose
the last one, to emphasize that we treat the objects as frames rather than combinatorial shapes, hence their reflexivity
matters.
\begin{Def}\label{def:frame-universal}
A logic $L\Sset\kiv$ is \emph{subframe-universal for simple frames} if for each finite rooted frame~$F$ without proper clusters,
there exists an $L$-frame that subreduces onto~$F$; equivalently, if $\vdash_L\reltr(\fii)$ implies
$\vdash_\lgc{K4Grz}\fii$ for all formulas~$\fii$.
\end{Def}

For example, $\lgc{D4.2}$ is subframe-universal for simple frames, but not cofinally subframe-universal for trees, and
parameter-free $\lgc{D4.2}$-unifiability is in~$\np$.
\begin{Thm}\label{thm:nexp-0par}
\
\begin{enumerate}
\item\label{item:3}
Unifiability of formulas without parameters and with $1$~variable is uniformly $\NEXP$-hard in sublogics of\/
$\lgc{K4.2GrzBB_2}$.
\item\label{item:15}
Unifiability of formulas with $1$~parameters and $1$~variable is uniformly $\NEXP$-hard in the class of logics
subframe-universal for simple frames.
\end{enumerate}
\end{Thm}
\begin{Pf}

\ref{item:3}: We follow similar strategy as in the proof of Theorem~\ref{thm:nexp-1par}. We construct $\beta_i$ as formulas
defining points in a certain antichain in~$\unifr L\nul$; instead of the parameter~$p$, we will distinguish points
by their reflexivity.

\begin{figure}
\centering
\magicparoff
\unitlength=1pt
\begin{picture}(110,85)

\put(0,80){$\scriptscriptstyle0$}
\put(55,80){\circle*{4}}

\put(0,64){$\scriptscriptstyle1$}
\put(40,65){\circle{4}}
\put(41.5,66.5){\vector(1,1){12}}
\put(70,65){\circle*{4}}
\put(68.5,66.5){\vector(-1,1){12}}

\put(0,44){$\scriptscriptstyle2$}
\put(40,45){\circle*{4}}
\put(40,45){\vector(0,1){18}}
\put(40,45){\vector(3,2){28.5}}
\put(70,45){\circle*{4}}
\put(70,45){\vector(0,1){18}}

\put(0,24){$\scriptscriptstyle3$}
\put(20,25){\circle*{4}}
\put(20,25){\vector(1,1){18.5}}
\put(40,25){\circle*{4}}
\put(40,25){\vector(0,1){18}}
\put(40,25){\vector(3,2){28.5}}
\put(70,25){\circle*{4}}
\put(70,25){\vector(0,1){18}}
\put(70,25){\vector(-3,4){28.5}}
\put(90,25){\circle*{4}}
\put(90,25){\vector(-1,1){18.5}}

\put(14,2){$\iddots$}
\put(53,2){$\vdots$}
\put(82,2){$\ddots$}
\end{picture}
\caption{The subframe of $\unifr{\lgc{K4.2GrzBB_2}}\nul$ used in the proof of Theorem~\ref{thm:nexp-0par}.}
\label{fig:nexp-0par}
\end{figure}

We will define antichains $\{a^d_i:i<n_d\}\sset\unifr L\nul$ by induction on~$d$. There is more than one way to do it.
For conciseness of notation, if $X$ is a finite subset of~$\unifr L\nul$, and $\RI\in\{\I,\R\}$, let $\tp_\RI(X)$ denote the
unique $\RI$-tp of $X\Up$ in~$\unifr\kiv\nul$. We will only use this notation if it is guaranteed that actually
$\tp_\RI(X)\in\unifr L\nul$ (in particular, $X$ is generated by at most $2$~points), and if $X\Up$ is not rooted with a
reflexive root, so that always $\tp_\RI(X)\notin X\Up$.

We put $n_0=1$, $n_1=n_2=2$, $n_3=4$, and $n_{d+1}=\binom{n_d+1}2$ for $d\ge3$. Let $a^0_0=\tp_\I(\nul)$ be the
irreflexive leaf of~$\unifr L\nul$, $a^1_0=\tp_\R(a^0_0)$, and~$a^1_1=\tp_\I(a^0_0)$. All remaining points will be
irreflexive: $a^2_0=\tp_\I(a^1_0,a^1_1)$, $a^2_1=\tp_\I(a^1_1)$, $a^3_0=\tp_\I(a^2_0)$, $a^3_1=\tp_\I(a^2_0,a^2_1)$,
$a^3_2=\tp_\I(a^1_0,a^2_1)$, and $a^3_3=\tp_\I(a^2_1)$. For $d\ge3$, we identify $i<\binom{n_d+1}2$ with sets $\{j,k\}$
for $j,k<n_d$ (not necessarily distinct), and we put $a^{d+1}_{\{j,k\}}=\tp_\I(a^d_j,a^d_k)$. (See
Fig.~\ref{fig:nexp-0par}.)

We define $\beta^d_i$ as the formulas $\beta_{a^d_i}$ from~\eqref{eq:60}. As in the proof of Theorem~\ref{thm:nexp-1par}, we
have $n_d=2^{2^{\Omega(d)}}$. In order to see that $\lh{\beta^d_i}=2^{O(d)}$, it is better to present the formulas in a
slightly different way. Let $\eta_u=\LOR_{v\ge u}\beta_v$; we may define $\beta_u$ and $\eta_u$ by simultaneous
recursion
\begin{align*}
\beta_u&=\ET_{i<m_u}\dia\beta_{u_i}\land\Box\LOR_{i<m_u}\eta_{u_i},\\
\eta_u&=\beta_u\lor\LOR_{i<m_u}\eta_{u_i},
\end{align*}
where $\{u_i:i<m_u\}$ are the immediate successors of~$u$, and $u$ is irreflexive. (We leave the reflexive case to the
reader, as with our choice of antichains, it is irrelevant.) Since in our case $m_u\le2$, the formulas $\beta_u$
and~$\eta_u$ are built from $O(1)$~instances of $\beta_v$ and~$\eta_v$ for $v$ of lower depth, hence the bound
$\lh{\beta^d_i}\le2^{O(d)}$ follows by induction on~$d$.

As in the proof of Theorem~\ref{thm:nexp-1par}, we take $D$ minimal such that $n_D\ge n+m+2$, and define the
formulas~\eqref{eq:42} by $\beta_i=\beta^D_i$ for~$i<n$, $\gamma_\alpha=\beta^D_{n+\alpha}$ for $\alpha<m$,
$\delta=\beta^D_{n+m}$, and
\begin{align*}
\theta_{\phantom\alpha}&=\dia\delta\land\neg\delta,\\
\theta_\alpha&=\dia\delta\land\dia\gamma_\alpha\land\ET_{\alpha'\ne\alpha}\neg\dia\gamma_{\alpha'}.
\end{align*}
We have $D=O(\log\log(n+m))$, thus $\lh{\beta^D_i}=(\log(n+m))^{O(1)}$, and it is routine to verify that the formulas are
logspace-constructible. We denote $a_{\beta_i}=a^D_i$ for $i<n$, $a_{\gamma_\alpha}=a^D_{n+\alpha}$ for $\alpha<m$,
$a_\delta=a^D_{n+m}$, and $a_-=a^D_{n+m+1}$.

For each $t\sset n$, we define $a_t\in\unifr L\nul$ as a predecessor of $\{a_-\}\cup\{a_{\beta_i}:i\in t\}$ arranged in a
binary tree of tp's: that is, if $t=\{i_j:j<l\}$, we put
\[a_t=\tp_\I(a_{\beta_{i_{l-1}}},\tp_\I(\dots,\tp_\I(a_{\beta_{i_0}},a_-)\dots)).\]
For any $\alpha<m$, we define
\[a_{t,\alpha}=\tp_\I(a_\delta,\tp_\I(a_{\gamma_\alpha},a_t)).\]
Finally, for each $\vec t\in(\pw n)^m$, let
\[a_{\vec t}=\tp_\I(a_{t_{m-1},m-1},\tp_\I(\dots,\tp_\I(a_{t_1,1},a_{t_0,0})\dots)).\]

We have to show that if $\vdash_L\sigma(\xi_\Phi)$, then $\Phi$ is true. We may assume that
$\sigma(x)\in\Form(\nul,\nul)$. We define a set $X\sset\pw n$ by
\[t\in X\iff a_{t,\alpha}\model\sigma(x)\]
for any $t\sset n$ and $\alpha<m$. This definition is independent of the choice of~$\alpha$: putting
$u=\tp_\I(a_{t,\alpha},a_{t,\alpha'})$, the only element of $u\Up$ that satisfies~$\delta$ is $a_\delta$, hence
$\theta$ only holds in the three points $u,a_{t,\alpha},a_{t,\alpha'}$. These all see $a_{\beta_i}$ for $i\in t$, and
for no other $i<n$. Thus, $u$ satisfies the premise of~$\sigma(\ref{eq:40})$, hence its conclusion, which gives
\[a_{t,\alpha}\model\sigma(x)\iff a_{t,\alpha'}\model\sigma(x).\]

To check that $X$ is a witness for~$\Phi$, let $\vec t\in(\pw n)^m$. As above, it is easy to see that $\theta_\alpha$
is satisfied in~$a_{\vec t}\Up$ only in the point $a_{t_\alpha,\alpha}$, hence $a_{\vec t}$ satisfies the premise
of~$\sigma(\ref{eq:41})$, hence
\[a_{\vec t}\model\fii\bigl(\dots,\dia(\theta_\alpha\land\dia\beta_i),\dots,\dia(\theta_\alpha\land\sigma(x)),\dots\bigr).\]
Moreover,
\begin{alignat*}{2}
a_{\vec t}&\model\dia(\theta_\alpha\land\dia\beta_i)&&\iff i\in t_\alpha,\\
a_{\vec t}&\model\dia(\theta_\alpha\land\sigma(x))&&\iff t_\alpha\in X,
\end{alignat*}
hence \eqref{eq:51} is true.

\ref{item:15}: We relativize the formulas used in~\ref{item:3}, i.e., we define $\beta_i=p\land(\beta^D_i)^p$,
$\gamma_\alpha=p\land(\beta^D_{n+\alpha})^p$, $\delta=p\land(\beta^D_{n+m})^p$, and we define $\theta,\theta_\alpha$ as
in \eqref{eq:61} and~\eqref{eq:62}.

Continuing the argument from~\ref{item:3}, let $a\in\unifr\kiv\nul$ be a common predecessor of all the points $a_{\vec
t}$ for $\vec t\in(\pw n)^m$. By assumption, there exists an $L$-frame $W$, and a subreduction $f\colon W\onto a\Up$.
We make $W$ into a parametric frame by putting
\[u\model p\iff u\in\dom(f).\]
Recall that by Lemma~\ref{lem:relat}, we have
\[u\model\fii^p\iff f(u)\model\fii\]
for all $u\in\dom(f)$ and $\fii\in\Form(\nul,\nul)$.

Let $\sigma$ be an $L$-unifier of~$\xi_\Phi$ such that $\sigma(x)\in\Form(p,\nul)$. We define $X\sset\pw n$ by
\begin{equation}\label{eq:63}
t\in X\iff\exists u\in W\,\exists\alpha<m\,\bigl(f(u)=a_{t,\alpha}\land u\model\sigma(x)\bigr).
\end{equation}
Using \eqref{eq:40} and a modified valuation that makes $p$~true only in $f^{-1}[a_{t,\alpha}\Up\cup
a_{t,\alpha'}\Up]$, we may check that the definition of $X$ does not depend on the choice of $\alpha$ or~$u$
in~\eqref{eq:63}.

Then, for any $\vec t\in(\pw n)^m$, we take $u\in f^{-1}[a_{\vec t}]$, and using~\eqref{eq:41} we obtain
\[u\model\fii(\dots,\dia(\theta_\alpha\land\dia\beta_i),\dots,\dia(\theta_\alpha\land\sigma(x)),\dots\bigr),\]
from which \eqref{eq:51} follows as usual.
\end{Pf}

As we already mentioned, parameter-free unification is in~$\np$ for a large number of logics of interest. One way to
make it harder is to use parameters, but another way is to consider \emph{admissibility} rather than unification:
parameter-free admissibility is $\coNEXP$-hard for typical logics of unbounded branching, as shown in~\cite{ej:admcomp}. We will
generalize this result to a class of logics that includes all $\p{\R,2}$-extensible or $\p{\I,2}$-extensible logics,
and in particular, all nonlinear clx logics.

While we can prove hardness of unification using ``static'' conditions requiring the presence of suitable subframes, it
seems that hardness of admissibility requires ``dynamic'' conditions expressing closure properties of the class of
$L$-frames. Thus, the condition we employ is a little peculiar to formulate:
\begin{Def}\label{def:weak-tree-ext}
Let $T$ be a (finite) tree. A logic $L\Sset\kiv$ has the \emph{weak $T$-extension property} if for every labelling of the
leaves of~$T$ by finite clusters that are $L$-frames, there exists an $L$-frame $W$ with skeleton~$T$ such that the
final clusters of~$W$ are isomorphic to their labels in~$T$.

If $\mathcal T$ is a family of trees, $L$ has the \emph{weak extension property wrt~$\mathcal T$} if if has the weak
$T$-extension property for each $T\in\mathcal T$.

A family of trees $\mathcal T$ is \emph{depth-$3$-universal} if every tree of depth~$3$ embeds (as a poset) in a tree
$T\in\mathcal T$.
\end{Def}
\begin{Exm}\label{exm:dp3-univ}
The \emph{depth-$3$ weak extension property}, proved to imply $\coNEXP$-hardness of parameter-free admissibility in
\cite[Thm.~4.13]{ej:admcomp}, is nothing else than the weak extension property wrt the family of all trees of depth~$3$.

The family of all depth-$3$ trees is trivially depth-$3$-universal. The family of all binary trees is also
depth-$3$-universal.

Recall that a \emph{caterpillar} is a tree that becomes a single path if we remove all leaves; a binary caterpillar
tree thus consists of a central path $x_0<x_1<\dots<x_n$, with at most one additional leaf vertex attached to
each~$x_i$. Let us define a \emph{squared caterpillar} to be a tree obtained from a caterpillar by replacing each of
its leaves with a caterpillar. Then the family of binary squared caterpillars is depth-$3$-universal.
\end{Exm}
\begin{Thm}\label{thm:nexp-0par-adm}
Single-conclusion admissibility without parameters is uniformly $\coNEXP$-hard in the class of all logics satisfying
the weak extension property wrt a depth-$3$-universal family of trees (including all $\p{\R,2}$-extensible or
$\p{\I,2}$-extensible logics).
\end{Thm}
\begin{Pf}
Given a sentence~$\Phi$ as in~\eqref{eq:38}, we define formulas
\begin{alignat*}{2}
\beta_i&=\boxdot\Bigl(z_i\land\ET_{\substack{i'<n+m+4\\i'\ne i}}\neg z_{i'}\Bigr),&\qquad& i<n+m+4,\\
\gamma_\alpha&=\beta_{n+\alpha},&&\alpha<m,\\
\delta_p&=\beta_{n+m+p},&&p<4,\\
\theta_{\phantom\alpha}&=\dia\delta_1\land\dia\delta_2\land\dia\delta_3\land\neg\dia\delta_0,\\
\theta_\alpha&=\theta\land\dia\gamma_\alpha\land\ET_{\alpha'\ne\alpha}\neg\dia\gamma_{\alpha'},&&\alpha<m,\\
\zeta_\Phi&=\LOR_{i<n+m+4}\Box\neg\beta_i
\end{alignat*}
using variables $\{z_i:i<n+m+4\}$, and we define $\xi_\Phi$ as $(\ref{eq:40})\land(\ref{eq:41})$. We aim to prove
\begin{equation}\label{eq:52}
\Phi\text{ is true}\iff\xi_\Phi\nadm_L\zeta_\Phi.
\end{equation}
For the left-to-right implication, Lemma~\ref{lem:nexp-bdpar-true} and Remark~\ref{rem:next-bdpar-true-expl} show that there is a
$\kiv$-unifier $\sigma$ of~$\xi_\Phi$ such that $\sigma(z_i)=z_i$ for all $i<n+m+4$, thus
$\sigma(\zeta_\Phi)=\zeta_\Phi$. It remains to observe that $\nvdash_L\zeta_\Phi$: it suffices to take a finite rooted
$L$-frame with $\ge n+m+4$ final clusters, where we make $z_i$ (and consequently, $\beta_i$) true only in the $i$th
cluster; the existence of such an $L$-frame follows from the weak extension property.

For the right-to-left implication, let $\sigma$ be an $L$-unifier of~$\xi_\Phi$ such that
$\nvdash_L\sigma(\zeta_\Phi)$. For each $i<n+m+4$, $\sigma(\beta_i)$ is satisfiable in a finitely generated
descriptive $L$-frame~$W_i$. Since it has the form $\boxdot(\dots)$, it holds on a cone; using descriptivity, this cone
contains a final cluster, which is finite due to $W_i$ being finitely generated. Thus, we may fix a finite
single-cluster model $B_i\model\sigma(\beta_i)$ based on an $L$-frame. We will write $C_\alpha=B_{n+\alpha}$ for
$\alpha<m$, and $D_p=B_{n+m+p}$ for $p<4$.

At least two of the clusters $D_1,D_2,D_3$ have the same reflexivity; w.l.o.g.\ $\refl(D_1)=\refl(D_2)$. We may also
assume $\lh{D_1}=\lh{D_2}$ by replicating some point from the smaller cluster if necessary.

Let $S$ be the depth-$3$ tree consisting of a root with $m2^{n+1}$ immediate successors, each of which has $n+4$
successors. By assumption, $S$ embeds in a tree~$T$ such that $L$ has the weak $T$-extension property. This means that
$T$ includes an antichain $A$ of size $m2^{n+1}$ such that each element $a\in A$ sees at least $n+4$ leaves. If an
element $a\in A$ has only one immediate successor~$a'$, we may replace $a$ with~$a'$ while keeping the number of
leaves. Thus, we may assume w.l.o.g.\ that every $a\in A$ splits. Going in the opposite direction, let $a^-\le a$ be
the minimal element below~$a$ such that no element of the chain $[a^-,a]$ splits except~$a$. 

We can find a subset $A_0\sset A$ of size $\ge m2^n$ such that for each $a\in A_0$, $a^-$ has a sibling that sees
a leaf in $T\bez A_0\Up$. Indeed, let $\sim$ be the equivalence relation on~$A$
defined by $a_0\sim a_1$ iff $a_0^-$ is a sibling of~$a_1^-$, and let $A_0$ be constructed from~$A$ by removing one
element from each $\sim$-equivalence class of size $\ge2$. Since we removed at most half of the elements, we have
$\lh{A_0}\ge\lh A/2=m2^n$, and it is easy to see that $A_0$ has the required property. We name $m2^n$ elements
of~$A_0$ as $\{r_{t,\alpha}:t\sset n,\alpha<m\}$.

We will label each leaf of~$T$ with one of the clusters $B_i$, $i<n+m+4$, as follows. Let $t\sset n$ and $\alpha<m$.
There are $\ge n+4$ leaves in $r_{t,\alpha}\Up$. For each $p=1,2,3$, we choose one leaf above~$r_{t,\alpha}$ and label
it with~$D_p$; we do it in such a way that the leaves labelled with $D_1$ and~$D_2$ are above different immediate
successors of~$r_{t,\alpha}$, thus no element strictly above $r_{t,\alpha}$ sees both. For each $i\in t$, we label one
leaf above $r_{t,\alpha}$ with~$B_i$. We label the remaining leaves
above~$r_{t,\alpha}$ with $C_\alpha$. Finally, we label every leaf of~$T$ that is not above any~$r_{t,\alpha}$
with~$D_0$.

By the weak $T$-extension property, there exists an $L$-frame~$W$ obtained from~$T$ by choosing the reflexivity of each
inner node, and replacing each leaf with (the underlying frame of) its label. We will identify the corresponding inner nodes of $T$ and~$W$. We
consider $W$ not just as a frame, but also as a model: the valuation of variables in leaf clusters is taken from
the corresponding $B_i$~models, and the valuation in inner nodes is arbitrary (say, all variables are false).

Notice that $\vdash_\kiv\beta_i\to\boxdot\neg\beta_{i'}$ for $i\ne i'$. Each leaf cluster of~$W$ is labelled with
some~$B_i$, and it then satisfies~$\sigma(\beta_i)$, and $\neg\sigma(\beta_{i'})$ for all $i'\ne i$. An inner node
of~$W$ that sees leaves with two different labels cannot satisfy any~$\sigma(\beta_i)$.

It follows that for any $u\in W$,
\begin{alignat}{2}
\label{eq:55}
W,u&\model\sigma(\theta)&&\iff\exists t\sset n\,\exists\alpha<m\,(r_{t,\alpha}^-\le u\le r_{t,\alpha}),\\
\intertext{and for each~$\alpha<m$,}
\label{eq:54}
W,u&\model\sigma(\theta_\alpha)&&\iff\exists t\sset n\,(r_{t,\alpha}^-\le u\le r_{t,\alpha}).
\end{alignat}
Indeed, the right-to-left implications follow easily from the definition. On the other hand, if $u$ is not above
any~$r_{t,\alpha'}^-$, the construction of~$A_0$ ensures that $u$ sees a leaf labelled~$D_0$, hence
$u\model\sigma(\diadot\delta_0)$, which implies $u\nmodel\sigma(\theta)$. If $u\gnsim r_{t,\tilde\alpha}$ for some
$t\sset n$ and~$\tilde\alpha<m$, then $u\nmodel\sigma(\dia\delta_1\land\dia\delta_2)$, hence $u\nmodel\sigma(\theta)$
again. And if $r_{t,\alpha'}^-\le u\le r_{t,\alpha'}$ for some $t$ and $\alpha'\ne\alpha$, then
$u\model\dia\gamma_{\alpha'}$, which implies $u\nmodel\sigma(\theta_\alpha)$.
\begin{Cl}\label{cl:nexp-0par-adm-X}
Let $t\sset n$, $\alpha,\alpha'<m$, and $u,u'\in W$ be such that $r_{t,\alpha}^-\le u\le r_{t,\alpha}$ and
$r_{t,\alpha'}^-\le u'\le r_{t,\alpha'}$. Then
\begin{equation}\label{eq:56}
W,u\model\sigma(x)\iff W,u'\model\sigma(x).
\end{equation}
\end{Cl}
\begin{Pf*}
We define a modified valuation $\model^t$ in~$W$ as follows: for any leaf cluster labelled~$D_1$ above some
$r_{\tilde t,\tilde\alpha}$, where $\tilde t\ne t$, we change the valuation to match~$D_2$. (This is possible because
$D_1$ and~$D_2$ have isomorphic underlying frames.) This ensures that no point $v\ge r_{\tilde t,\tilde\alpha}^-$ satisfies
$\sigma(\dia\delta_1)$, and consequently, $v\nmodel\sigma(\theta)$. Thus, the only points that
satisfy~$\sigma(\theta)$ in~$\p{W,{\model^t}}$ are the elements of
$\bigcup_{\tilde\alpha<c}[r_{t,\tilde\alpha}^-,r_{t,\tilde\alpha}]$. All these points satisfy the same formulas of the form
$\sigma(\dia\beta_i)$, namely those such that $i\in t$.

This means $\p{W,{\model^t}}$ satisfies the premise of $\sigma(\ref{eq:40})$ in the root, hence
\[W\model^t\Box\bigl(\sigma(\theta)\to\sigma(x)\bigr)\quad\text{or}\quad W\model^t\Box\bigl(\sigma(\theta)\to\neg\sigma(x)\bigr).\]
In particular, \eqref{eq:56} follows, as the valuation in $r_{t,\alpha}^-\Up\cup r_{t,\alpha'}^-\Up$ is unchanged,
and $u,u'\model\sigma(\theta)$.
\end{Pf*}

In view of Claim~\ref{cl:nexp-0par-adm-X}, we may define a set $X\sset\two^n$ such that
\[r_{t,\alpha}^-\le u\le r_{t,\alpha}\implies\bigl(u\model\sigma(x)\iff t\in X\bigr)\]
for all $t\sset n$, $\alpha<m$, and~$u\in W$. We claim that $X$ is a witness for the truth of~$\Phi$. Let $\vec
t\in(\pw n)^m$, we need to show that
\begin{equation}\label{eq:57}
\fii\bigl(\dots,i\in t_\alpha,\dots,t_\alpha\in X,\dots\bigr)
\end{equation}
holds. Similarly to the proof of Claim~\ref{cl:nexp-0par-adm-X}, we consider a modified valuation $\model^{\vec t}$ where
we ``disable'' $D_1$~leaf clusters that are not above $r_{t_\alpha,\alpha}$ for any~$\alpha<m$ by turning them
into a~$D_2$. The new valuation satisfies $\sigma(\theta_\alpha)$ only in the interval
$[r_{t_\alpha,\alpha}^-,r_{t_\alpha,\alpha}]$, hence the premise of $\sigma(\ref{eq:41})$ holds in the root~$r$ of~$W$, hence
also
\[W,r\model^{\vec t}\fii\bigl(\dots,\sigma\bigl(\dia(\theta_\alpha\land\dia\beta_i)\bigr),
           \dots,\sigma\bigl(\dia(\theta_\alpha\land x)\bigr),\dots\bigr).\]
Moreover, we have
\begin{alignat*}{2}
r&\model^{\vec t}\sigma\bigl(\dia(\theta_\alpha\land\dia\beta_i)\bigr)&&\iff i\in t_\alpha,\\
r&\model^{\vec t}\sigma\bigl(\dia(\theta_\alpha\land x)\bigr)&&\iff t_\alpha\in X,
\end{alignat*}
hence \eqref{eq:57} follows.
\end{Pf}
\begin{Rem}
We could slightly relax the weak $T$-extension property: it is enough to ask that there exists an $L$-frame~$W$, and a
weak p-morphism $f\colon W\onto T$ (i.e., a p-morphism $f\colon W^\R\onto T$) such that for each leaf $a\in T$ with
label~$C$, there is a p-morphism $f_C\colon f^{-1}[a]\onto C$. Since the class of $L$-frames is stable under p-morphic
images, this is almost equivalent to Definition~\ref{def:weak-tree-ext}; the only difference arises because a single node
of~$T$ may have both reflexive and irreflexive preimages in~$W$, hence it may not be possible to choose reflexivity of
inner nodes of~$T$ in a consistent way so that $f$ becomes a true p-morphism from~$W$. However, we do not know of any
interesting application where this would make a difference, hence we prefer the simpler formulation of
Definition~\ref{def:weak-tree-ext}.
\end{Rem}

\subsection{$\psp$ and below}\label{sec:psp-below}
\begin{Thm}\label{thm:lb-infdp}
\ \begin{enumerate}
\item\label{item:1}
For any~$d>0$, unifiability in logics of depth at least~$d$ is uniformly $\Pi^p_{2d}$-hard.
\item\label{item:2}
Unifiability of formulas with $2$~parameters and $1$~variable in logics of unbounded depth is uniformly $\psp$-hard.
\end{enumerate}
\end{Thm}
\begin{Pf}
Let us define auxiliary formulas
\begin{align*}
\delta_{0,e}&=q^e,&
\delta_{i+1,e}&=q^e\land\dia\delta_{i,1-e},\\
\theta_{i,e}&=\delta_{i,e}\land\neg\delta_{i+1,e}=\delta_{i,e}\land\Box\neg\delta_{i,1-e},&
\theta_i&=\theta_{i,i\bmod2}
\end{align*}
for $i\in\omega$ and~$e\in\two$. The meaning is that $w\model\delta_{i,e}$ iff there is a chain $w=w_i<w_{i-1}<\dots<w_0$
where $q$ is alternately true and false, with it being true (false) in~$w_i$ if $e=1$ ($0$, resp.), and
$w\model\theta_{i,e}$ if, in addition, there is no longer chain. The formula $\theta_i$ is normalized so that the chain
has $w_0\nmodel q$ on top. (If $w\model q^e$ sees a proper cluster in which both $q$ and~$\neg q$ are realized, then
$w\model\delta_{i,e}$ and $w\nmodel\theta_{i,e}$ for all $i\in\omega$.)

In particular, $w\model\theta_i$ (or $w\model\delta_{i,e}$) implies that $w$ has depth $>i$, and in a model where $q$
holds exactly in points of even depth, we have $w\model\theta_i$ iff the depth of~$w$ is~$i+1$. It follows from the
description that
\begin{align}
\label{eq:16}&\vdash_\kiv\theta_i\to\neg\theta_j\quad\text{for $i\ne j$,}\\
\label{eq:17}&\vdash_\kiv\theta_i\to\dia\theta_j\quad\text{for $i>j$.}
\end{align}

Let
\begin{equation}\label{eq:18}
\Phi=\forall\vec a_0\,\exists\vec b_0\dots\forall\vec a_{d-1}\,\exists\vec b_{d-1}\,
          \fii(\vec a_0,\dots,\vec a_{d-1},\vec b_0,\dots,\vec b_{d-1})
\end{equation}
be a given quantified Boolean sentence with $\fii$ quantifier-free, where each $\vec a_i$ stands for an $m$-tuple
$a_{i,0},\dots,a_{i,m-1}$, and similarly for~$\vec b_i$. We define a formula~$\xi_\Phi$ in $m+1$~parameters
$p_0,\dots,p_{m-1},q$, and $m$~variables $x_0,\dots,x_{m-1}$, as the conjunction of the formulas
\begin{align}
\label{eq:4}
\gamma&\to\ET_{i<d}\ET_{j<m}\bigl(\boxdot(\theta_i\to x_j)\lor\boxdot(\theta_i\to\neg x_j)\bigr),\\
\label{eq:5}
\gamma\land\theta_{d-1}&\to\fii\bigl(\dots,a_{i,j}/\diadot(\theta_i\land p_j),\dots,b_{i,j}/\diadot(\theta_i\land x_j),\dots\bigr),
\end{align}
where
\[\gamma=\ET_{i<d}\ET_{j<m}\bigl(\boxdot(\theta_i\to p_j)\lor\boxdot(\theta_i\to\neg p_j)\bigr).\]
Notice that $\lh{\theta_i}=O(i)$, hence $\lh{\xi_\Phi}=O\bigl(d^2m+d\lh\fii\bigr)=O\bigl(\lh\Phi^2\bigr)$; it is
also easy to see that the mapping $\Phi\mapsto\xi_\Phi$ is logspace-computable.
\pagebreak[2]
\begin{Cl}\label{cl:psptrue}
If $\Phi$ is true, then $\xi_\Phi$ is $\kiv$-unifiable.
\end{Cl}
\begin{Pf*}
For $i\le d$, denote
\begin{equation}\label{eq:21}
\Phi_i(\vec a_0,\dots,\vec a_{i-1},\vec b_0,\dots,\vec b_{i-1})=
      \forall\vec a_i\,\exists\vec b_i\dots\forall\vec a_{d-1}\,\exists\vec b_{d-1}\,
           \fii(\vec a_0,\dots,\vec a_{d-1},\vec b_0,\dots,\vec b_{d-1}).
\end{equation}
Note that $\Phi_0=\Phi$ and~$\Phi_d=\fii$. For every $i<d$ and $\vec a_0,\dots,\vec a_i,\vec b_0,\dots,\vec
b_{i-1}\in\two^m$ such that $\exists\vec b_i\,\Phi_{i+1}(\vec a_0,\dots,\vec a_i,\vec b_0,\dots,\vec b_i)$ is true, fix
a witness $F_i(\vec a_0,\dots,\vec a_i,\vec b_0,\dots,\vec b_{i-1})\in\two^m$ for the $\exists\vec b_i$ quantifier
block. Also, define
\[G_i(\vec a_0,\dots,\vec a_i)=
   F_i\bigl(\vec a_0,\dots,\vec a_i,G_0(\vec a_0),\dots,G_{i-1}(\vec a_0,\dots,\vec a_{i-1})\bigr)\]
by induction on~$i<d$. Since $\Phi_0$ is true, we can show by induction on~$i$ that
\begin{equation}\label{eq:2}
\Phi_i\bigl(\vec a_0,\dots,\vec a_{i-1},G_0(\vec a_0),\dots,G_{i-1}(\vec a_0,\dots,\vec a_{i-1})\bigr)
\end{equation}
is true for every $i\le d$ and every $\vec a_0,\dots,\vec a_{i-1}\in\two^m$. 

Let $W=\unifr\kiv{\vec p,q}$;
by \citi{Thm.~5.18}, it suffices to construct a valuation of $\vec x$ in~$W$ that makes $W\model\xi_\Phi$. Let
$w\in W$. If $w\model\theta_i\land\gamma$ for some $i<d$ (which is unique by~\eqref{eq:16}), let
$\vec a_0(w),\dots,\vec a_i(w)\in\two^m$ be the unique Boolean vectors such that $v\model p_j^{a_{i',j}(w)}$ for every
$i'\le i$, $j<m$, and~$v\ge w$ such that $v\model\theta_{i'}$: their existence and uniqueness follows from
\eqref{eq:17} and $w\model\gamma$. Putting $\vec b_i(w)=G_i\bigl(\vec a_0(w),\dots,\vec a_i(w)\bigr)$, we define
valuation of $\vec x$ in~$w$ so that $w\model x_j^{b_{i,j}(w)}$ for each $j<m$. If $w\nmodel\gamma$, or
$w\nmodel\theta_i$ for any $i<d$, we define $w\model x_j$ arbitrarily.

Assume that $w\model\gamma$, and fix $i<d$. Then every $v,v'\ge w$ satisfying~$\theta_i$ also
satisfy~$\gamma$, and we have $\vec a_{i'}(v)=\vec a_{i'}(v')$ for every $i'\le i$, hence $v$ and~$v'$ agree on the
satisfaction of~$\vec x$. It follows that $w\model\boxdot(\theta_i\to x_j)\lor\boxdot(\theta_i\to\neg x_j)$ for each
$j<m$. As $i$ was arbitrary, \eqref{eq:4} is true in~$W$.

By the same argument, if $w\model\theta_{d-1}\land\gamma$, then
\begin{align*}
w\model\diadot(\theta_i\land p_j)&\iff a_{i,j}(w)=1,\\
w\model\diadot(\theta_i\land x_j)&\iff b_{i,j}(w)=1,
\end{align*}
for each $i<d$ and $j<m$. Since $\Phi_d=\fii$, we have
$\fii\bigl(\vec a_0(w),\dots,\vec a_{d-1}(w),\vec b_0(w),\dots,\vec b_{d-1}(w)\bigr)$ by \eqref{eq:2} and the choice
of~$\vec b_i(w)$, thus \eqref{eq:5} holds in~$W$ as well.
\end{Pf*}
\begin{Cl}\label{cl:pspfalse}
If a logic $L$ has depth at least~$d$, and $\xi_\Phi$ is $L$-unifiable, then $\Phi$ is true.
\end{Cl}
\begin{Pf*}
By the assumption, there exists a finite $L$-frame~$F$ of depth~$d$ with root~$r$. Let $\sigma$ be an $L$-unifier of
$\xi_\Phi$, w.l.o.g.\ chosen so that its target formulas contain no variables, and only the parameters $\vec p,q$. Assume for
contradiction that $\Phi$ is false. Put $w\model q$ iff $w\in F$ has even depth, so that $w\model\theta_i$ iff $w$ has
depth~$i+1$.

By induction on~$i\le d$, we will fix valuation of~$\vec p$ in points of~$F$ of depth at most~$i$, and
$\vec a_0,\dots,\vec a_{i-1},\vec b_0,\dots,\vec b_{i-1}\in\two^m$ such that
$\Phi_i(\vec a_0,\dots,\vec a_{i-1},\vec b_0,\dots,\vec b_{i-1})$ is false. The valuation will be defined so that
$w\model p_j^{a_{i',j}}$ and $w\model\sigma(x_j)^{b_{i',j}}$ for every $w\in F$ of depth~$i'+1$, where $i'<i$; notice that
$\sigma(x_j)$ is a formula in the parameters $\vec p,q$, and its value in~$w$ is only affected by the valuation
of~$p_j$ in points of depth $i'+1\le i$ or less, hence it will not change in later steps.

The base step $i=0$ is trivial---we only observe that $\Phi_0$ is false by assumption. Assume the
construction has been carried out up to~$i<d$. Since $\Phi_i(\vec a_0,\dots,\vec a_{i-1},\vec b_0,\dots,\vec b_{i-1})$
is false, we can fix $\vec a_i\in\two^m$ such that
\begin{equation}\label{eq:6}
\forall\vec b_i\,\neg\Phi_{i+1}(\vec a_0,\dots,\vec a_i,\vec b_0,\dots,\vec b_i).
\end{equation}
Define $w\model p_j^{a_{i,j}}$ for all $j<m$ and $w\in F$ of depth~$i+1$, and temporarily put (say) $w\nmodel p_j$ for
all $w$ of depth more than~$i+1$. By the definition and the induction hypothesis, this makes $\gamma$ true in~$r$.
Since $\vdash_L\sigma(\xi_\Phi)$, we have
$r\model\boxdot(\theta_i\to\sigma(x_j))\lor\boxdot(\theta_i\to\neg\sigma(x_j))$. Thus, there exists a
unique~$\vec b_i\in\two^m$ such that $w\model\sigma(x_j)^{b_{i,j}}$ for every $w$ of depth~$i+1$ and $j<m$. As
explained above, this is independent of the temporary definition of valuation of~$\vec p$, as the valuation in points
of depth at most~$i+1$ has already been fixed. The formula $\Phi_{i+1}(\vec a_0,\dots,\vec a_i,\vec b_0,\dots,\vec b_i)$ is false
by~\eqref{eq:6}.

When the construction is finished, we have $r\model\theta_{d-1}\land\gamma$, and the valuation of $\diadot(\theta_i\land
p_j)$ and $\diadot(\theta_i\land\sigma(x_j))$ in~$r$ agrees with $a_{i,j}$ and~$b_{i,j}$, respectively, for each $i<d$
and $j<m$, hence $\fii(\vec a_0,\dots,\vec a_{d-1},\vec b_0,\dots,\vec b_{d-1})$ is true due
to~$\vdash_L\sigma(\xi_\Phi)$. However, this is a contradiction, as $\fii=\Phi_d$.
\end{Pf*}

For fixed $d$, validity of quantified Boolean sentences of the form~\eqref{eq:18} is a $\Pi^p_{2d}$-complete problem,
hence Claims \ref{cl:psptrue} and~\ref{cl:pspfalse} imply~\ref{item:1}. When $d$ is unrestricted, the validity problem
for~\eqref{eq:18} is $\psp$-complete, and we may arrange $m=1$ by inserting dummy quantifiers to make universal and
existential quantifiers alternate. This shows~\ref{item:2}.
\end{Pf}

We may strengthen Theorem~\ref{thm:lb-infdp} by combining it with our other lower bounds, observing that $\psp$ (and
$\Pi^p_{2d}$) is included in both $\NEXP$ and~$\coNEXP$:
\begin{Cor}\label{cor:lb-infdp}
\ \begin{enumerate}
\item\label{item:4}
Unifiability in all logics except tabular logics of width~$1$ is uniformly $\psp$-hard.
\item\label{item:5}
For every~$d>0$, unifiability in all logics except tabular logics of width~$1$ and depth at most~$d-1$ is uniformly
$\Pi^p_{2d}$-hard.

In particular, unifiability in all consistent logics is uniformly $\Pi^p_2$-hard.
\end{enumerate}
\end{Cor}
\begin{Pf}
Let $A$ be a language in $\psp$ or $\Pi^p_{2d}$ as appropriate. Let $w\mapsto\xi^0_w$, $w\mapsto\xi^1_w$, and
$w\mapsto\xi^2_w$ be the reductions supplied by Theorems \ref{thm:lb-nexp}, \ref{thm:lb-conexp}, and~\ref{thm:lb-infdp} (respectively). We may
assume that the formulas $\xi^0_w$, $\xi^1_w$, and $\xi^2_w$ use disjoint sets of variables. Then
$\xi_w:=\xi^0_w\land\xi^1_w\land\xi^2_w$ is $L$-unifiable if and only if each $\xi^i_w$ is, thus:
\begin{itemize}
\item if $w\in A$, then $\xi_w$ is $\kiv$-unifiable;
\item if $w\notin A$, then $\xi_w$ is not $L$-unifiable for any logic $L$ that is nonlinear, or of unbounded cluster
size, or of sufficiently large depth.
\end{itemize}
The three classes of logics in the second point together cover the class from the statement of this corollary.
\end{Pf}
\begin{Cor}\label{cor:psp-compl}
Let $L$ be a consistent linear clx logic of bounded cluster size. If $L$ has branching at least~$1$ (hence unbounded
depth), then $L$-unifiability and $L$-admissibility are $\psp$-complete; otherwise, $L$-unifiability is
$\Pi^p_2$-complete, and $L$-admissibility $\Sigma^p_2$-complete.

If $L$ is a linear tabular logic of depth (exactly) $d\ge1$, then $L$-unifiability is $\Pi^p_{2d}$-complete, and
$L$-admissibility is $\Sigma^p_{2d}$-complete.
\end{Cor}
\begin{Pf}
By Theorems \ref{thm:ub-lin-clx}, \ref{thm:ub-lin-bddp}, and~\ref{thm:lb-infdp}.
\end{Pf}
\begin{Exm}\label{exm:psp-compl}
$L$-unifiability and $L$-admissibility are $\psp$-complete for the logics $L=\lgc{GL.3}$, $\lgc{S4.3Grz}$, and~$\lgc{K4.3Grz}$.
Unifiability in $L\oplus\lgc{BD}_d$ is $\Pi^p_{2d}$-complete for $d\ge1$.

Unifiability is also $\Pi^p_2$-complete in
the logics $\lgc{Verum}=\kiv\oplus\Box\bot$ or $\lgc{S5}\oplus\lgc{Alt}_k$ for~$k\ge1$.
\end{Exm}

In general, two parameters are necessary in Theorem~\ref{thm:lb-infdp}~\ref{item:2}:
\begin{Exm}\label{exm:two-par}
The logic $L=\lgc{S4.2Grz}$ or the clx logic $L=\lgc{S4Grz.3}$ have unbounded depth, but $L$-unifiability of formulas
with one parameter is $\np$-complete, as shown in Example~\ref{exm:s42grz}.

Unifiability with one parameter in~$\lgc{S4.3}$ is also $\np$-complete: the universal frame is only modified by
attaching copies of the two-element cluster realizing both $p$ and~$\neg p$ below each original node, and including an
additional copy as a third connected component. We leave the details to the reader.
\end{Exm}

However, we will show for completeness that one parameter is sufficient under a stronger hypothesis, namely that the
logic admits arbitrarily long \emph{irreflexive} chains (possibly embedded among reflexive points). Note that at least
one parameter is needed anyway, as parameter-free unifiability is in~$\np$ even for irreflexive logics such
as~$\lgc{GL}$.

\begin{Def}\label{def:irr-dp}
A logic $L$ has \emph{irreflexive depth at least~$n$} if there exists an $L$-frame that subreduces onto an $n$-element
irreflexive chain. A logic has \emph{unbounded irreflexive depth} if it has irreflexive depth at least~$n$ for all
$n\in\omega$.
\end{Def}

\begin{Thm}\label{thm:lb-infirrdp}
Unifiability of formulas with $1$~parameter and $1$~variable in logics of unbounded irreflexive depth is uniformly
$\psp$-hard.
\end{Thm}
\begin{Pf}
As in the proof of Theorem~\ref{thm:lb-infdp}~\ref{item:1}, we will provide a reduction from validity of quantified Boolean
sentences of the form~\eqref{eq:18} with $m=1$, but instead of $p_0$ and~$q$, we will need to reuse a single
parameter~$q$ for detecting the depth and for specifying~$a_i$. (Since $m=1$, we will omit the vector signs, and write
just $a_i,b_i$ instead of $a_{i,0},b_{i,0}$.) Roughly, the idea is that each quantifier pair $\forall a_i\,\exists b_i$
corresponds to three irreflexive points in a chain: the top one will satisfy~$\neg q$, the bottom one~$q$, and the
middle one will depend on~$a_i$. In this way, we get two alternations in the value of $q$ for each~$i$ irrespective of
the value of~$a_i$, while we can read off $a_i$ by checking if the $i$th layer of points satisfying~$q$ contains a
nontrivial chain $u<v$. (Here we are using the irreflexivity.) The value of $b_i$ will be given using a variable~$x$,
one level lower still (so that $a_i$ is reliably fixed).

Formally, let $\theta_i$ denote the formulas defined in the proof of Theorem~\ref{thm:lb-infdp}, and let $\xi_\Phi$ be
the conjunction of the formulas
\begin{align}
\label{eq:19}
\gamma&\to\ET_{i<d}\bigl(\boxdot(\theta_{2i+2}\to x)\lor\boxdot(\theta_{2i+2}\to\neg x)\bigr),\\
\label{eq:20}
\gamma\land\theta_{2d}&\to\fii\bigl(\dots,a_i/\dia(\theta_{2i+1}\land\dia\theta_{2i+1}),\dots,b_i/\diadot(\theta_{2i+2}\land x),\dots\bigr),
\end{align}
where
\[\gamma=\ET_{i<d}\bigl(\boxdot(\theta_{2i+1}\to\Box\neg\theta_{2i+1})
              \lor\boxdot(\theta_{2i+2}\to\dia(\theta_{2i+1}\land\dia\theta_{2i+1}))\bigr).\]
Again, $\Phi\mapsto\xi_\Phi$ is a logspace-computable mapping.
\begin{Cl}\label{cl:psptrue-1p}
If $\Phi$ is true, then $\xi_\Phi$ is $\kiv$-unifiable.
\end{Cl}
\begin{Pf*}
Define the functions $F_i$ and~$G_i$ as in Claim~\ref{cl:psptrue}. Let $W=\unifr\kiv q$; we will construct a valuation of the
variable~$x$ in~$W$ as follows. Let $w\in W$. If $w\model\theta_{2i+2}$ for some $i<d$, let
$a_0(w),\dots,a_i(w)\in\two$ be defined by
\[a_{i'}(w)=1\iff w\model\dia(\theta_{2i'+1}\land\dia\theta_{2i'+1}).\]
Put $b_i(w)=G_i\bigl(a_0(w),\dots,a_i(w)\bigr)$, and define $w\model x^{b_i(w)}$. For other $w$, define $w\model x$ in
an arbitrary way.

A similar argument as in Claim~\ref{cl:psptrue-1p} shows that under this valuation, all points of~$W$ satisfy \eqref{eq:19}
and~\eqref{eq:20}. Thus, $\xi_\Phi$ is $\kiv$-unifiable by \citi{Thm.~5.18}.
\end{Pf*}
\begin{Cl}\label{cl:pspfalse-1p}
If $L$ has irreflexive depth at least~$3d+1$, and $\xi_\Phi$ is $L$-unifiable, then $\Phi$ is true.
\end{Cl}
\begin{Pf*}
Let us fix an $L$-frame~$W$, and a subreduction $f$ from $W$ onto the irreflexive chain of length $3d+1$, whose elements
we label as $c_0>c_1>\dots>c_{3d}$. We may assume that $W$ is rooted, and its root $r$ maps to~$c_{3d}$. For any
$i\le3d$, put $D_i=f^{-1}[c_i]$. The sets $D_i$ are pairwise disjoint admissible subsets of the frame~$W$; each $D_i$
is a nonempty irreflexive antichain, and for each $i'<i$, $D_{i'}$ is a cover of~$D_i$ (i.e., every element of $D_i$
sees some element of $D_{i'}$). We have $D_{3d}=\{r\}$.

For each $a=\p{a_0,\dots,a_{d-1}}\in\two^d$, let $W_a$ denote the parametric frame for parameter~$q$ based on~$W$ such
that
\[\{w\in W_a:w\model q\}=\bigcup_{i<d}D_{3i+2}\cup\bigcup_{\substack{i<d\\a_i=1}}(D_{3i+1}\Down\bez D_{3i+2}\Down).\]
It is straightforward to check that in~$W_a$:
\begin{itemize}
\item For each $i<d$, points of~$D_{3i+2}$ satisfy~$\theta_{2i+1}$.
\item Points of $W\bez D_1\Down$ satisfy~$\theta_0$, points of $D_{3d-1}\down$ satisfy~$\theta_{2d}$, and for each
$0<i<d$, points of $D_{3i-1}\down\bez D_{3i+1}\Down$ satisfy~$\theta_{2i}$. In particular, for each $i\le d$, points
of~$D_{3i}$ satisfy~$\theta_{2i}$.
\item For each $i<d$, points of $D_{3i+1}\Down\bez D_{3i+2}\Down$ (which includes~$D_{3i+1}$) satisfy $\theta_{2i}$
or~$\theta_{2i+1}$ according to the value of~$a_i$.
\end{itemize}
As a consequence, $W_a\model\gamma$: specifically, for any $i<d$, $W_a\model\theta_{2i+1}\to\Box\neg\theta_{2i+1}$ if
$a_i=0$, and $W_a\model\theta_{2i+2}\to\dia(\theta_{2i+1}\land\dia\theta_{2i+1})$ if $a_i=1$.

Now, assume that $\xi_\Phi$ is $L$-unifiable, and let us fix a unifier~$\sigma$ of~$\xi_\Phi$ whose target
formulas contain no variables, and no parameters save~$q$. We define a function $G\colon\two^d\to\two^d$ as follows:
given $a\in\two^d$, we have $W_a\model\sigma(\xi_\Phi)$ as $\sigma$ is a unifier, and we know $W_a\model\gamma$,
thus using~\eqref{eq:19}, there is for each $i<d$ a unique $b_i\in\two$ such that
$W_a\model\theta_{2i+2}\to\sigma(x)^{b_i}$. We put $G(a)=\p{b_0,\dots,b_{d-1}}$.

By the breakdown above, points $w\in D_{3i+3}$ satisfy $\theta_{2i+2}$ in~$W_a$, and the valuation of~$q$ in~$w\Up$ is
determined by $a_0,\dots,a_i$, hence so is the valuation of $\sigma(x)$. That is, $b_i$ only depends on
$a_0,\dots,a_i$: there are functions $G_i\colon\two^{i+1}\to\two$, $i<d$, such that
\[G(a_0,\dots,a_{d-1})=\p{G_0(a_0),G_1(a_0,a_1),\dots,G_{d-1}(a_0,\dots,a_{d-1})}\]
for each $a\in\two^d$. Since $W_a\model\sigma(\xi_\Phi)$, we have
\[\fii\bigl(a_0,\dots,a_{d-1},G_0(a_0),\dots,G_{d-1}(a_0,\dots,a_{d-1})\bigr)=1\]
using~\eqref{eq:20}. This makes $G_0,\dots,G_{d-1}$ Skolem functions witnessing the truth of~$\Phi$: formally, we may
show
\[\forall a_0,\dots,a_{i-1}\in\two\,\Phi_i\bigl(a_0,\dots,a_{i-1},G_0(a_0),\dots,G_{i-1}(a_0,\dots,a_{i-1})\bigr)=1\]
by reverse induction on $i\le d$, with $\Phi_i$ defined by~\eqref{eq:21}.
\end{Pf*}

The result now follows from Claims \ref{cl:psptrue-1p} and~\ref{cl:pspfalse-1p}.
\end{Pf}
\begin{Cor}\label{cor:psp-1par}
Unifiability of formulas with $1$~parameter and $1$~variable is $\psp$-hard in all consistent $\p{\I,1}$-extensible logics.
\noproof\end{Cor}

\section{Conclusion}\label{sec:conclusion}

We have undertaken a thorough investigation of the complexity of admissibility and unifiability with parameters in
transitive modal logics. In particular, we obtained a classification of these problems for cluster-extensible logics, a
number of results for logics of bounded depth and width, and several very general lower bounds on hardness of
unifiability in broad classes of transitive logics.

When we reflect on the effect of parameters on the complexity, one recurring observation is that the power of
having an unlimited supply of parameters can be fully exploited only for logics that allow arbitrary large clusters,
but in that case it blows up the complexity of unifiability to $\coNEXP$ (or from $\NEXP$ to~$\Sigma^{\exp}_2$). This is
perhaps most striking for~$\lgc{S5}$, which is at the same time one of the most fundamental and one of the most simple
modal logics; the derivability problem for~$\lgc{S5}$ is $\conp$-complete, and this also holds for parameter-free
admissibility, or even admissibility with a constant number of parameters. However, $\lgc{S5}$-unifiability with
arbitrary parameters is $\coNEXP$-complete, i.e., exponentially harder.

A perhaps more intriguing effect is that already the addition of $1$--$2$ parameters may raise the complexity of
unifiability from $\np$ to~$\psp$ in logics of unbounded depth. Even better, we saw that for some logics such as
$\lgc{GL}$ or~$\lgc{D4}$, unifiability with $1$ parameter is $\NEXP$-complete, but parameter-free unifiability is
$\np$-complete, hence again we have an exponential blow-up. In these cases, the complexity of parameter-free
unifiability is exceptionally low (e.g., parameter-free inadmissibility is already $\NEXP$-complete).

This brings us to one of the rough spots where our analysis is incomplete, even for clx logics: what is the complexity
of parameter-free unifiability in nonlinear clx logics? There are some cases where it is $\NEXP$-complete (see
Theorem~\ref{thm:nexp-0par}), and some cases where it is $\np$-complete (e.g., extensions of $\lgc{GL}$ or~$\lgc{D4}$, as
just mentioned), but we do not know where exactly is the dividing line, or even if there are cases of intermediate
complexity.

In the regime with $O(1)$ parameters, our results are enough to settle the complexity for clx logics and some other
logics of interest, nevertheless the results are far from satisfactory: the conditions used for the lower bounds have an
ad hoc feeling to them, and are unlikely to be exhaustive. We suspect that a more comprehensive classification might be
quite hard here.

Finally, in the regime with arbitrarily many parameters, our lower and upper bounds do not match for some
interesting classes of logics that are nonlinear and of unbounded cluster size, but do not satisfy the hypothesis of
Theorem~\ref{thm:lb-sig2exp}. As we already mentioned in the introduction, this gap will be addressed in the sequel.

\subsection*{Acknowledgements}

I want to thank the anonymous referee whose suggestions improved the presentation of the paper.

The research was supported by grant 19-05497S of GA~\v CR. The Institute of Mathematics of the Czech Academy of
Sciences is supported by RVO: 67985840.

\bibliographystyle{mybib}
\bibliography{mybib}
\end{document}